\documentclass[usenatbib,useAMS]{mn2e}

\usepackage{rotating}
\usepackage{graphicx}
\usepackage{amsmath}
\usepackage{amssymb}
\usepackage{bbold}
\usepackage{multirow}
 
\usepackage{xcolor}

\usepackage{float}

\newcommand{\mbi}[1]{\mbox{\boldmath$#1$}}

\newcommand{\mat}[1]{\mbox{\rm\bf #1}}
\newcommand{\lsim}{\mbox{${\,\hbox{\hbox{$ < $}\kern -0.8em \lower 1.0ex\hbox{$\sim$}}\,}$}}
\newcommand{\gsim}{\mbox{${\,\hbox{\hbox{$ > $}\kern -0.8em \lower 1.0ex\hbox{$\sim$}}\,}$}}

\newcommand{\dd}{{\rm d}}

\def\beqn{\vspace{2mm}
\begin{eqnarray}} 
\def\eeqn{\vspace{2mm} 
\end{eqnarray}}

 \def\la{\langle}

\newcommand{\be}{\begin{equation}}
\newcommand{\ee}{\end{equation}}
\newcommand{\ba}{\begin{eqnarray}}
\newcommand{\ea}{\end{eqnarray}}
\newcommand{\brr}{\begin{array}}
 
\newcommand{\err}{\end{array}}
\newcommand{\bc}{\begin{center}}
\newcommand{\ec}{\end{center}}

\begin{document}
\title[Linearisation with Cosmological PT]{Linearisation with Cosmological Perturbation Theory}

\author[F.~S.~Kitaura and R.~E.~Angulo]{Francisco-Shu Kitaura$^{1,2}$\thanks{E-mail: kitaura@aip.de, Karl-Schwarzschild fellow} and  Raul E.~Angulo$^{2}$\thanks{E-mail: rangulo@mpa-garching.mpg.de}\\
 $^{1}$ Leibniz-Institut f\"ur Astrophysik Potsdam (AIP), An der Sternwarte 16, D-14482 Potsdam, Germany \\
$^{2}$ Max-Planck Institut f\"ur Astrophysik (MPA), Karl-Schwarzschildstr.~1, D-85748 Garching, Germany 
}

\maketitle

\begin{abstract}
We propose a new method to linearise cosmological mass density fields using
higher order Lagrangian perturbation theory (LPT). We demonstrate that a given
density field can be expressed as the sum of a linear and a nonlinear component
which are tightly coupled to each other by the tidal field tensor within the
LPT framework. The linear component corresponds to the initial density field in
Eulerian coordinates, and its mean relation with the total field can be
approximated by a logarithm (giving theoretical support to recent attempts to
find such component).  We also propose to use a combination of the
linearisation method and the continuity equation to find the mapping between
Eulerian and Lagrangian coordinates. In addition, we note that this method
opens the possibility of use directly higher order LPT on nonlinear fields.  We
test our linearization scheme by applying it to the $z\sim0.5$ density field
from an $N$-body simulation. We find that the linearised version of the full
density field can be successfully recovered on $\gsim{5}$ $h^{-1}$Mpc, reducing
the skewness and kurtosis of the distribution by about one and two orders of
magnitude, respectively. This component can also be successfully traced back in
time, converging towards the initial unevolved density field at $z\sim100$. We
anticipate a number of applications of our results, from predicting velocity
fields to estimates of the initial conditions of the universe, passing by
improved constraints on cosmological parameters derived from galaxy clustering
via reconstruction methods.
\end{abstract}


\section{introduction}

The present-day mass density field contains information about the fundamental
pillars of modern cosmology. It is a mixture and cross-talk between the
primordial hierarchy of correlation functions of fluctuations, the law of
gravity and the value of cosmological parameters. Unfortunately, disentangling
all these ingredients and extracting useful information about them is not a
trivial task. On large scales this is still relatively simple; linear theory
applies and different Fourier modes evolve independently from each other. In
fact, thanks to these features, cosmological parameters are almost routinely
constrained using large-scale galaxy clustering  \citep[see
e.g.][]{2005MNRAS.362..505C,2005ApJ...633..560E,2006A&A...459..375H,2007MNRAS.374.1527B,2010MNRAS.401.2148P,2011MNRAS.415.2892B}.
The description of small scales is much more difficult; highly nonlinear
processes are in place, gravity couples perturbations on different scales and
additional complications arise from nonlinear galaxy biasing and redshift space
distortions.

Different approaches have been proposed to recover the primordial, linear and
Gaussian, density field on medium- or small-scales -- a process
usually referred to as ``Gaussianisation'' or ``Linearisation''. The majority
are based on a local rank-ordered mapping, where the $n$-th largest density
fluctuation in one field causes the $n$-th largest perturbation in the other.
Examples of this are;  the logarithm of the local density field
\citep{2009ApJ...698L..90N,2011ApJ...731..116N,2011arXiv1104.1399J}, a local
Gaussianisation assuming that the primordial probability distribution function
(PDF) of densities is known \citep{weinberg,2011arXiv1103.2858Y},
and applying linear or closely linear filters (by simple Gaussianisation
\citep{2011ApJ...731..116N} or more sophisticated Wiener-filtering with a
wavelet truncation \citep{2011ApJ...728...35Z}).  Although these methods have
shown to accomplish their goals (with different degrees of success), they still
lack a rigorous theoretical motivation and support.  Additionally, gravity is a
nonlocal process, where the evolution of a density fluctuation is not only
determined by its amplitude, but it also depends on the surrounding tidal field
\citep{2006MNRAS.371.1205R,2005MNRAS.360L..82R}. Therefore, a rank-preserving
mapping can not be correct in detail. 

In this work we propose and explore another way to recover the initial density
field. Our method takes advantage of the fact that an initial, linear field and
its gravitationally evolved counterpart are not independent from each other,
but are related through the tidal field tensor. This gives us an extra piece of
information to constrain and improve their mapping. Furthermore, the tidal
tensor can be predicted analytically, and be fully specified by a linear field,
using higher order Lagrangian perturbation theory \citep[LPT,
][]{1994A&A...288..349B,1995A&A...296..575B,1998MNRAS.299.1097S,2002PhR...367....1B}.
Putting these ingredients together allows us to uniquely identify the initial
density field which, evolved under 2LPT gravity, would give rise to the final
density field we aim to linearise.  We note that similar approaches have been
explored in the past but with limited success and different scopes
\citep[see][]{1993ApJ...405..449G,1999MNRAS.308..763M}. 

With this physically motivated Gaussianisation process we can interprete
rank-ordering mappings, in particular, the widely used logarithmic
transformation.  Explicitly, we find that the mean transformation between the
nonlinear and LPT linearised density fields can be approximated by a logarithmic
function, consistent with the solution of the linear version of the continuity
equation. In addition, we demonstrate that the linearised field is a good
estimate of the initial field, not at its respective (earlier) time but at the
present.  Thus, one needs to trace this linearised field back in time, or,
equivalently, to find the transformation from Eulerian to Lagrangian
coordinates, if this field is to be used for constrained simulations and/or
improved cosmological parameters constraints. One exception occurs on large
scales or high redshifts, where Eulerian and Lagrangian coordinates coincide.

We validate these ideas by applying our 2LPT linearisation method to a density
field extracted from a cosmological $N$-body simulation. The PDF of the
resulting linearised field is closely described by a Gaussian function. In
addition, this field correlates with the high redshift outputs of the
simulation, much more than the original field. The correlation increases further 
when the linearised field is traced back in time. All this
on scales even as small as $\sim 5\,h^{-1}$Mpc.

An useful consequence of a Gaussianisation, is that the linearised field in
Eulerian coordinates can be used as an input for Lagrangian perturbation theory
and consistently predict the associated velocity, displacement and future
density fields. 
A comparison between estimations of the displacement field in
Eulerian coordinates and the linearised field, as given by the logarithmic
transformation, was provided in a recent paper \citep[][]{2011arXiv1111.4466F}.
The latter field could be specially useful for reconstruction of the
large-scale density field in general, and of Baryonic Acoustic Oscillations
(BAO) in particular
\citep[see][]{2007ApJ...664..675E,2009PhRvD..80l3501N,2011ApJ...734...94M}.  In
a companion paper we show that an accurate estimation of peculiar velocities
can also be obtained in this way. In this case, the usage of our estimation of
the linear field yields to results superior to those obtained by using a
logarithmic transformation \citep[][]{kitvel}. In subsequent papers we will
address other applications of the linearised density field. 

The paper is structured as follows.  In section \S \ref{sec:theory} we present
the theoretical basis for our 2LPT and for the logarithmic linearisation. We
also derive the equations governing the time-reversal of the linearised field.
We discuss a practical implementation in \S \ref{sec:solution} and discuss its
performance once applied to a $N$-body simulation in \S \ref{sec:results}.
Finally, we present our conclusions.

\section{Theory}
\label{sec:theory}

In this section we recap Lagrangian perturbation theory and show how a
gravitationally evolved density field can be expanded into a linear and a
nonlinear component.  We also show that the widely used lognormal
transformation gives an estimate of this linear component in Eulerian
coordinates. Finally, we derive the equations to trace a density field back in
time, allowing us to express the linear component (or linearised field) in
Lagrangian coordinates, which then corresponds to the actual initial density
field.

\subsection{Lagrangian perturbation theory linearisation}
\label{sec:LPT}

Let us start by considering the mapping between the comoving coordinates of a set
of test particles at two redshifts $\mbi x(z)$ and $\mbi q(z_0)$, with $z < z_0$.  
In Lagrangian perturbation theory this relation is expressed via a displacement
field, ${\mbi\Psi}({\mbi q})$ \citep[see e.~g.~][]{2002PhR...367....1B}:

\begin{equation} 
\label{eq:lag}
 {\mbi x} = {\mbi q} + {\mbi\Psi}({\mbi q}) \, .
\end{equation}

\noindent which defines a unique mapping between $\mbi q$
and $\mbi x$ (usually referred to as Lagrangian and Eulerian coordinates). We
note that such a description of gravitational clustering starts breaking down when shell-crossing begins.    If we further assume that the test particles were initially
homogeneously distributed, then we can write the following mass conservation
relation:  

\be
\rho(\mbi x,z)\dd \mbi x=\langle \rho(z_0)\rangle \dd \mbi q \,.
\ee

The inverse of the Jacobian of the coordinate transformation defines the 
overdensity field, $\delta \equiv \rho/\langle \rho\rangle-1$, : 

\be
\label{eq:jac}
1+\delta(\mbi x(\mbi q,z))=\mat J(\mbi q,z)^{-1} \,,
\ee
with
\be
\mat J(\mbi q,z) \equiv \left|\frac{\partial\mbi x}{\partial\mbi q}\right| \,.
\ee

\noindent Combining Eqs.~(\ref{eq:lag}) and (\ref{eq:jac}) we obtain an expression for the density in Lagrangian coordinates $\mbi q$ using a relation for determinants of matrices and assuming curl-free velocity fields  \citep[$\mbi \Psi=-\nabla \Theta$, for a discussion on this see][]{kitvel}:

\ba
\label{eq:expa}
\delta(\mbi q,z)&=&\left|1+\nabla_{q}\cdot\mbi \Psi(\mbi q,z)\right|^{-1}-1\\
&\simeq& - \nabla_{q}\cdot\mbi \Psi(\mbi q,z) +\mu^{(2)}[\Theta](\mbi q,z)+\mu^{(3)}[\Theta](\mbi q,z) \nonumber\,,
\ea

\noindent where the subscripts
${\mbi q}$ refer to partial derivatives with respect to ${\mbi q}$.
{\color{black} Here one should note that we expand the inverse of the Jacobian. The importance of this will become clear below.}
 The second term in Jacobian expansion is given by;

\be
\mu^{(2)}[\Theta]({\mbi q},z)=\sum_{i>j} 
    \Big( \Theta_{,ii}({\mbi q},z)\Theta_{,jj}({\mbi q},z)-
    [\Theta_{,ij}({\mbi q},z)]^2\Big)\, ,
\ee

\noindent where we use  the abbreviation
$\Theta_{,ij} \equiv \partial^2\Theta/\partial q_i\partial q_j$.
The third term is:
\be
\mu^{(3)}[\Theta]=\det\left(\Theta_{,ij}\right)\,.
\ee

Note that the density field at redshift $z$ expressed in 
Lagrangian coordinates, $\delta(\mbi x(\mbi q,z))$, is fully
determined by the displacement field. This field in turn
can be calculated within 2nd order LPT (2LPT), in particular it is given in terms 
of two potentials:

\begin{equation}  \label{eq:psi} 
 \mbi \Psi(\mbi q,z)  =  - D(z)\nabla_q\phi^{(1)}(\mbi q) + D_2(z)\nabla_q\phi^{(2)}(\mbi q),
\end{equation}

\noindent and consequently;

\begin{equation}  \label{eq:theta} 
 \mbi \Theta (\mbi q,z) =  D(z)\phi^{(1)}(\mbi q) - D_2(z)\phi^{(2)}(\mbi q),
\end{equation}

\noindent where $D$ is the linear growth factor, and $D_2$ the second
order growth factor given by $D_2 =\alpha D^2$ and $\alpha\approx -3/7$. 
The linear $\phi^{(1)}$ and nonlinear potential $\phi^{(2)}$ are obtained by
solving a pair of Poisson equations: $\nabla^2_q\phi^{(1)}({\mbi q}) =  \delta^{(1)}({\mbi q})$, where $\delta^{(1)}({\mbi q})$ is the linear overdensity, and 
  $\nabla^2_q\phi^{(2)}({\mbi q}) =  \delta^{(2)}({\mbi q})$.

The term $\delta^{(2)}({\mbi q})$  includes the effects of tidal forces and represents the
`second-order overdensity' which is related to the
linear overdensity field by the following quadratic expression \citep[see e.g.][]{1995A&A...296..575B}:
\begin{equation}\label{twolpt_source}
 \delta^{(2)}({\mbi q})=\sum_{i>j} 
    \Big( \phi^{(1)}_{,ii}({\mbi q})\phi^{(1)}_{,jj}({\mbi q})-
    [\phi^{(1)}_{,ij}({\mbi q})]^2\Big)\,.
\end{equation}

\noindent Inserting these relations in Eq.~(\ref{eq:expa}) we get the desired decomposition of the field

\be
\label{eq:expa2}
\delta(\mbi q,z)=\delta^{\rm L}(\mbi q,z)+\delta^{\rm NL}(\mbi q,z) \,,
\ee

\noindent where $\delta^{\rm L}(\mbi q,z)=D(z)\delta^{(1)}({\mbi q})$ is the
linear component of the density field and the rest being the nonlinear part $\delta^{\rm NL}(\mbi q,z)=-D_2(z)\delta^{(2)}({\mbi q})+\mu^{(2)}[\Theta](\mbi q,z)+\mu^{(3)}[\Theta](\mbi q,z)$. From now on we will also use the
following notation for short $\delta_D=\delta/D(z)$.

Note that Eq.~\ref{eq:expa2} is only a function of the coordinates $\mbi q$ and 
the redshift $z$. The full nonlinear density field $\delta(\mbi q,z)$ is 
expressed in Lagrangian coordinates while it naturally should be expressed in
Eulerian coordinates.


{\color{black} Let us therefore derive the analougous expression in the Eulerian frame. We consider now the inverse transformation with respect to Eq.~\ref{eq:lag}:

\begin{equation} 
\label{eq:eul}
 {\mbi q} = {\mbi x} - {\mbi\Psi}({\mbi x}) \, .
\end{equation}

Mass conservation leads now to the following relation \citep[see][]{1991ApJ...379....6N}:

\be
\label{eq:jacE}
1+\delta(\mbi q(\mbi x,z))= \tilde{\mat J}(\mbi x,z) \,,
\ee
with
\be
 \tilde{\mat J}(\mbi x,z) \equiv \left|\frac{\partial\mbi q}{\partial\mbi x}\right|\,.
\ee

From which we get:

\ba
\label{eq:expaE}
\delta(\mbi x,z)&=&\left|1-\nabla_{x}\cdot\mbi \Psi(\mbi x,z)\right|-1\\
&\simeq& - \nabla_{x}\cdot\mbi \Psi(\mbi x,z) +\mu^{(2)}[\Theta](\mbi x,z)+\mu^{(3)}[\Theta](\mbi x,z) \nonumber\,.
\ea

We have thus found that the Eulerian and Lagrangian descriptions are equivalent (Eqs.~\ref{eq:expa} and \ref{eq:expaE}) when the Jacobian is expanded in the Eulerian frame and the inverse of the corresponding Jacobian is expanded in Lagrangian coordinates\footnote{One should note that this equivalency is not true when the Jacobian is expanded in the Lagrangian frame and then the inverse of that expansion is taken as it is done in \citet[][]{1999MNRAS.308..763M}.}.}
We
will use this result in either  formulation and leave therefore the coordinate
dependence out. {\color{black} One should note that the formulations given by Eqs.~\ref{eq:expa} and \ref{eq:expaE} do not transform the density fields from one frame to the other. This point will be further clarified in our numerical experiments presented in \S \ref{sec:lin}. }

Integrating Eq.~(\ref{eq:expa2}) we get an analogous expression for the full potential:

\be
\label{eq:main}
D(z)\phi_g = D(z)\phi^{(1)} + \phi^{\rm NL} 
\ee

\noindent with  $\phi^{\rm NL}=-D_2(z)\phi^{(2)}[\phi^{(1)}]+\phi^{(2)}[\Theta(\phi^{(1)})]+\phi^{(3)}[\Theta(\phi^{(1)})]$ and following operator notation $\phi^{(2)}[\phi]\equiv\nabla^{-2}\mu^{(2)}[\phi]$ and  $\phi^{(3)}[\phi]\equiv\nabla^{-2}\mu^{(3)}[\phi]$ with $\phi$ being some field.

This equation tells us how an evolved
gravitational potential is fully determined by its associated linear potential.
Therefore, linearising a field then becomes an inversion problem, which in
section \S\ref{sec:solution} we discuss how to solve.

\subsection{Lognormal linearisation}

Here we investigate the lognormal transformation as a mean to get an estimate of the linear component of the density field.
Let us follow \citet[][]{1991MNRAS.248....1C} and start with the continuity equation describing the matter content in the Universe as a fluid:

\be
\frac{\partial \rho}{\partial t}+\frac{1}{a}\nabla\left(\rho\cdot\mbi u\right)=0\,,
\ee

\noindent which can be expanded 

\be
\frac{\partial \rho}{\partial t}+\frac{1}{a}\left(\mbi u\cdot\nabla\right)\rho+\frac{1}{a}\rho\nabla\cdot\mbi u=0\,.
\ee

We can write this equation in Lagrangian coordinates introducing the total derivative

\be
\frac{\dd \rho}{\dd t}=\frac{\partial \rho}{\partial t}+\frac{1}{a}\left(\mbi u\cdot\nabla\right)\rho\,.
\ee

If we also switch to conformal time $a\dd \tau=\dd t$ then we find

\be
\frac{1}{\rho}\frac{\dd \rho}{\dd \tau}=-\nabla\cdot\mbi u
\ee

As long as we can follow particles (no shell-crossings) we may also write the continuity equation as

\be
\label{eq:lognormal}
\ln (1+\delta)=-\int\dd \tau \, \nabla\cdot\mbi u\,.
\ee

One must be especially careful at this point as the divergence of the peculiar
velocity field in the right hand side of the latter equation is in Eulerian
coordinates and not in Lagrangian coordinates \citep[for such an approach
see][]{1992MNRAS.259..437M}. The expansion of this term is not straightforward
for this reason. 

According to LPT (see Eq.~\ref{eq:jac}) we have yet another expression for the
logarithm of the density field, which can be Taylor expanded

 {\color{black}
\ba
\label{eq:declog}
\ln (1+\delta)&=&\ln(\tilde{J})\,,\nonumber\\
&&\hspace{-1.5cm}= \ln\left(1-\nabla_{x}\cdot\mbi \Psi +\mu^{(2)}[\Theta]+\mu^{(3)}[\Theta]\right)\,,\nonumber\\
&&\hspace{-1.5cm}\simeq \delta^{\rm L}+\delta^{+}(\delta^{\rm L})\,,
\ea
}

\noindent where the quantity $\delta^+$ summarises all the higher order terms.
{\color{black} The decomposition in Eq.~\ref{eq:declog} can always be done. 
The important point to be noticed is that} $\delta^{+}$ is in general a nonlocal and nonlinear function of
$\delta^{\rm L}$.  Taking the ensemble average\footnote{\color{black}The ensemble average is taken over all possible linear fields (in Eulerian coordinates) $\langle  \dots  \rangle\equiv\langle \dots \rangle_{\delta^{\rm L}(\mbi x)}$. Assuming a {\it fair} sample the ensemble average is reduced to a volume average in Eulerian coordinates $\langle  \dots  \rangle\equiv\langle \dots \rangle_{\mbi x}$.} of the previous equation we find that 

\be
\langle \delta^{+}\rangle =\mu=\langle\ln (1+\delta)\rangle\,,
\ee

\noindent since $\langle \delta^{\rm L}\rangle=0$, and

\be
\delta^{\rm L}\simeq \ln (1+\delta)-\mu\,.
\ee

In this way we have demonstrated that first order Taylor expansion of the
higher order corrections are given by the mean field:
$\langle\delta^{+}\rangle=\mu$. Note, that in reality the term
$\delta^{+}=\delta^{+}(\mbi x,z)$ will not be a homogeneous  field. In order to
improve this one has to make higher order expansions.  The importance of
computing the mean field $\mu$ was especially emphasized in
\citet[][]{kitaura_log}. A way to compute this field from the linear field
$\delta^{\rm L}$ was presented in \citet[][]{kitaura_lyman}. Note that the lognormal transformation will be equal to minus
the divergence of the displacement field $\delta^{\rm L}=-\nabla\cdot\mbi\Psi$
{\it only} in linear Lagrangian perturbation theory.

\subsection{Time-reversal evolution equations: the Eulerian-Lagrangian approach}
\label{sec:evo}

In this section we investigate different formulations of the continuity
equation  which permit to trace the structures back in time. In particular we
find an expression which shows how the linear component can be iteratively
traced back in time. For the derivation of such an equation we will combine the
results from Lagrangian perturbation theory (based on the equation of motion)
with an Eulerian formulation of the continuity equation.

Let us express the continuity equation as a function of the overdensity
$\delta$ and a scaled peculiar velocity  given by $\mbi v\equiv \mbi
u/\dot{D}=\mbi u/(fHD)$ 

\ba
\label{eq:continuity}
&&\frac{\partial \delta}{\partial D}+\nabla\cdot\left(\left(1+\delta\right)\mbi v\right)=0\,,\nonumber\\
&&\frac{\partial \delta}{\partial D}+\left(1+\delta\right)\nabla\cdot\mbi v+(\mbi v\cdot\nabla)\delta=0\,.
\ea

We could try to directly integrate this equation back in time computing in
each iteration the peculiar velocity field from the updated density field.
However, let us derive a formulation of the continuity equation which can be
better compared with previous works and ensures time-reversibility.  Following
\citet[][]{1993ApJ...405..449G} we define the deviation from linear theory as 

$\delta_{gv}\equiv\delta/D-\delta_{\rm LPT}/D=\delta/D+\nabla\cdot\mbi v$ with 
$\delta_{\rm LPT}\equiv -D\,\nabla\cdot\mbi v$.

We can then rewrite Eq.~\ref{eq:continuity} as

\be
\label{eq:continuity2}
{\frac{\partial \delta}{\partial D}-\frac{\delta}{ D}-D\,\left(\left(\nabla\cdot\mbi v\right)^2+\left(\mbi v\cdot\nabla\right)\nabla\cdot\mbi v\right)}
+\delta_{gv}+D\,\nabla\cdot\left(\delta_{gv}\mbi v\right)=0\,.
\ee

Under the assumption that flows are  irrotational: 

$\left(\nabla\cdot\mbi v\right)^2+\left(\mbi v\cdot\nabla\right)\nabla\cdot\mbi v=\frac{1}{2}\nabla^2v^2+2\delta^{(2)}[\phi_v]$ (with $\mbi v=-\nabla\phi_v$), Eq.~\ref{eq:continuity2} is simplified to
\be
{\frac{\partial \delta}{\partial D}-\frac{\delta}{ D}-D\,\left(\frac{1}{2}\nabla^2v^2+2\mu^{(2)}[\phi_v]\right)}+\delta_{gv}+D\,\nabla\cdot\left(\delta_{gv}\mbi v\right)=0\,.
\ee

Integrating the latter equation we obtain

\be
\frac{\partial \phi_g}{\partial D}-\frac{1}{2}v^2-2\phi^{(2)}[\phi_v]+\frac{1}{D}\phi_{gv}+\nabla^{-2}\nabla\cdot\left(\delta_{gv}\mbi v\right)=0\,,
\ee

\noindent with $\delta_{gv}=\nabla^{2}\phi_{gv}$.

\subsubsection{2nd order continuity equation}

Let us consider only terms up to second order, i.~e. neglecting terms involving
${\mathcal O}(D^3)$.  The velocity is given by $\mbi
v^{[2]}\equiv-\nabla\phi^{(1)}+\frac{f_2D_2}{fD}\nabla\phi^{(2)}[\phi^{(1)}]$.
Hereafter the numbers in brackets denote the order of the expansion.
Accordingly, the gravitational potential is given by
$\phi_g^{[2]}\equiv\phi^{(1)}+
\frac{1}{D}\left(D^2-D_2\right)\phi^{(2)}[\phi^{(1)}]$.  Note, that the
peculiar velocity $\mbi u$ needs to be linear: $\mbi
v^{[1]}\equiv-\nabla\phi^{(1)}$ in the quadratic term in the continuity
equation. However, it will have a second order contribution in the deviation
term

\be
\delta_{gv}^{[2]}\equiv\left(\left(\frac{f_2}{f}-1\right)\frac{D_2}{D}+D\,\right)\delta^{(2)}[\phi^{(1)}]\,.
\ee

Putting all together we get

\be
\frac{\partial \phi_g^{[2]}}{\partial D}-\frac{1}{2}\left(v^{[1]}\right)^2-\left(1+\left(\frac{f_2}{f}-1\right)\frac{D_2}{D^2}\right)\phi^{(2)}[\phi^{(1)}]=0\,.
\ee

If we neglect the contribution of 2LPT ($D_2=0$) we recover the formula derived by
\citet[][]{1993ApJ...405..449G}. Neglecting tidal forces
($\phi^{(2)}[\phi^{(1)}]=0$) we get the formula by
\citet[][]{1992ApJ...391..443N}.

\subsubsection{Higher order continuity equation}

To go beyond the Zeldovich approximation in the velocity term, say to 2nd order
in Lagrangian perturbation theory, one needs to consider at least 4th order
terms in the continuity equation.  The gravitational potential can be written
according to \S \ref{sec:LPT} as
$\phi_g^{\tilde{[6]}}\equiv\phi^{(1)}-\frac{D_2}{D}\phi^{(2)}[\phi^{(1)}]+\frac{1}{D}\left(\phi^{(2)}[\Theta]+\phi^{(3)}[\Theta]\right)$
with the symbol $\tilde{[6]}$ indicating that the 6th order is incomplete. Note
that the term $\phi^{(3)}[\phi]\equiv\nabla^{-2}\mu^{(3)}[\phi]$ includes sixth
order terms involving $D^6$. However a proper sixth order formulation would
require including third order Lagrangian perturbation theory. To obtain the 4th
order equation one would need to truncate that term.  The deviation term is
correspondingly given by

\be
\phi_{gv}^{\tilde{[6]}}\equiv\left(\frac{f_2}{f}-1\right)\frac{D_2}{D}\phi^{(2)}[\phi^{(1)}]+\frac{1}{D}\left(\phi^{(2)}[\Theta]+\phi^{(3)}[\Theta]\right)\,.
\ee
Finally, the continuity equation yields
\be
\label{eq:timer}
\frac{\partial \phi_g^{\tilde{[6]}}}{\partial D}-\frac{1}{2}\left(v^{[2]}\right)^2-2\phi^{(2)}[\phi_v^{[2]}]+\frac{1}{D}\phi_{gv}^{\tilde{[6]}}+\nabla^{-2}\nabla\cdot\left(\delta_{gv}^{\tilde{[6]}}\mbi v^{[2]}\right)=0\,.
\ee

\begin{figure*}
\begin{tabular}{cc}
\includegraphics[width=6.0cm]{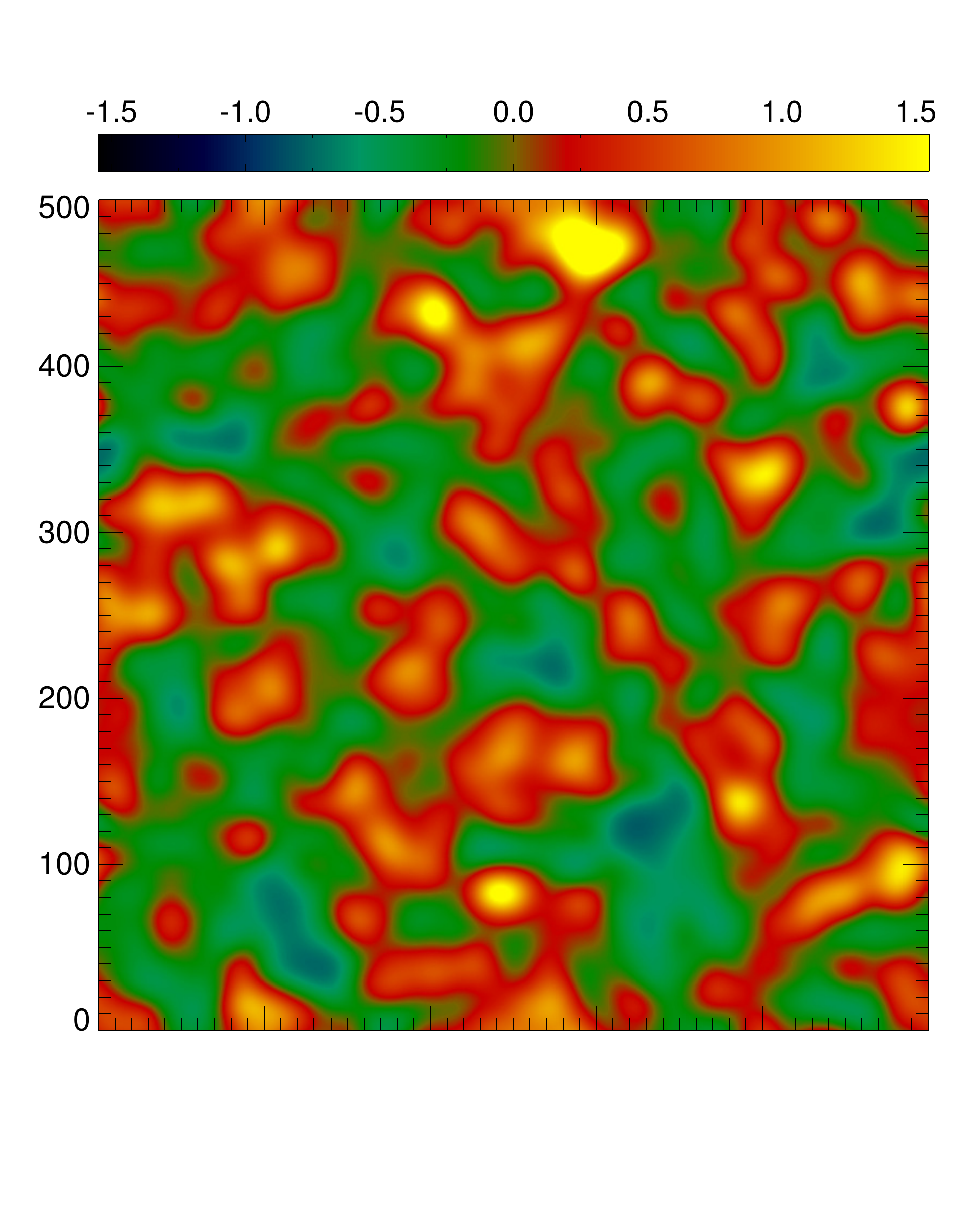}
\put(-175,117){\rotatebox[]{90}{{$Z$ [$h^{-1}\,$Mpc]}}}
\put(-100,210){{$\delta(\mbi q,z=0)$}}
\put(5,210){{$=$}}
\includegraphics[width=6.0cm]{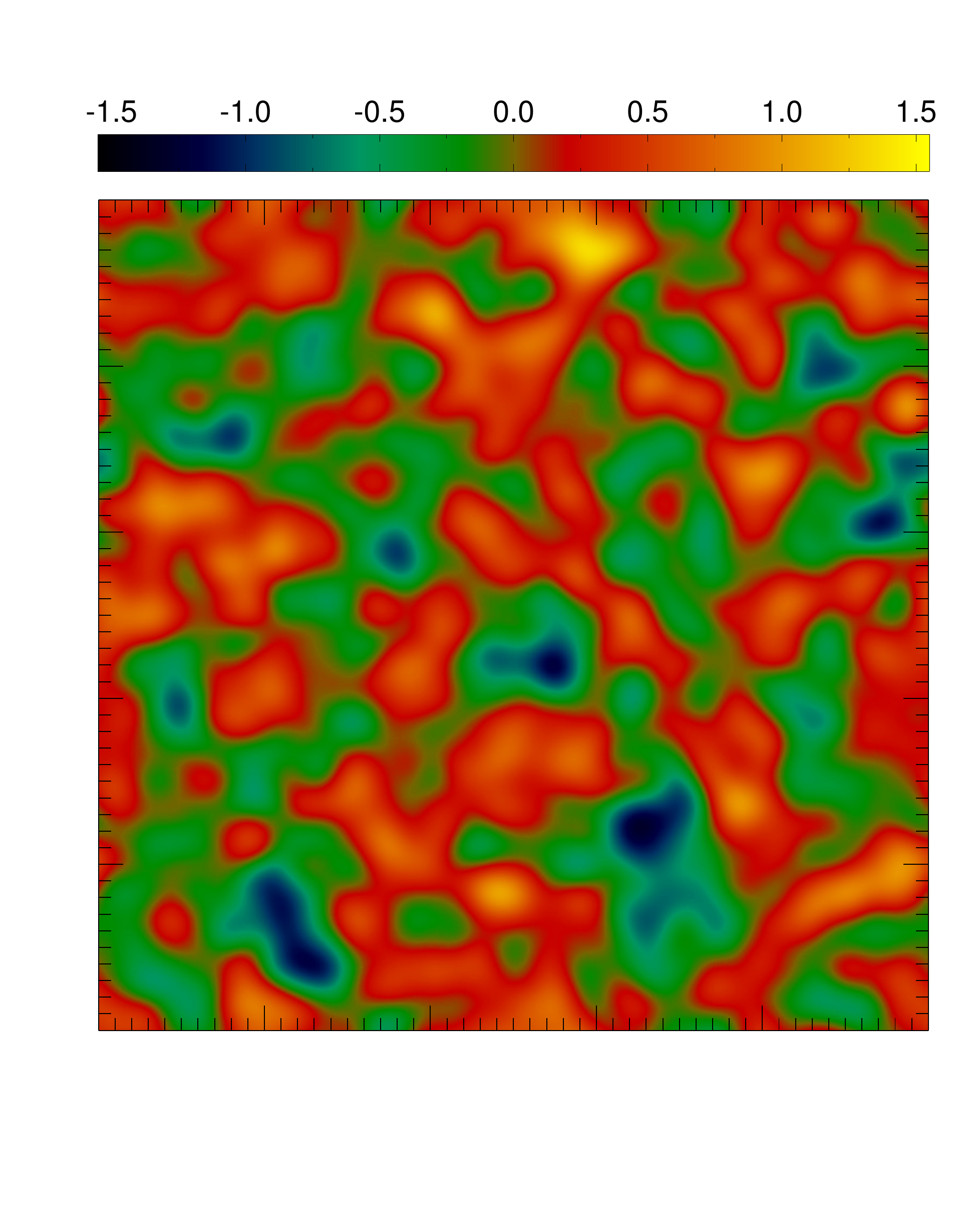}
\put(-155,210){{$\delta^{\rm L}(\mbi q,z=0)[=\delta_D^{\rm Nbody}(\mbi x=\mbi q,z=127)]$}}
\put(5,210){{$+$}}
\includegraphics[width=6.0cm]{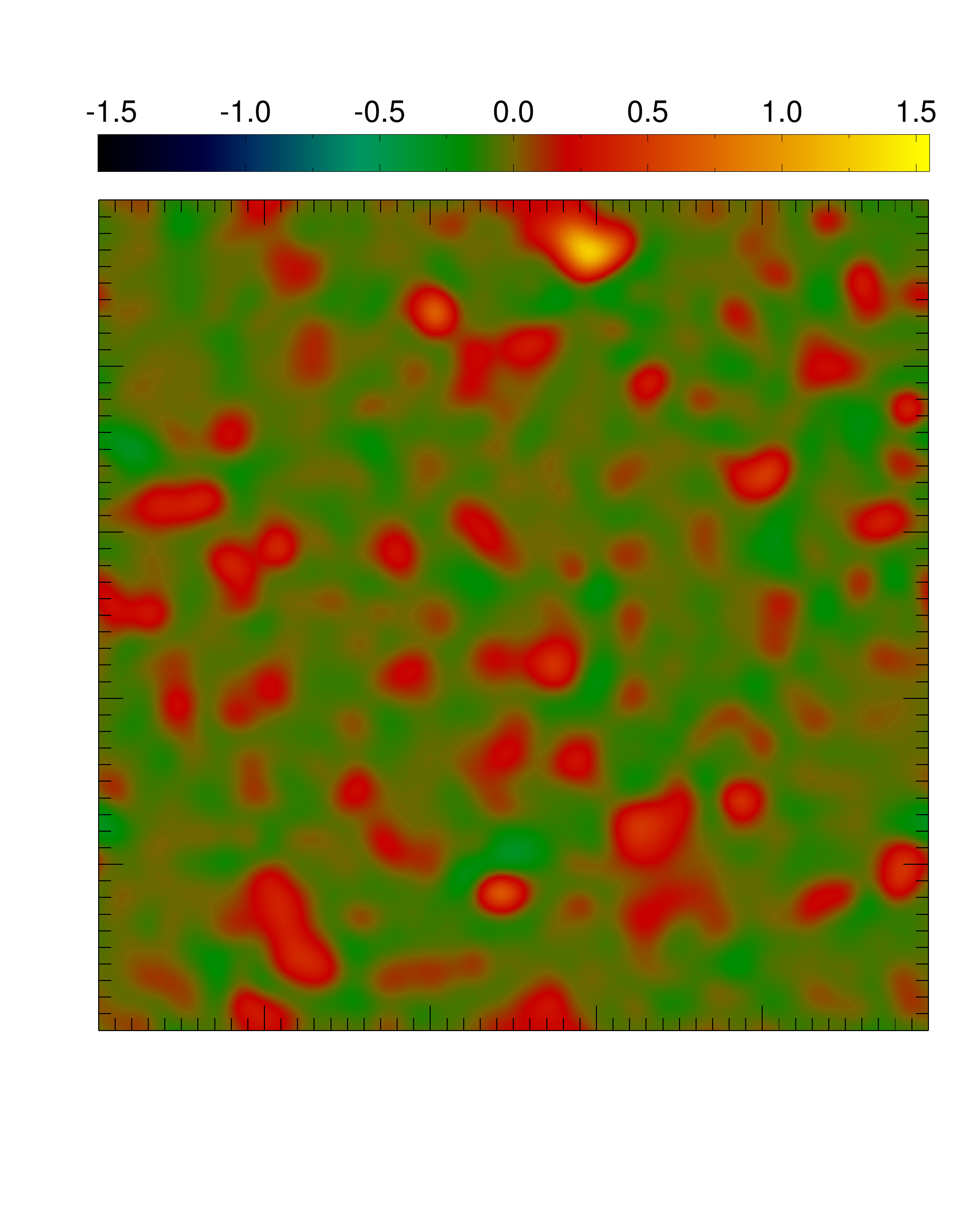}
\put(-100,210){{$\delta^{\rm NL}(\mbi q,z=0)$}}
\vspace{-1.3cm}
\\
\includegraphics[width=6.0cm]{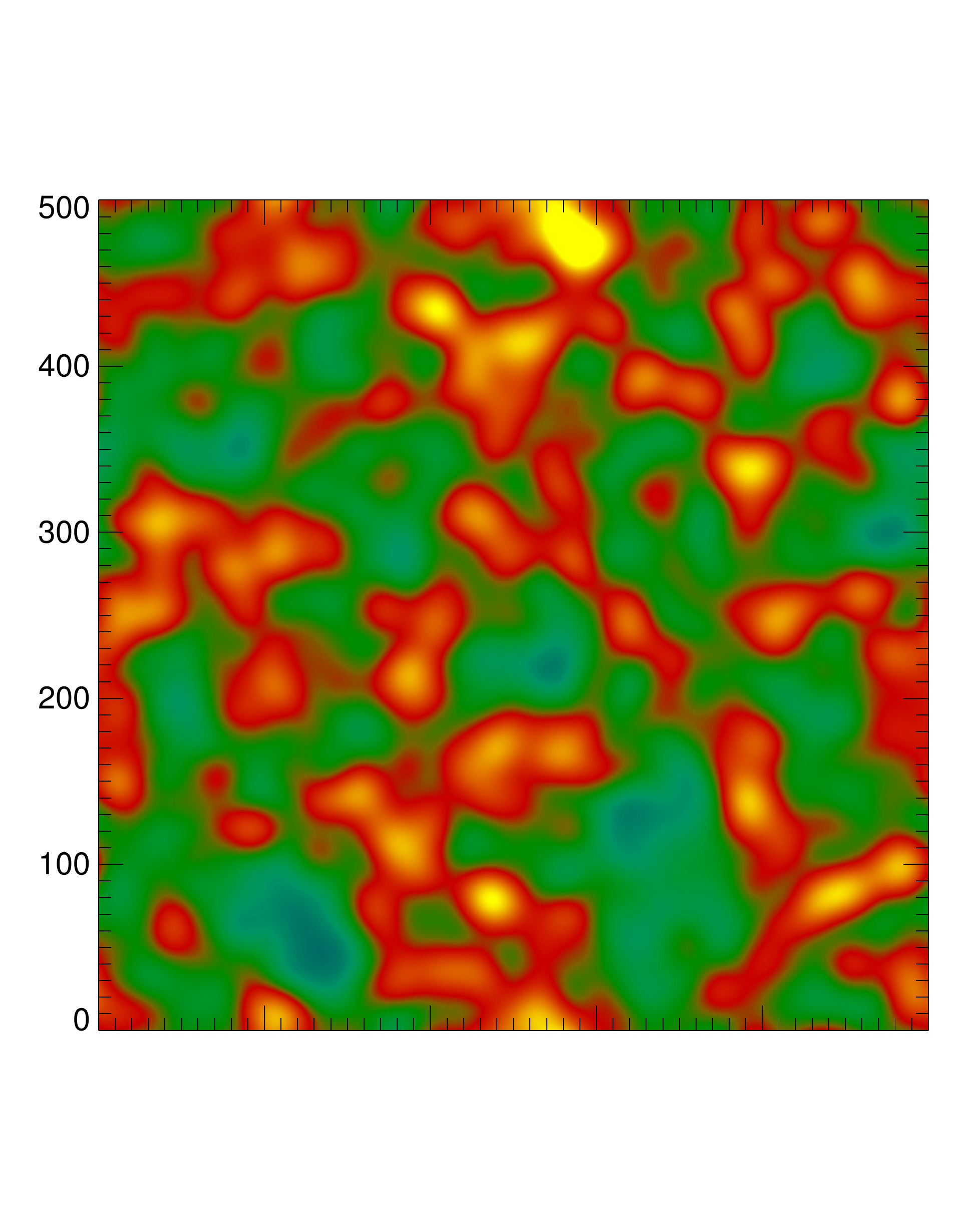}
\put(-175,117){\rotatebox[]{90}{{$Z$ [$h^{-1}\,$Mpc]}}}
\put(-140,200){{$\delta(\mbi x,z=0)[=\delta^{\rm Nbody}(\mbi x,z=0)]$}}
\put(5,200){{$=$}}
\includegraphics[width=6.0cm]{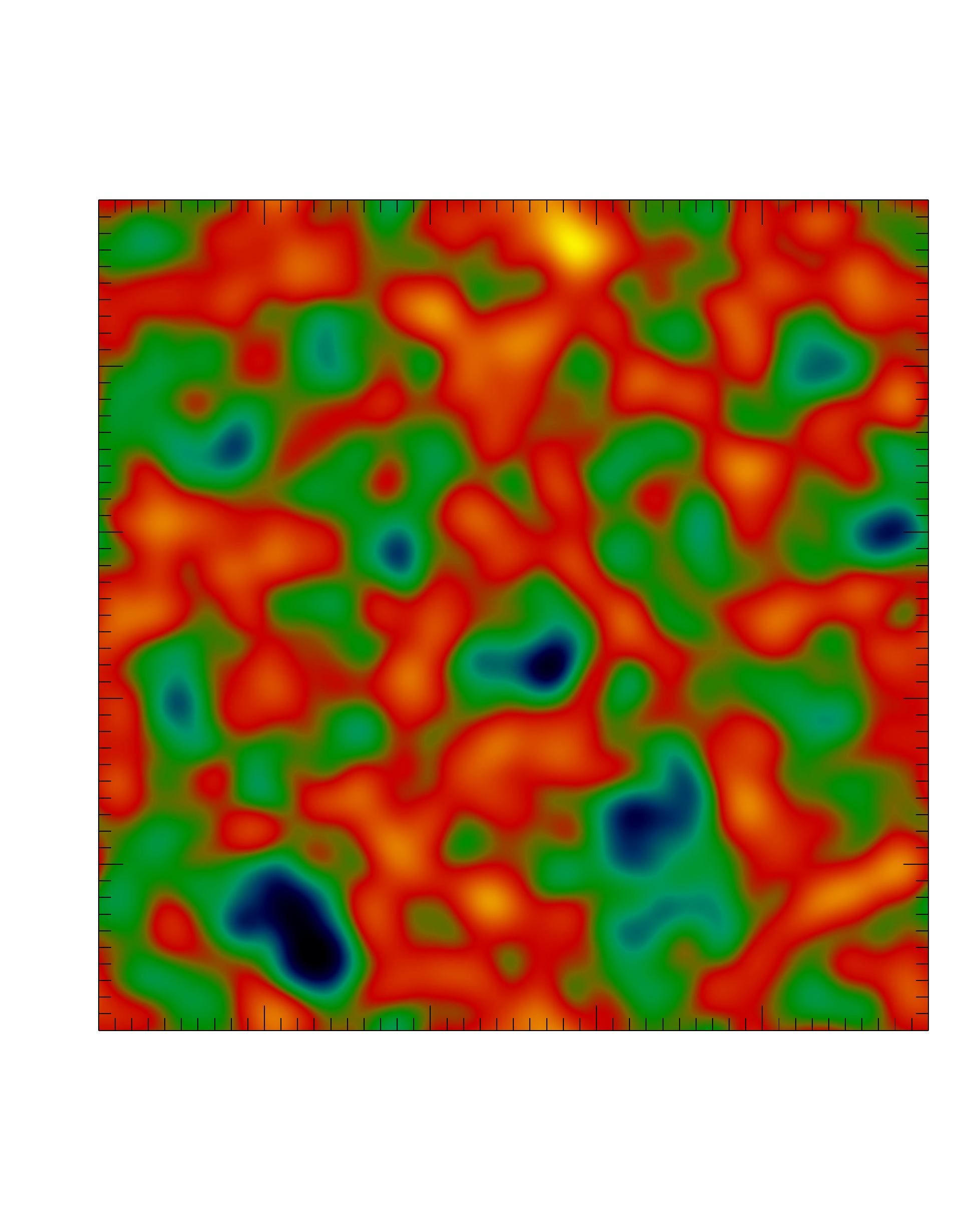}
\put(-100,200){{$\delta^{\rm L}(\mbi x,z=0)$}}
\put(5,200){{$+$}}
\includegraphics[width=6.0cm]{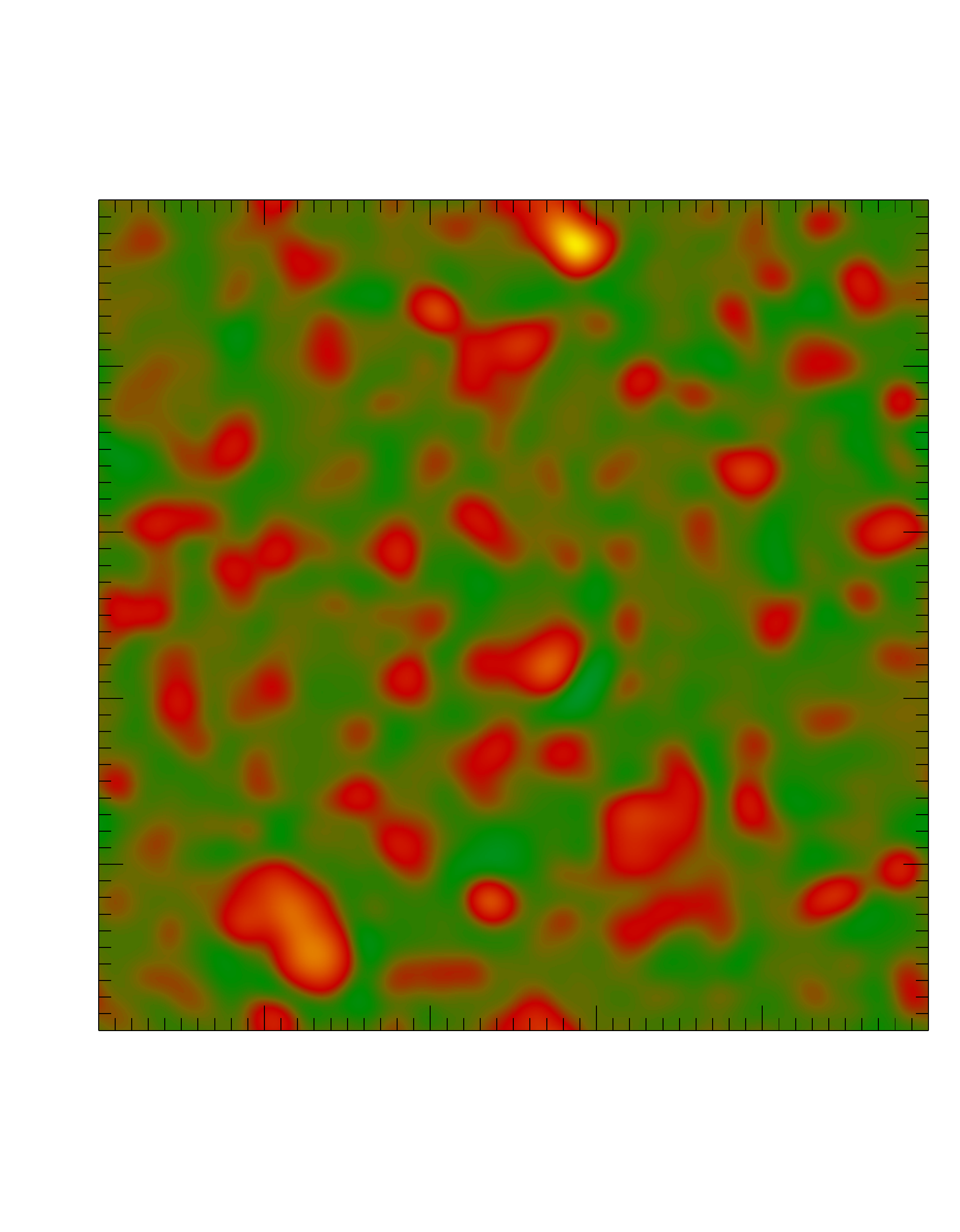}
\put(-100,200){{$\delta^{\rm NL}(\mbi x,z=0)$}}
\vspace{-1.3cm}
\\
\includegraphics[width=6.0cm]{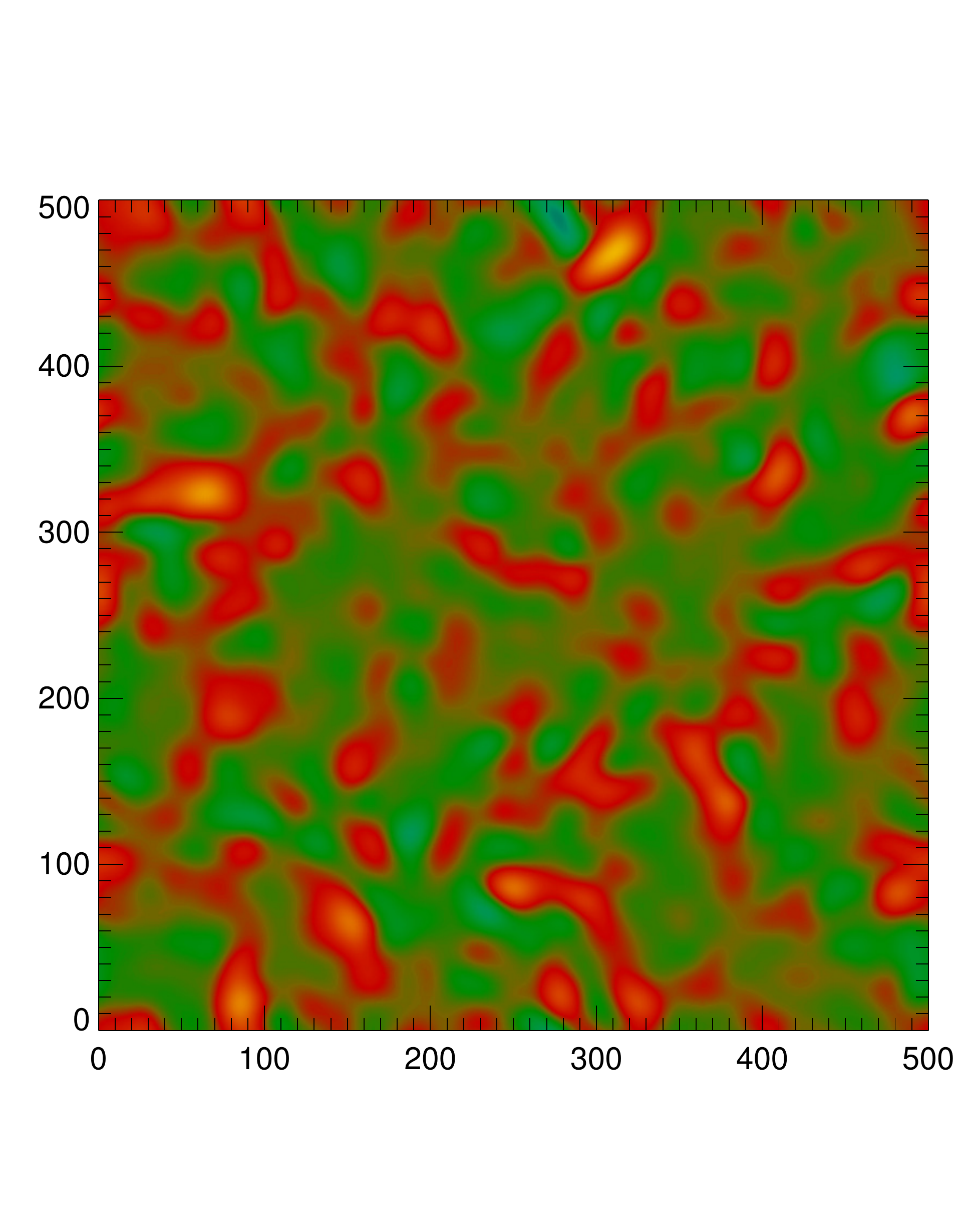}
\put(-175,117){\rotatebox[]{90}{{$Z$ [$h^{-1}\,$Mpc]}}}
\put(-100,205){{$\delta(\mbi q,z=0)$}}
\put(-110,193){{$-\delta^{\rm Nbody}(\mbi x,z=0)$}}
\put(-105,15){{$X$ [$h^{-1}\,$Mpc]}}
\includegraphics[width=6.0cm]{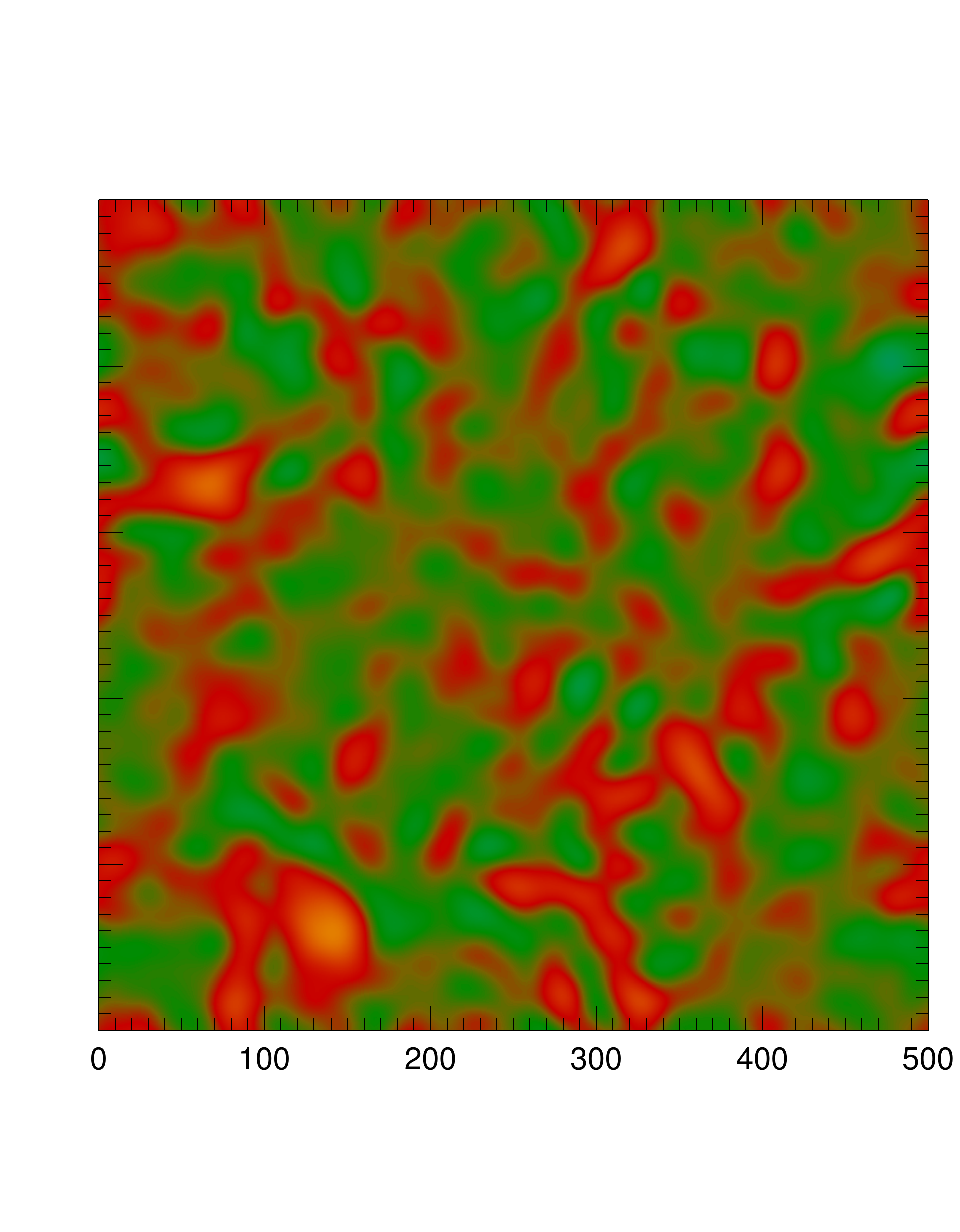}
\put(-155,205){{$\delta^{\rm L}(\mbi q,z=0)[=\delta_D^{\rm Nbody}(\mbi x=\mbi q,z=127)]$}}
\put(-105,193){{$-\delta^{\rm L}(\mbi x,z=0)$}}
\put(-105,15){{$X$ [$h^{-1}\,$Mpc]}}
\includegraphics[width=6.0cm]{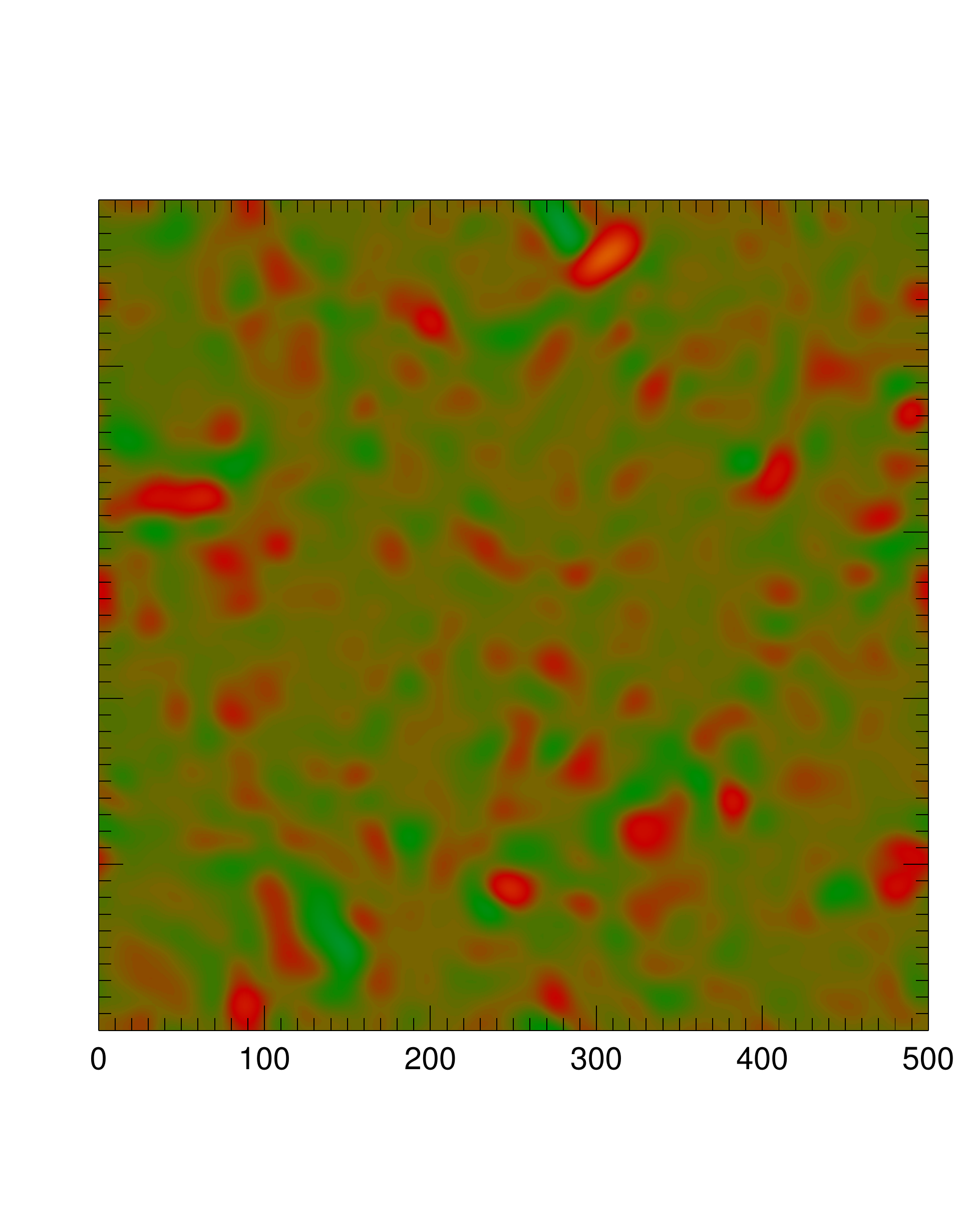}
\put(-100,205){{$\delta^{\rm NL}(\mbi q,z=0)$}}
\put(-105,193){{$-\delta^{\rm NL}(\mbi x,z=0)$}}
\put(-105,15){{$X$ [$h^{-1}\,$Mpc]}}
\end{tabular}
\caption{ \label{fig:dec} 
Slice through the density field of the Millennium Run after Gaussian smoothing
with $r^0_{\rm S}=10$ $h^{-1}\,$Mpc. Upper left panel: forward solution of
Eq.~6 taking as the linear field the Millennium Run at $z=127$. Upper middle
panel: Millennium Run at $z=127$. Upper right panel: nonlinear component
corresponding to the field in the middle panels. Central left panel: Millennium
Run at $z=0$. Central middle panel: iterative solution of the linear component.
Central right panel: nonlinear component. Lower panels: differences between the
corresponding fields in the upper panels and in the central panels.}
\end{figure*}

\begin{figure*}
\begin{tabular}{cc}
\includegraphics[width=8.0cm]{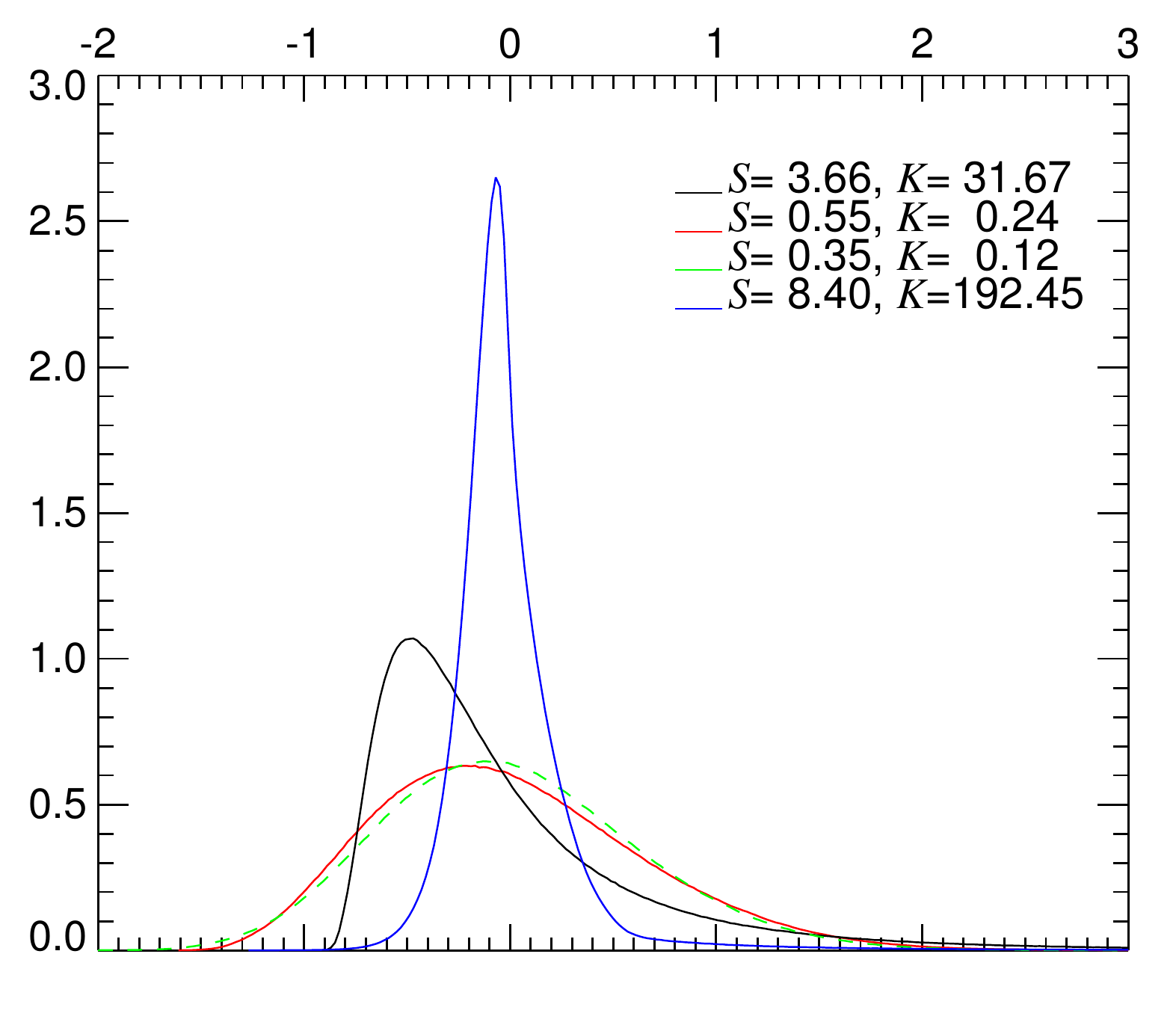}
\put(-90,70){{\large $r_{\rm S}^0$=5 $h^{-1}\,$Mpc}}
\put(-240,105){\rotatebox[]{90}{{PDF}}}
\put(-90,90){{\large $z=0$}}
\includegraphics[width=8.0cm]{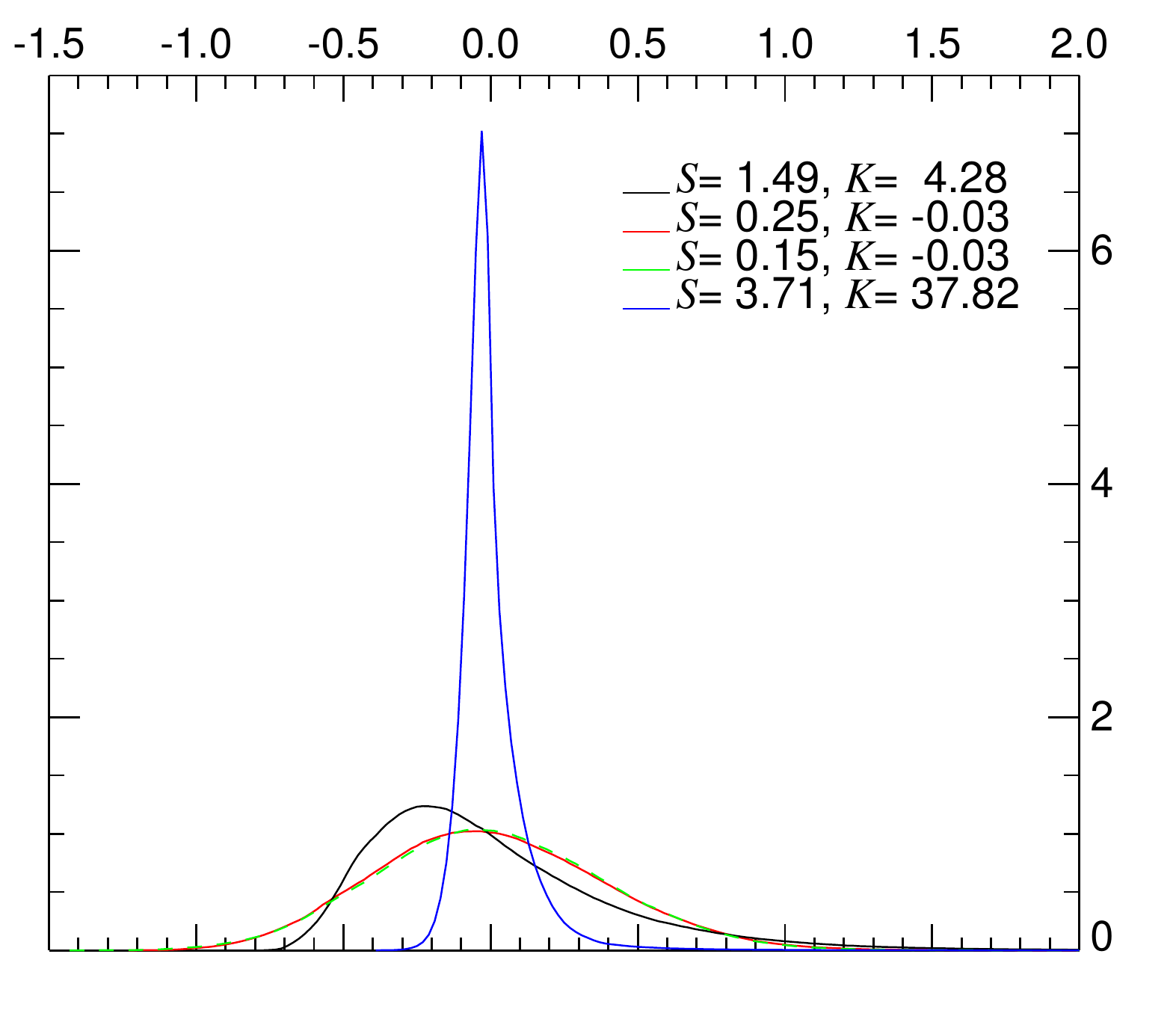}
\put(-100,90){{\large $z=0$}}
\put(-100,70){{\large $r_{\rm S}^0$=10 $h^{-1}\,$Mpc}}
\vspace{-0.5cm}
\\
\includegraphics[width=8.0cm]{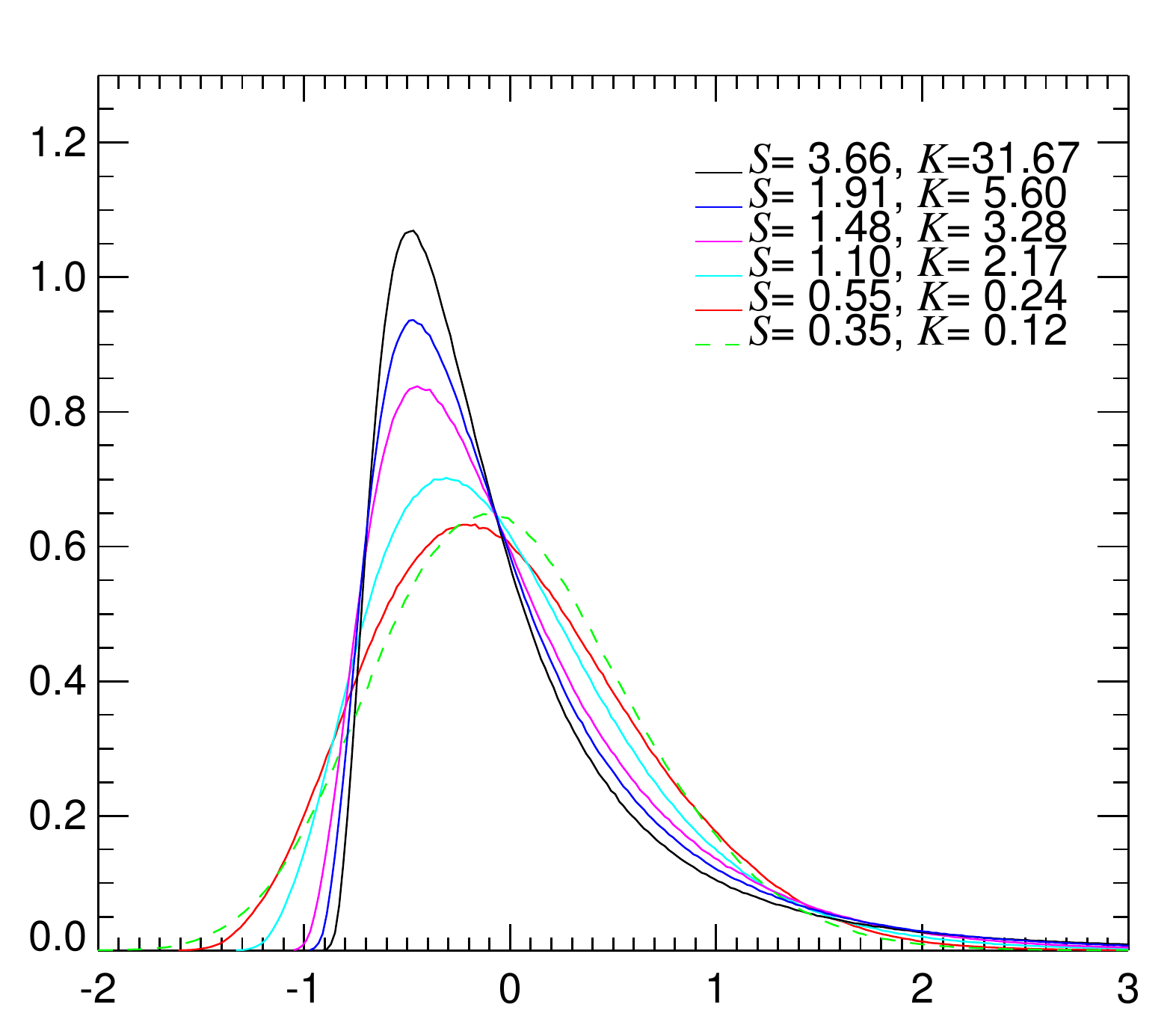}
\put(-90,90){{\large $z=0$}}
\put(-90,70){{\large $r_{\rm S}^0$=5 $h^{-1}\,$Mpc}}
\put(-240,105){\rotatebox[]{90}{{PDF}}}
\put(-120,-5){{$\delta_k$}}
\includegraphics[width=8.0cm]{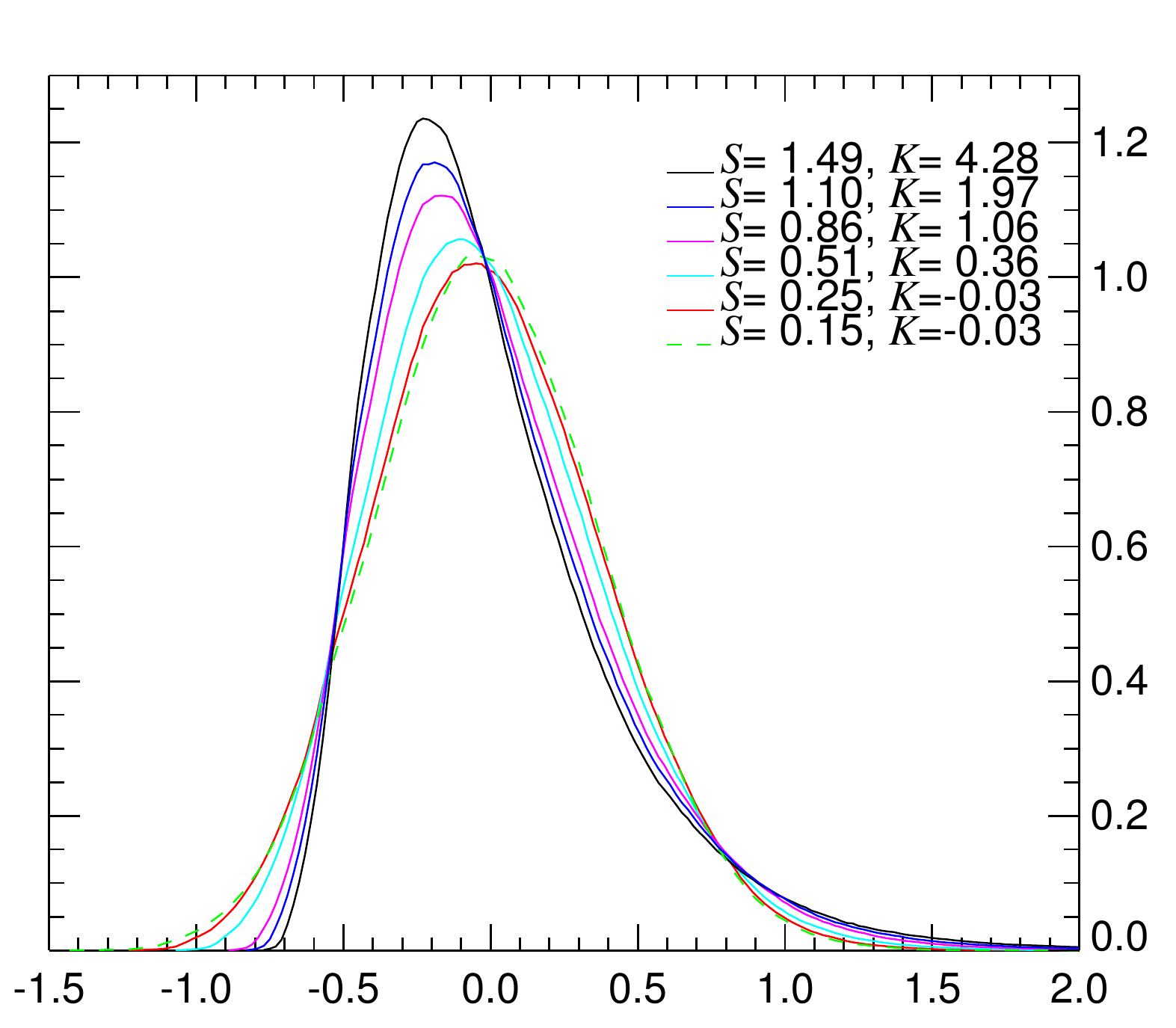}
\put(-100,90){{\large $z=0$}}
\put(-100,70){{\large $r_{\rm S}^0$=10 $h^{-1}\,$Mpc}}
\put(-120,-5){{$\delta_k$}}
\end{tabular}
\caption{\label{fig:stats} Matter probability distribution function (PDF) and corresponding skewness and kurtosis for $\delta_k: \delta^{\rm Nbody}(\mbi x,z=0), \ln(1+\delta^{\rm Nbody}(\mbi x,z=0))-\mu, \delta^{\rm L}(\mbi x,z=0)$ and $\delta^{\rm NL}(\mbi x,z=0)$. Upper panels show the decomposition into a linear and a nonlinear component with an initial smoothing of $r^0_{\rm S}=$5 $h^{-1}\,$Mpc (left) and 10 $h^{-1}\,$Mpc (right), black: total field, red: linear component, blue: nonlinear component, green dashed: lognormal transformation. Lower panels show subsequent steps demonstrating the convergence of the linearisation process. Corresponding skewness S and kurtosis K are also indicated. }
\end{figure*}

\begin{figure*}
\begin{tabular}{cc}
\includegraphics[width=6.5cm]{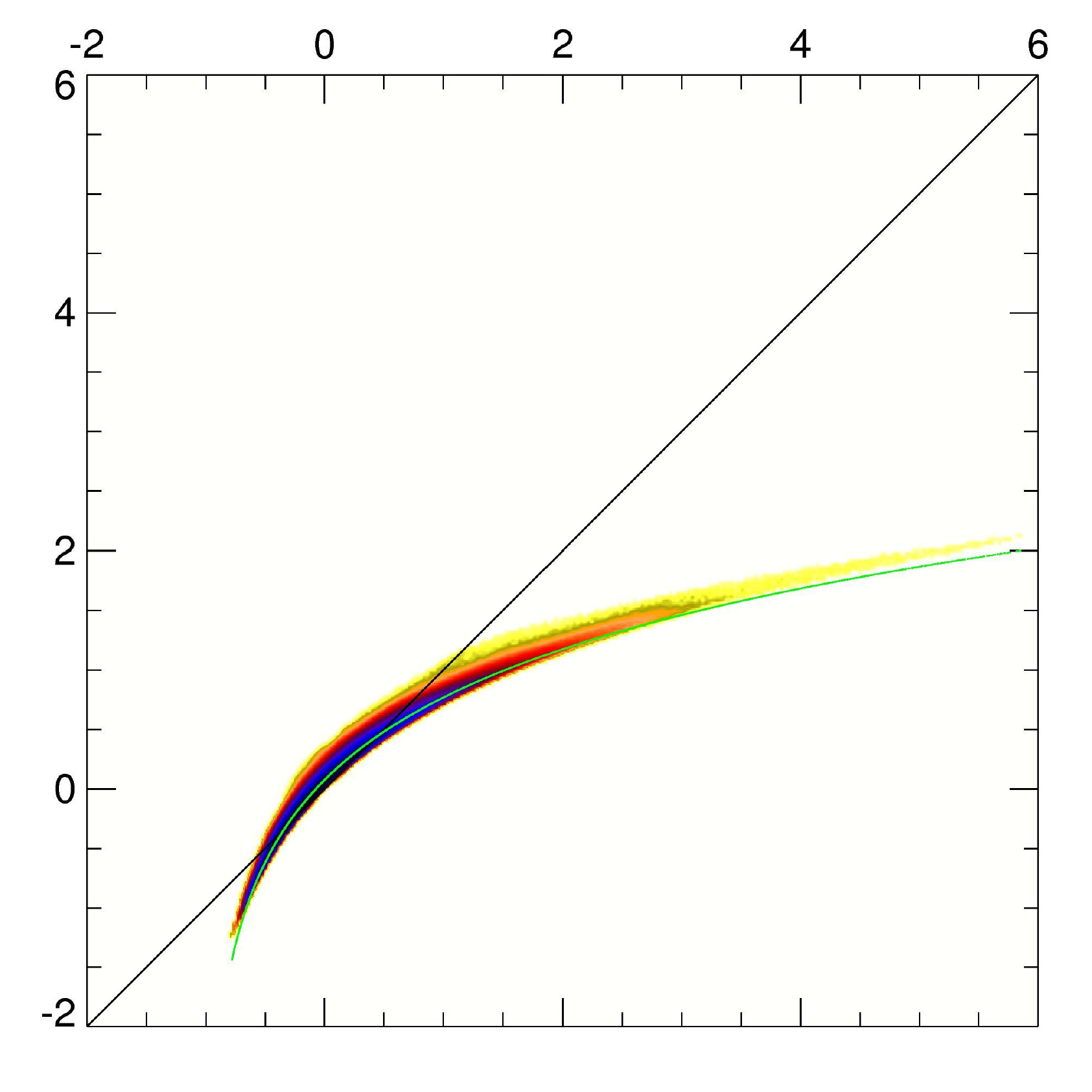}
\put(-195,90){\rotatebox[]{90}{$\delta^{\rm L}(\mbi x,z=0)$}}
\put(-160,150){{\large $r_{\rm S}^0$=10 $h^{-1}\,$Mpc}}
\put(-160,130){\color{green}$\ln(1+\delta^{\rm Nbody}(\mbi x,z=0))-\mu$}
\hspace{0.cm}
\includegraphics[width=6.5cm]{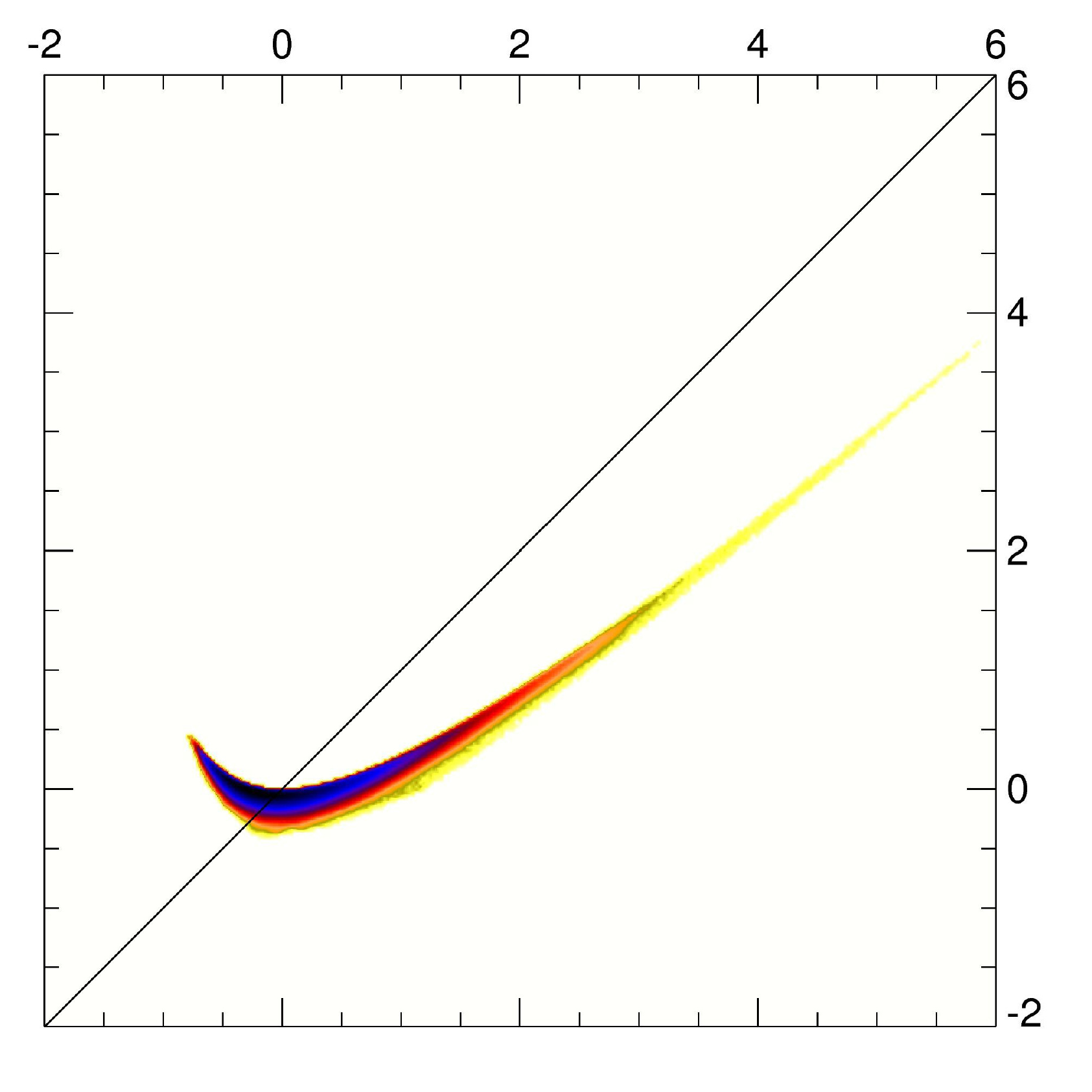}
\put(5,90){\rotatebox[]{-90}{$\delta^{\rm NL}(\mbi x,z=0)$}}
\put(-160,150){{\large $r_{\rm S}^0$=10 $h^{-1}\,$Mpc}}
\vspace{-.0cm}
\\
\includegraphics[width=6.5cm]{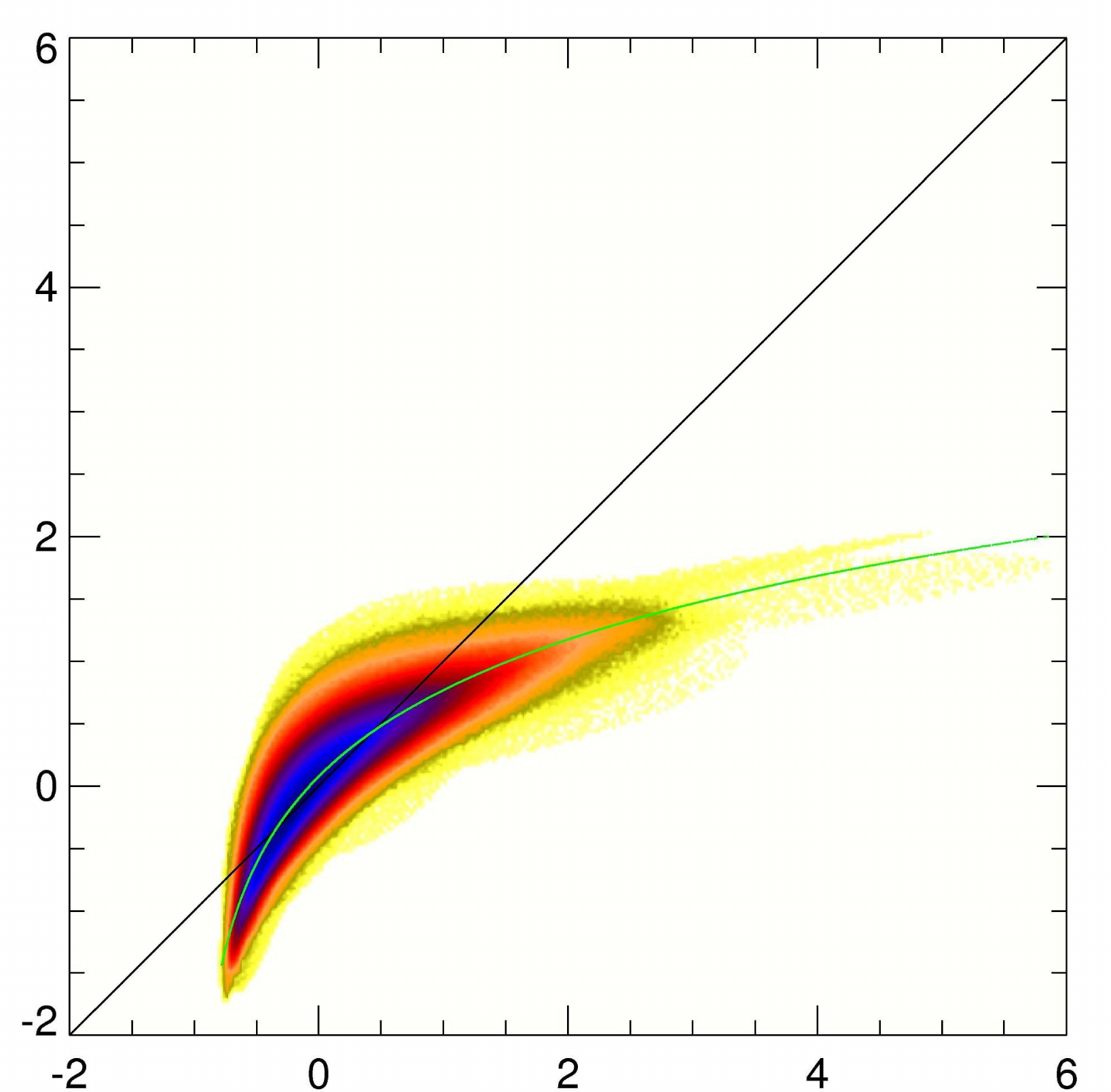}
\put(-115,-10){{$\delta^{\rm Nbody}(\mbi x,z=0)$}}
\put(-195,100){\rotatebox[]{90}{$\delta^{\rm L}(\mbi q,z=0)=\delta_D^{\rm Nbody}(\mbi x,z=127)$}}
\put(-160,150){{\large $r_{\rm S}^0$=10 $h^{-1}\,$Mpc}}
\put(-160,130){\color{green}$\ln(1+\delta^{\rm Nbody}(\mbi x,z=0))-\mu$}
\hspace{0.cm}
\includegraphics[width=6.5cm]{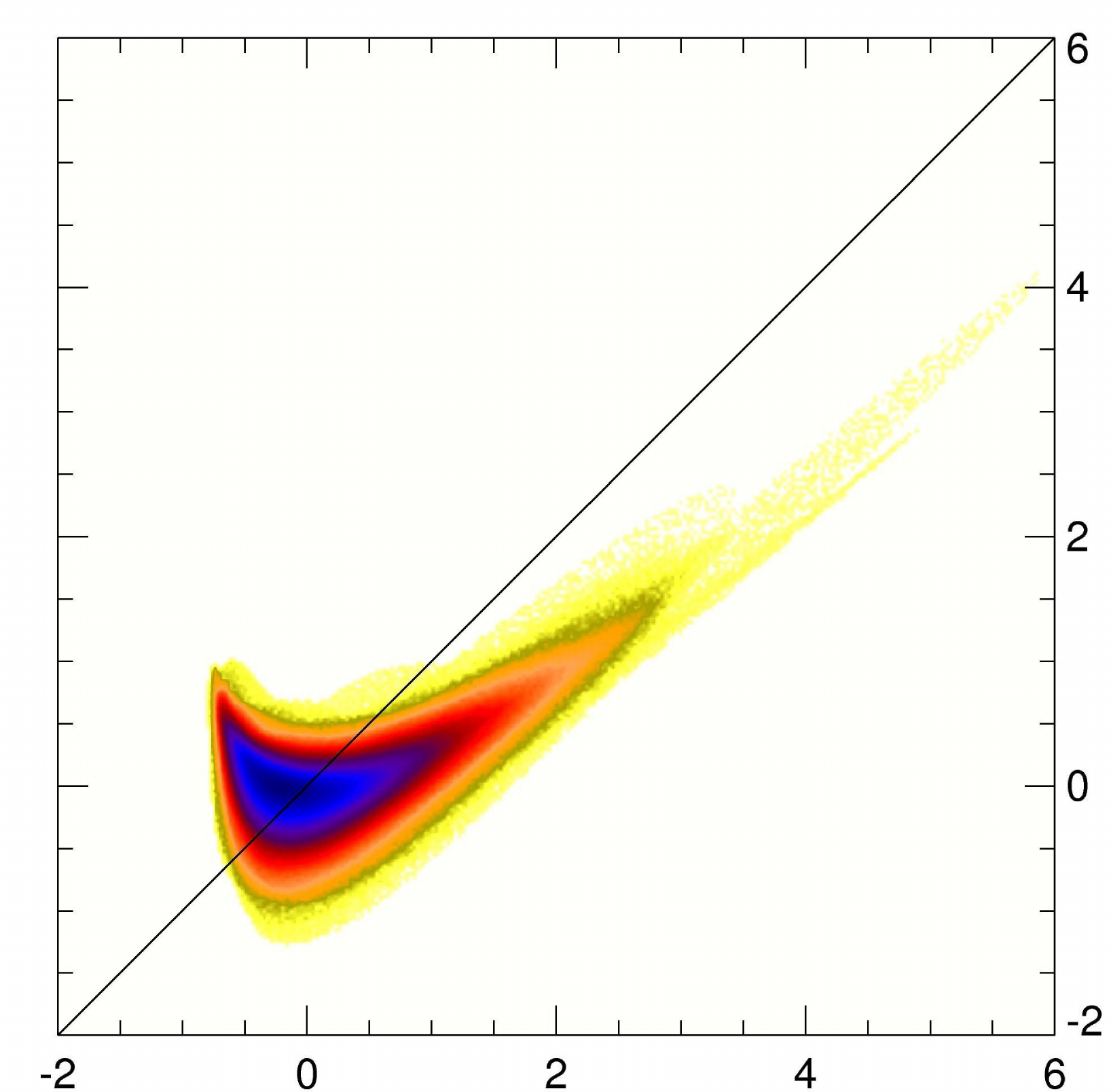}
\put(-115,-10){{$\delta^{\rm Nbody}(\mbi x,z=0)$}}
\put(5,85){\rotatebox[]{-90}{$\delta^{\rm Nbody}(\mbi x,z=0)-\delta_D^{\rm Nbody}(\mbi x,z=127)$}}
\put(-160,150){{\large $r_{\rm S}^0$=10 $h^{-1}\,$Mpc}}
\end{tabular}
\caption{\label{fig:c2c} Cell-to-cell comparison  after Gaussian smoothing with $r^0_{\rm S}=10$ $h^{-1}\,$Mpc between the matter field $\delta^{\rm Nbody}(\mbi x,z=0)$ of the simulation at $z=0$ and Upper panels: Left: the iterative solution of the linear component at $z=0$ in Eulerian coordinates Right: the nonlinear component in Eulerian coordinates, Lower panels: Left: the simulation at $z=127$ representing the linear component in Lagrangian coordinates, Right: the difference between the simulation at $z=0$ and the field at $z=127$. The green curve represents the lognormal transformation. Dark colour code indicates a larger number of cells and light colour code a lower number.}
\end{figure*}

\begin{figure*}
\begin{tabular}{ccc}
\includegraphics[width=5.5cm]{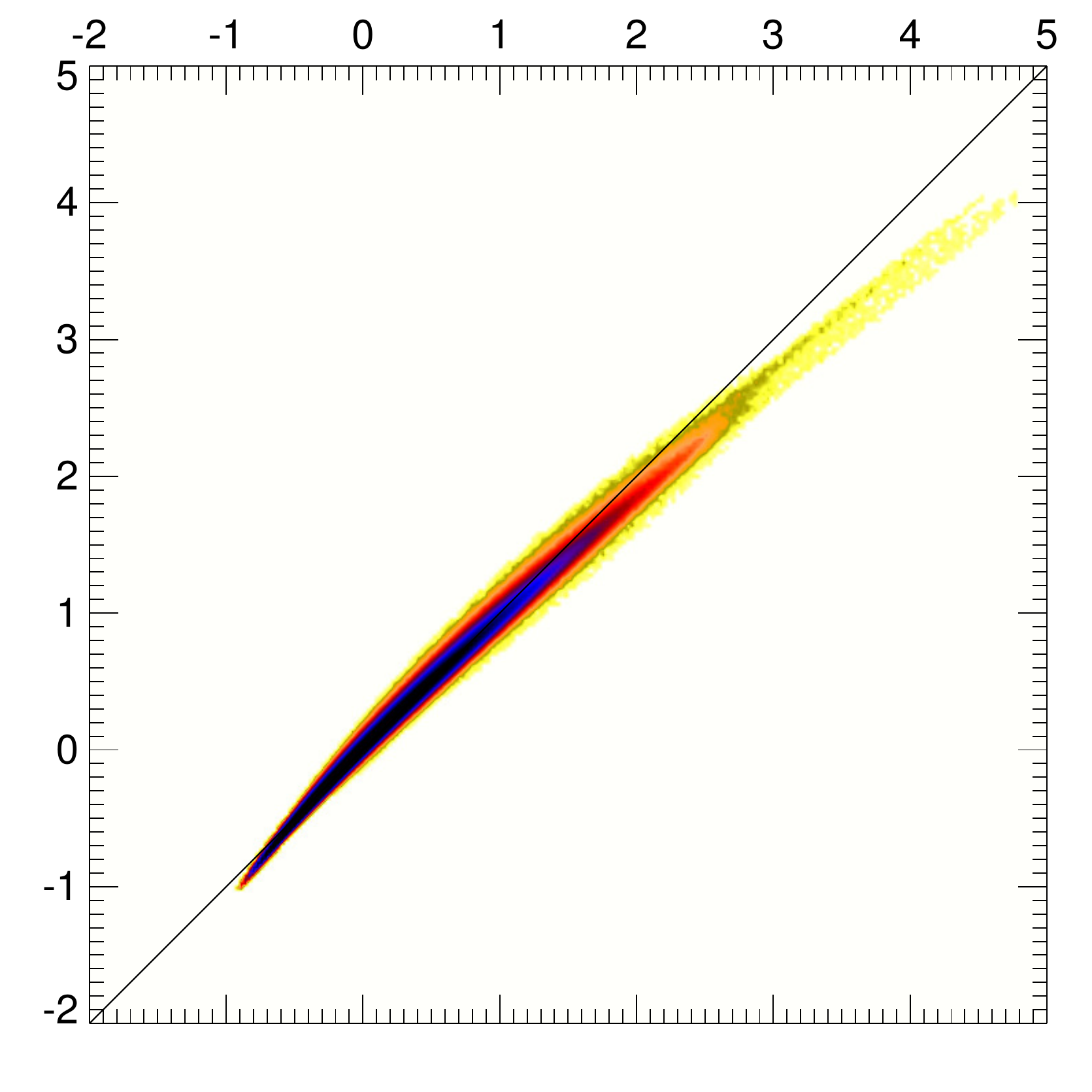}
\put(-120,160){{$\delta_D^{\rm Nbody}(\mbi x,z=0.5)$}}
\put(-160,80){\rotatebox[]{90}{$\delta_D^{\rm Nbody}(\mbi x,z=1)$}}
\put(-135,130){{\large $r_{\rm S}^0$=10 $h^{-1}\,$Mpc}}
\hspace{-0.5cm}
\includegraphics[width=5.5cm]{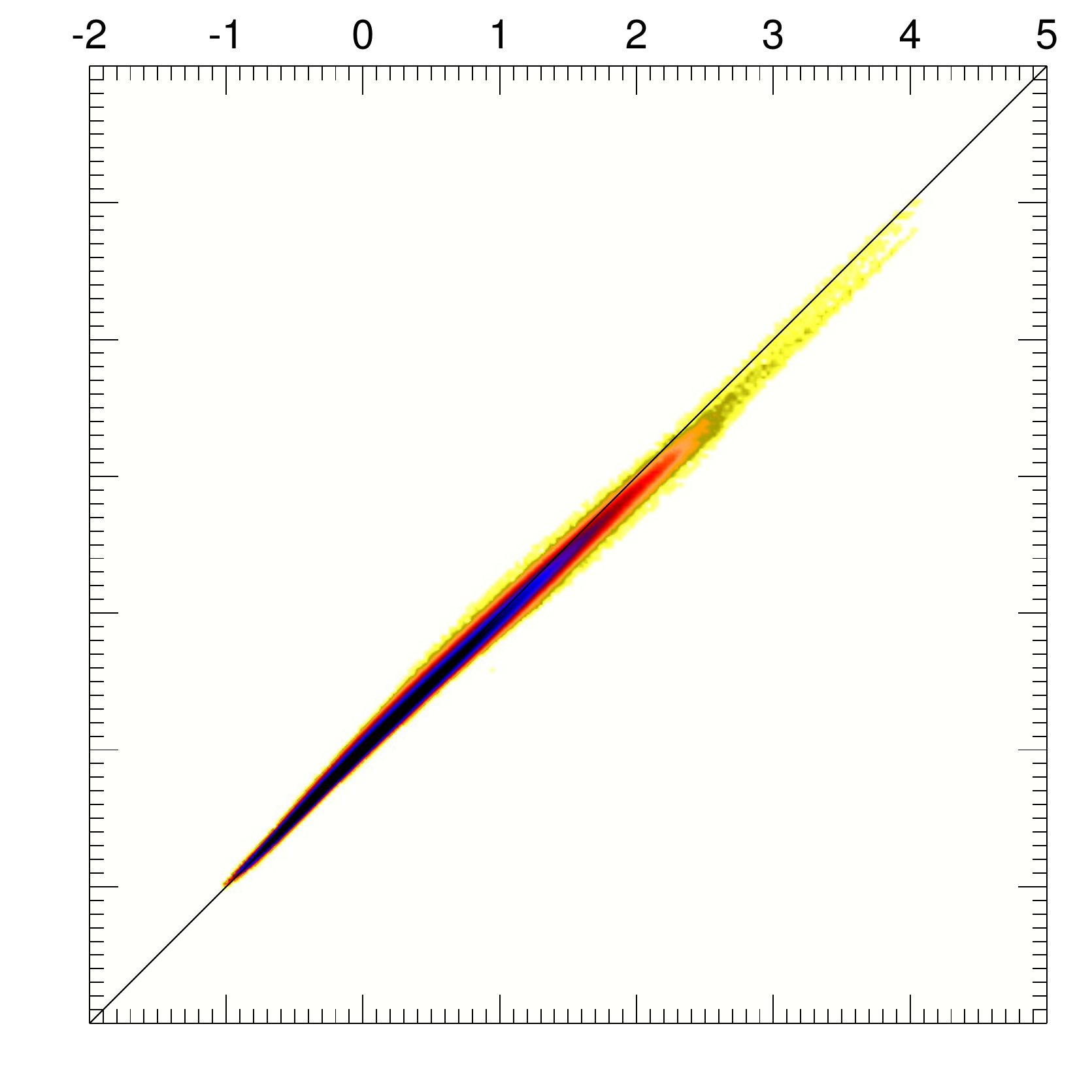}
\put(-120,160){{$\delta_D^{\rm Nbody}(\mbi x,z=z_j)$}}
\put(0,80){\rotatebox[]{-90}{$\delta_D^{\rm ELPT}(\mbi x,z=z_j)$}}
\put(-137,130){{\large $r_{\rm S}^0$=10 $h^{-1}\,$Mpc}}
\put(-130,100){{\large $z_j$=1}}
\hspace{0.5cm}
\includegraphics[width=5.5cm]{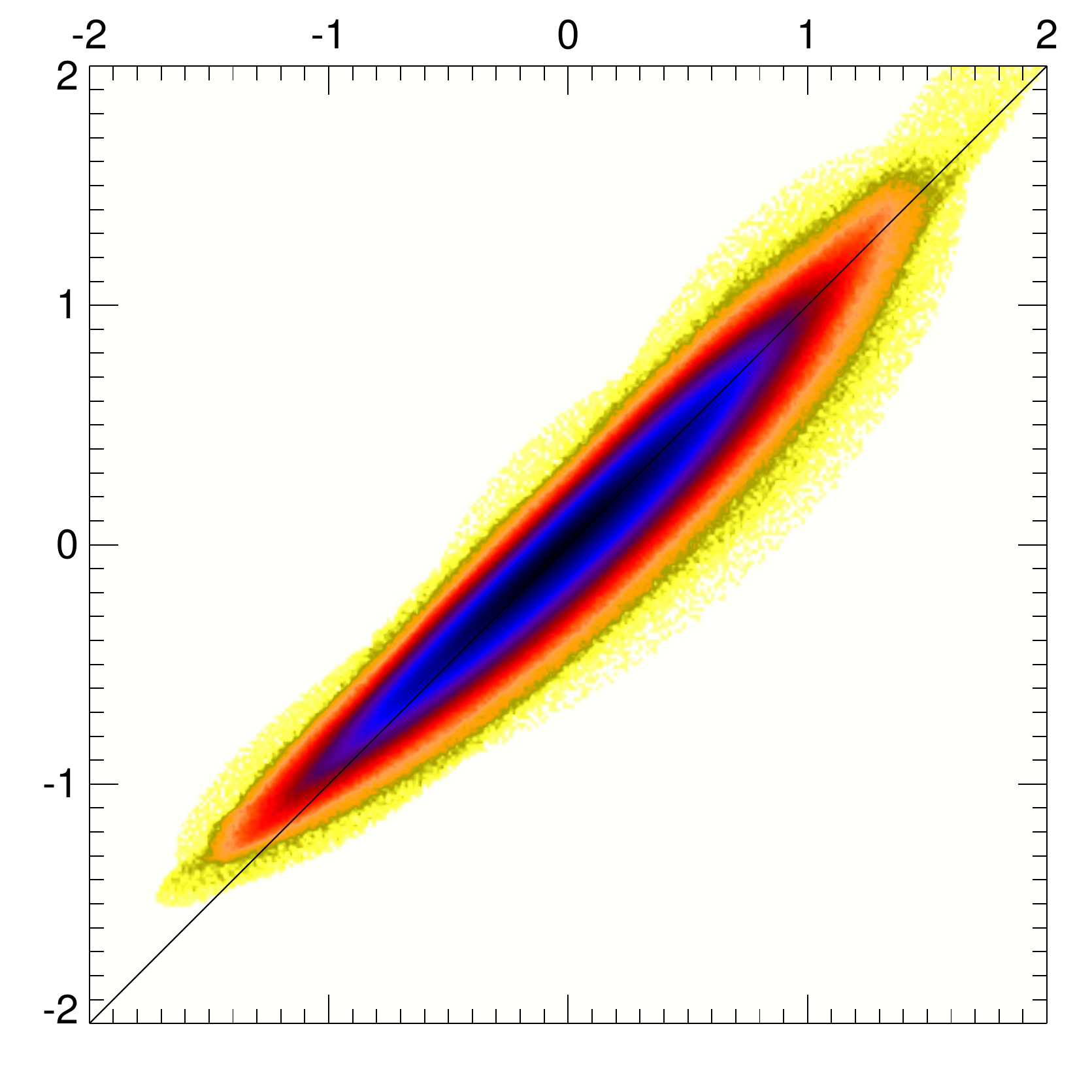}
\put(-130,160){{$\delta^{\rm L}(\mbi q,z=0)=\delta_D^{\rm Nbody}(z=127)$}}
\put(-160,80){\rotatebox[]{90}{$\delta^{\rm L}_{\rm D,ELPT}(\mbi x,z=1)$}}
\put(-135,130){{\large $r_{\rm S}^0$=10 $h^{-1}\,$Mpc}}
\vspace{-0.5cm}
\\
\includegraphics[width=5.5cm]{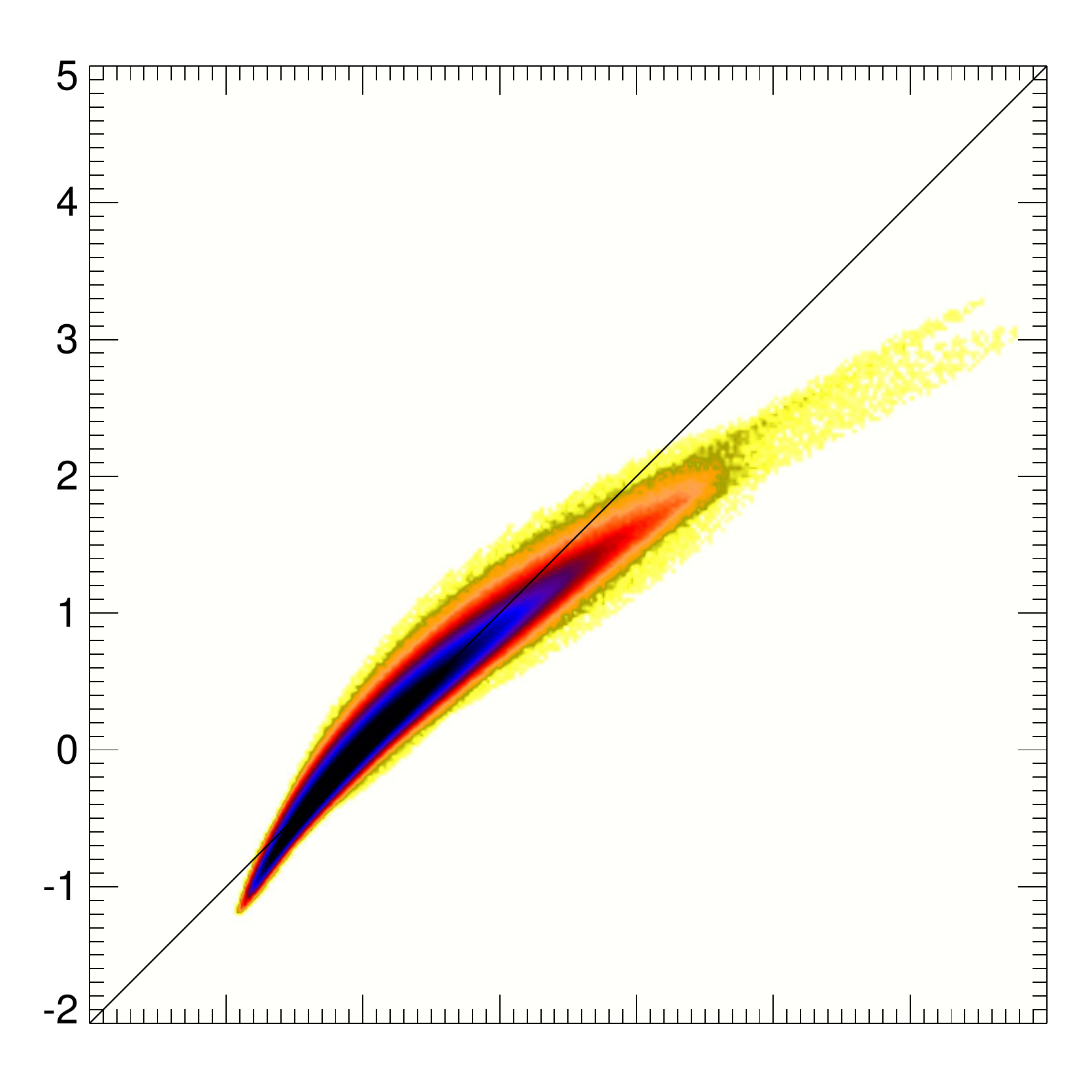}
\put(-160,80){\rotatebox[]{90}{$\delta_D^{\rm Nbody}(\mbi x,z=2)$}}
\put(-135,130){{\large $r_{\rm S}^0$=10 $h^{-1}\,$Mpc}}
\hspace{-0.5cm}
\includegraphics[width=5.5cm]{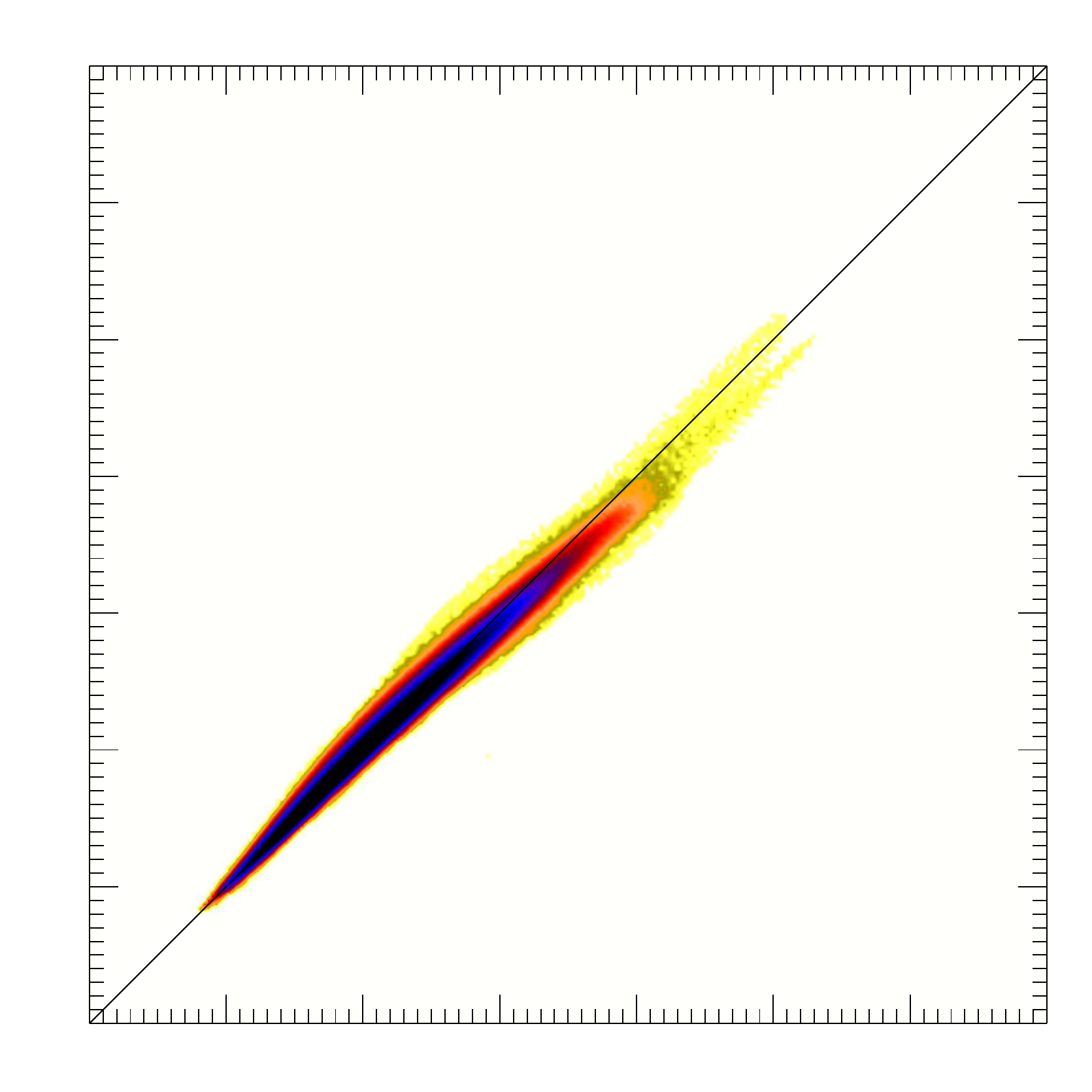}
\put(0,80){\rotatebox[]{-90}{$\delta_D^{\rm ELPT}(\mbi x,z=z_j)$}}
\put(-130,100){{\large $z_j$=2}}
\put(-137,130){{\large $r_{\rm S}^0$=10 $h^{-1}\,$Mpc}}
\hspace{0.5cm}
\includegraphics[width=5.5cm]{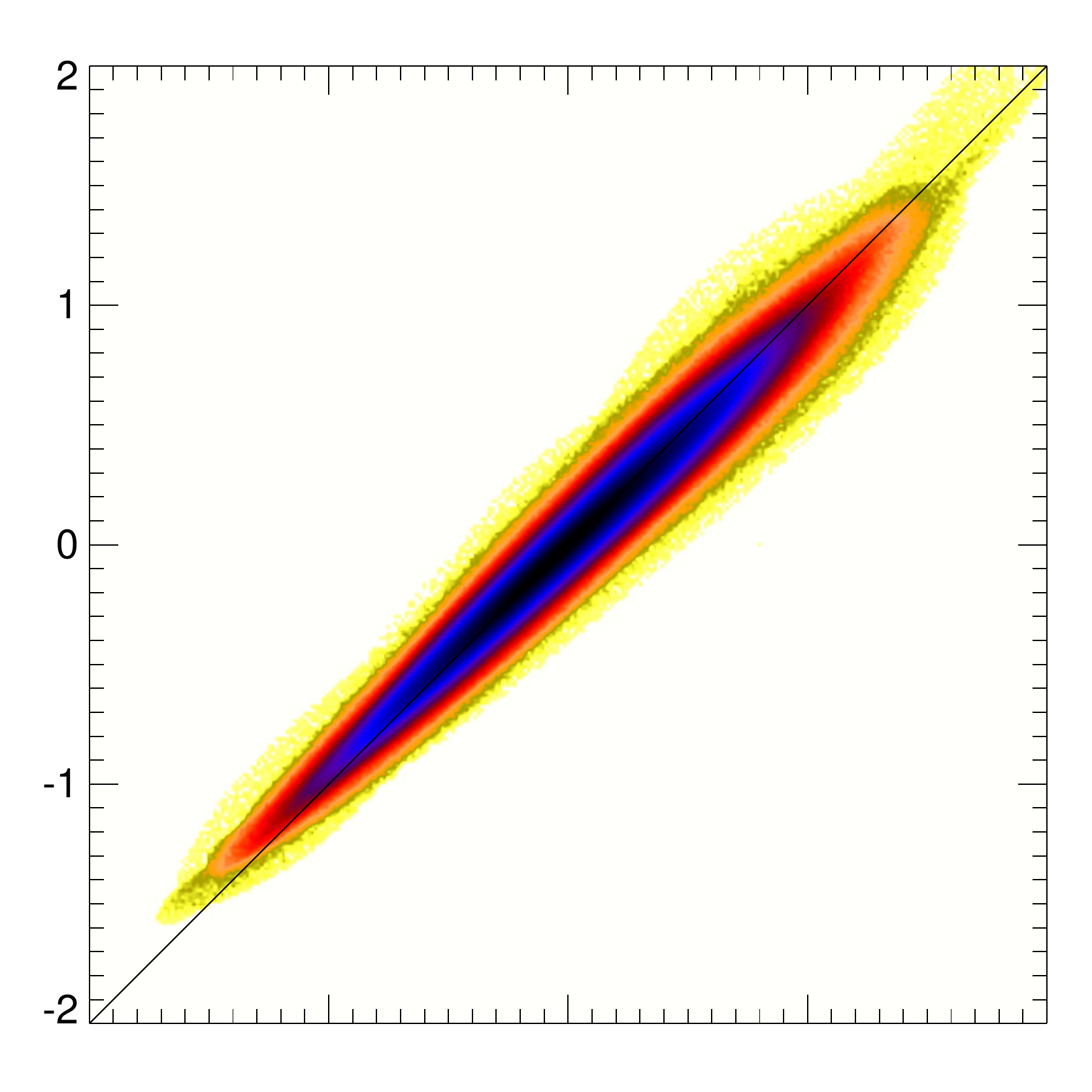}
\put(-160,80){\rotatebox[]{90}{$\delta^{\rm L}_{\rm D,ELPT}(\mbi x,z=2)$}}
\put(-135,130){{\large $r_{\rm S}^0$=10 $h^{-1}\,$Mpc}}
\vspace{-0.5cm}
\\
\includegraphics[width=5.5cm]{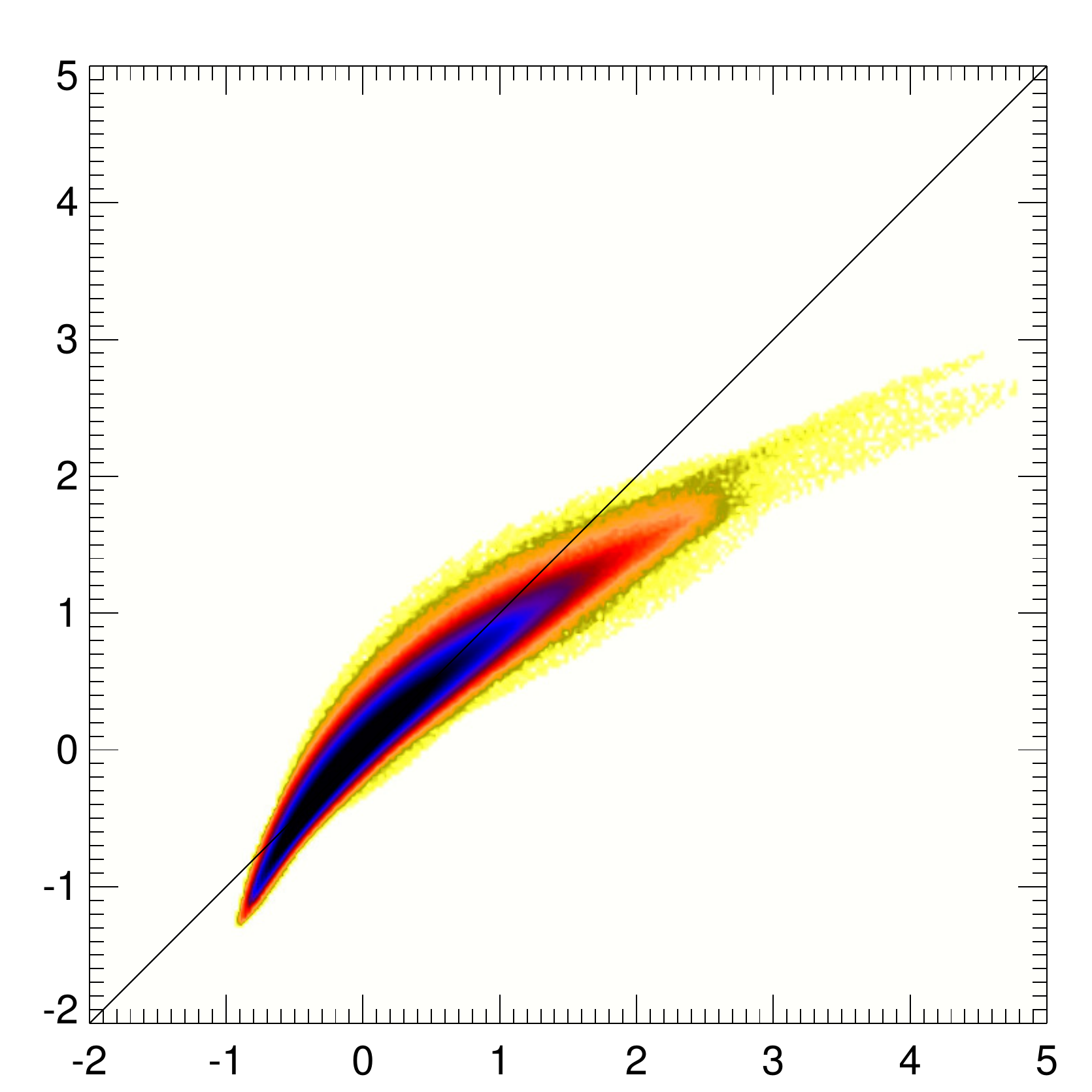}
\put(-160,80){\rotatebox[]{90}{$\delta_D^{\rm Nbody}(\mbi x,z=3)$}}
\put(-135,130){{\large $r_{\rm S}^0$=10 $h^{-1}\,$Mpc}}
\put(-120,-10){{$\delta_D^{\rm Nbody}(\mbi x,z=0.5)$}}
\hspace{-0.5cm}
\includegraphics[width=5.5cm]{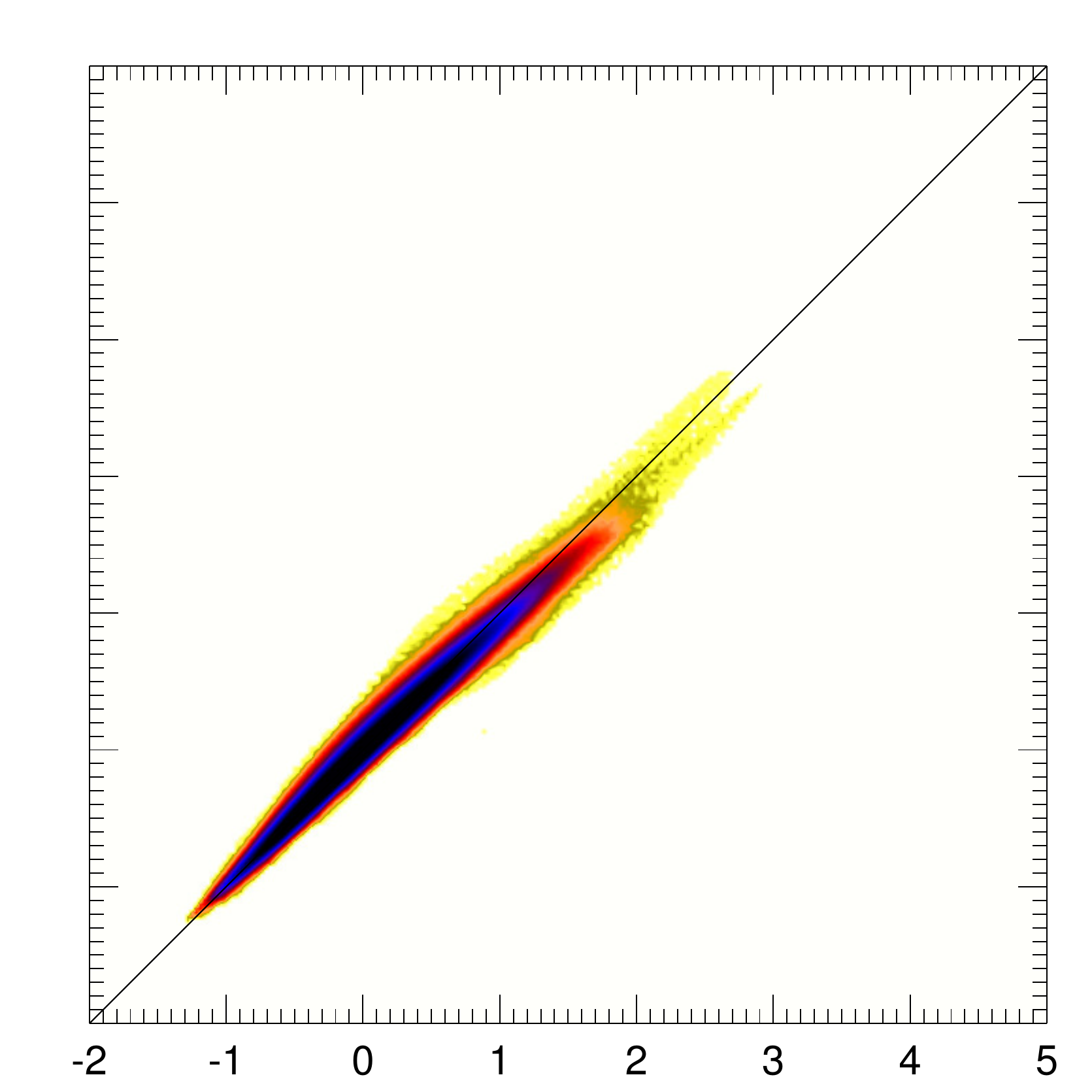}
\put(0,80){\rotatebox[]{-90}{$\delta_D^{\rm ELPT}(\mbi x,z=z_j)$}}
\put(-130,100){{\large $z_j$=3}}
\put(-120,-10){{$\delta_D^{\rm Nbody}(\mbi x,z=z_j)$}}
\put(-137,130){{\large $r_{\rm S}^0$=10 $h^{-1}\,$Mpc}}
\hspace{0.5cm}
\includegraphics[width=5.5cm]{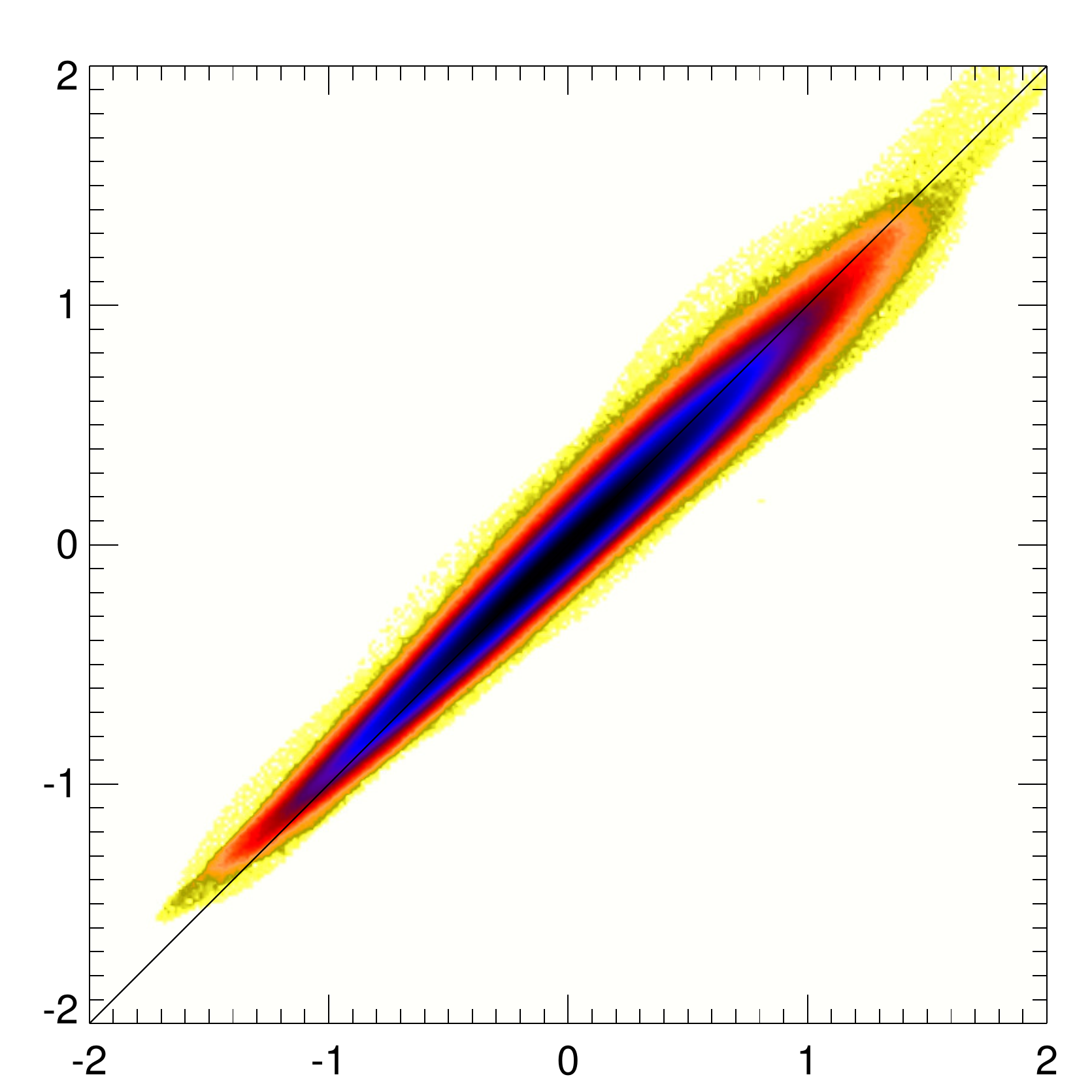}
\put(-160,80){\rotatebox[]{90}{$\delta^{\rm L}_{\rm D,ELPT}(\mbi x,z=3)$}}
\put(-130,-10){{$\delta^{\rm L}(\mbi q,z=0)=\delta_D^{\rm Nbody}(z=127)$}}
\put(-135,130){{\large $r_{\rm S}^0$=10 $h^{-1}\,$Mpc}}
\end{tabular}
\caption{\label{fig:c2c2} Left panels: cell-to-cell comparison  after Gaussian smoothing with $r^0_{\rm S}=10$ $h^{-1}\,$Mpc between the simulation at $z=0.5$: $\delta_D^{\rm Nbody}(\mbi x,z=0.5)$ and the simulation at different redshifts   $\delta_D^{\rm Nbody}(\mbi x,zj)$. Middle panels: cell-to-cell comparison between the simulation at different redshifts  $\delta_D^{\rm Nbody}(\mbi x,zj)$ and the time-reversal reconstruction of the full nonlinear field at the same redshift $\delta_D^{\rm ELPT}(\mbi x,z=z_j)$. Right panels: cell-to-cell comparison between the simulation at $z=127$ $\delta^{\rm L}(\mbi q,z=0)=\delta_D^{\rm Nbody}(z=127)$ and the linear component of the reconstruction at different redshifts: $\delta^{\rm L}_{\rm D,ELPT}(\mbi x,zj)$. Note that $z_j$ runs for the following redshifts $zj=1,2,3$. }
\end{figure*}

\begin{figure*}
\begin{tabular}{ccc}
\includegraphics[width=6.0cm]{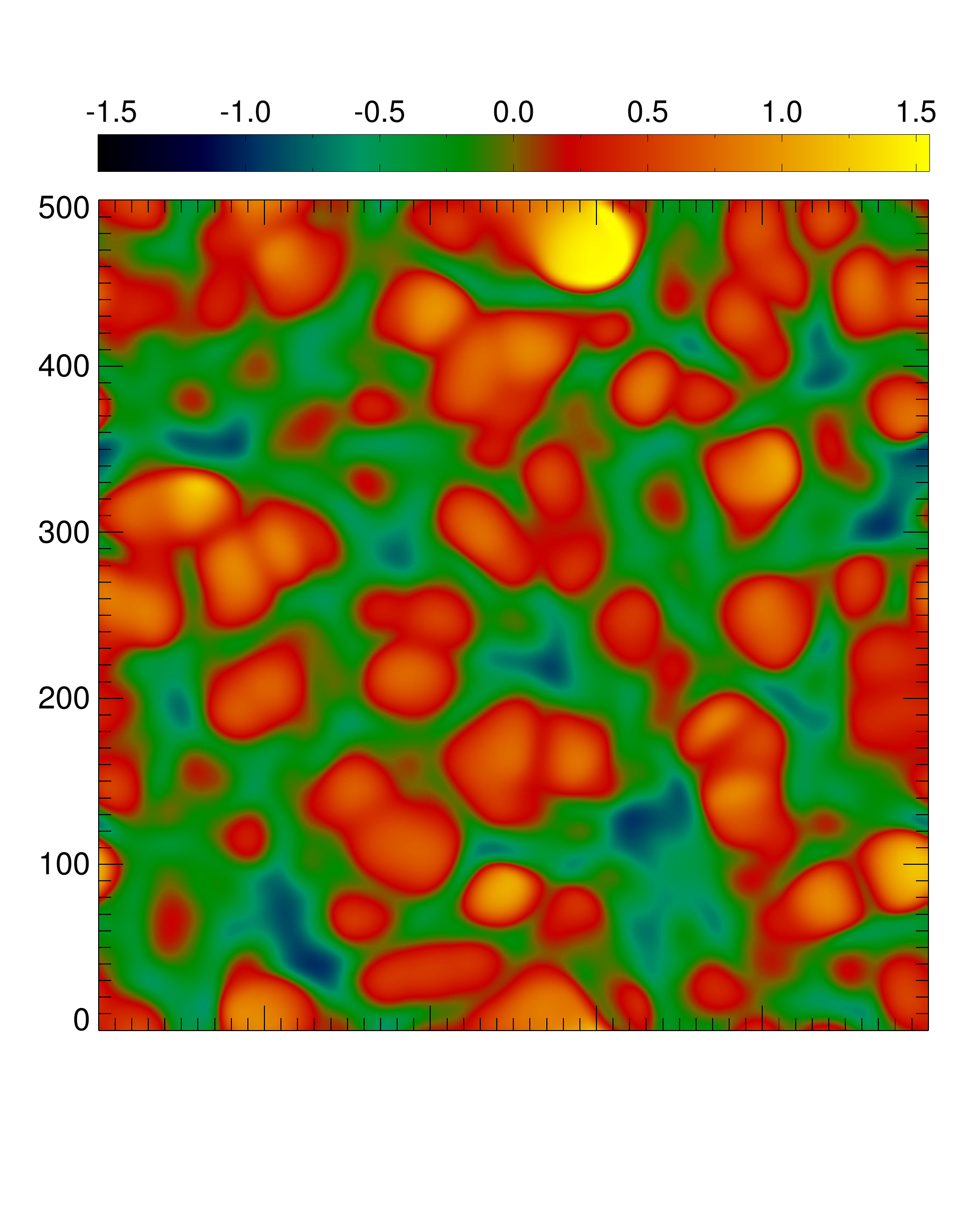}
\put(-175,117){\rotatebox[]{90}{{$Z$ [$h^{-1}\,$Mpc]}}}
\put(-110,210){{$\delta^{\rm L}_{\rm ZELD}(\mbi q,z=0)$}}
\includegraphics[width=6.0cm]{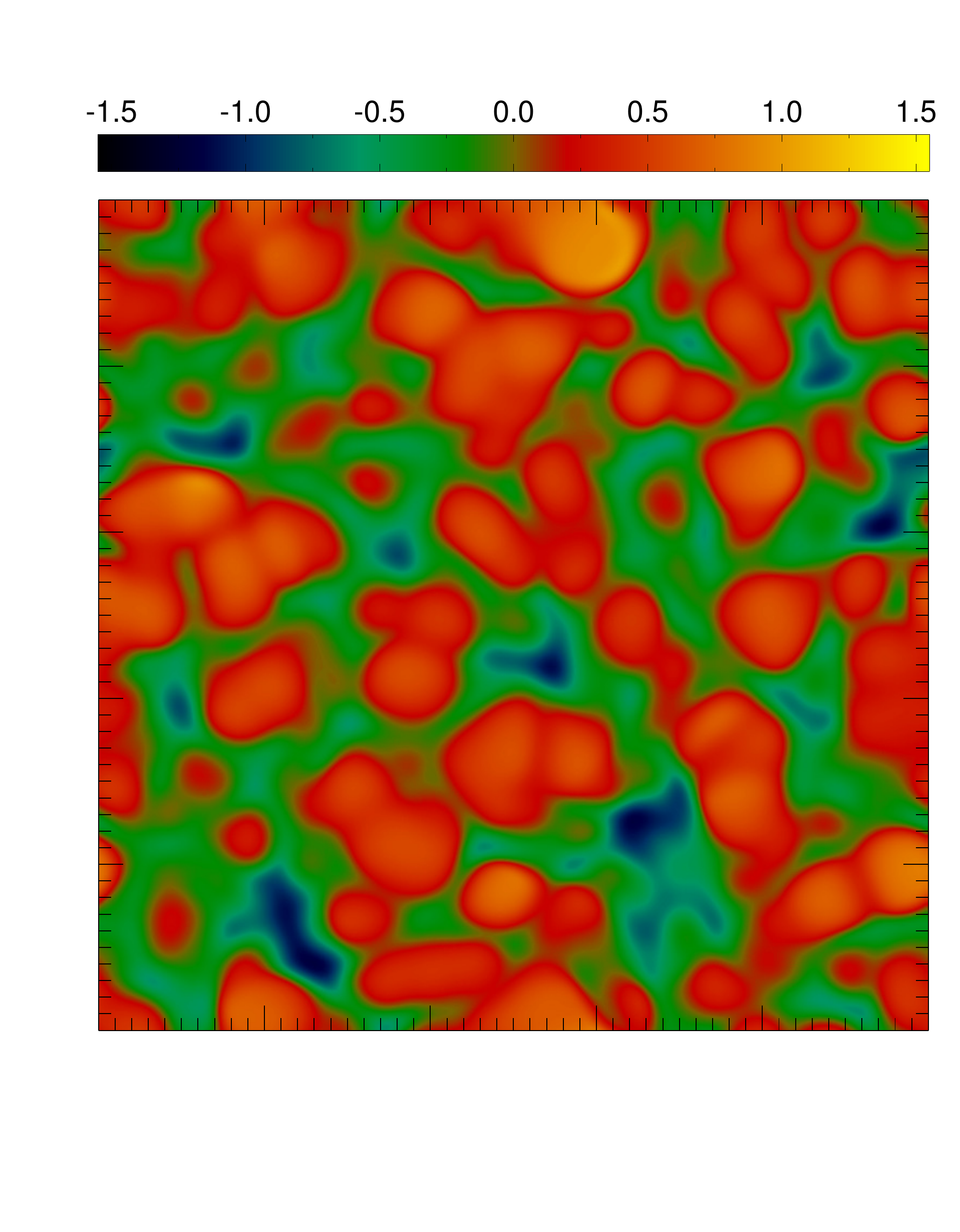}
\put(-110,210){{$\delta^{\rm L}_{\rm GRAM}(\mbi q,z=0)$}}
\includegraphics[width=6.0cm]{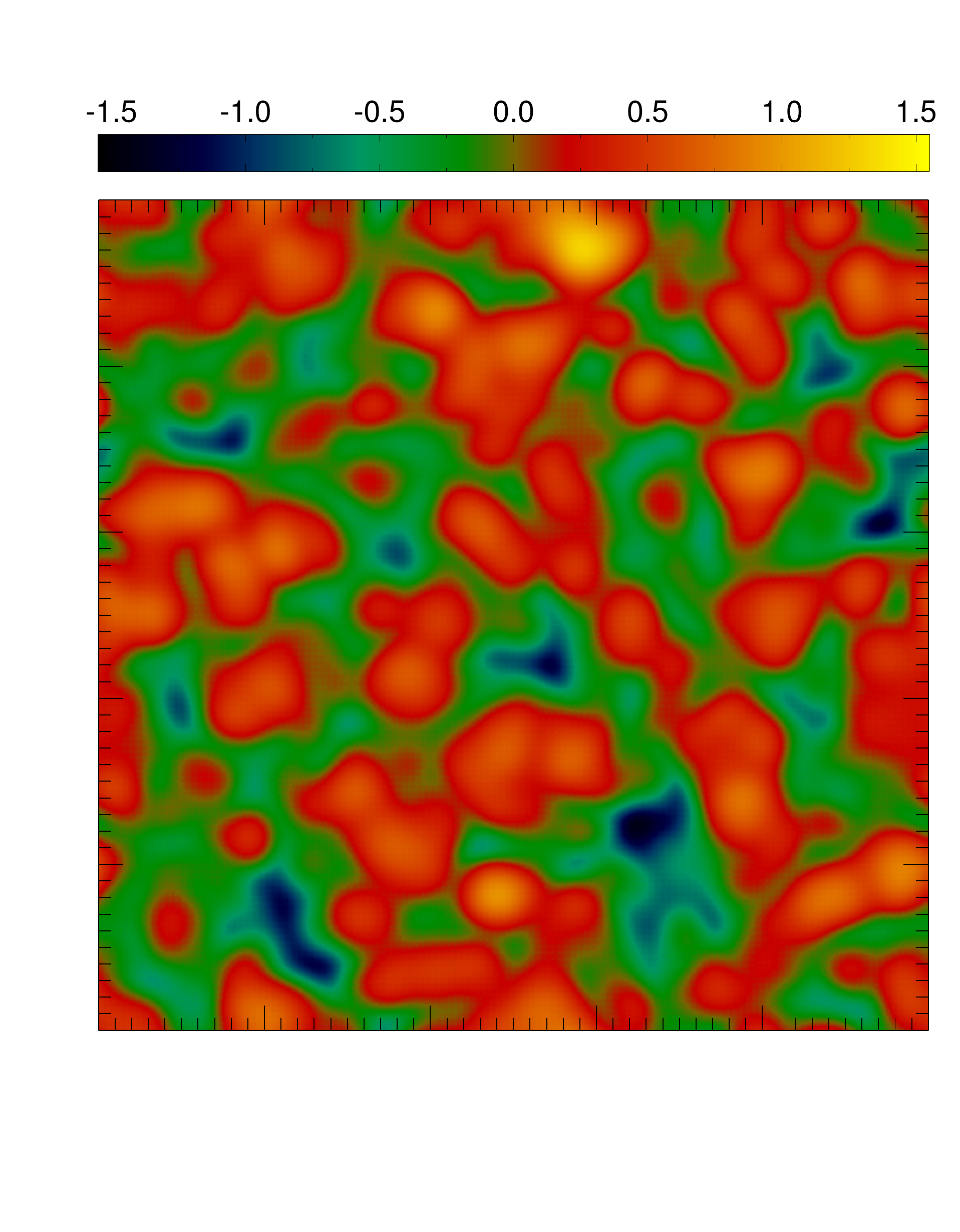}
\put(-110,210){{$\delta^{\rm L}_{\rm ELPT}(\mbi q,z=0)$}}
\vspace{-1.cm}
\\
\includegraphics[width=6.0cm]{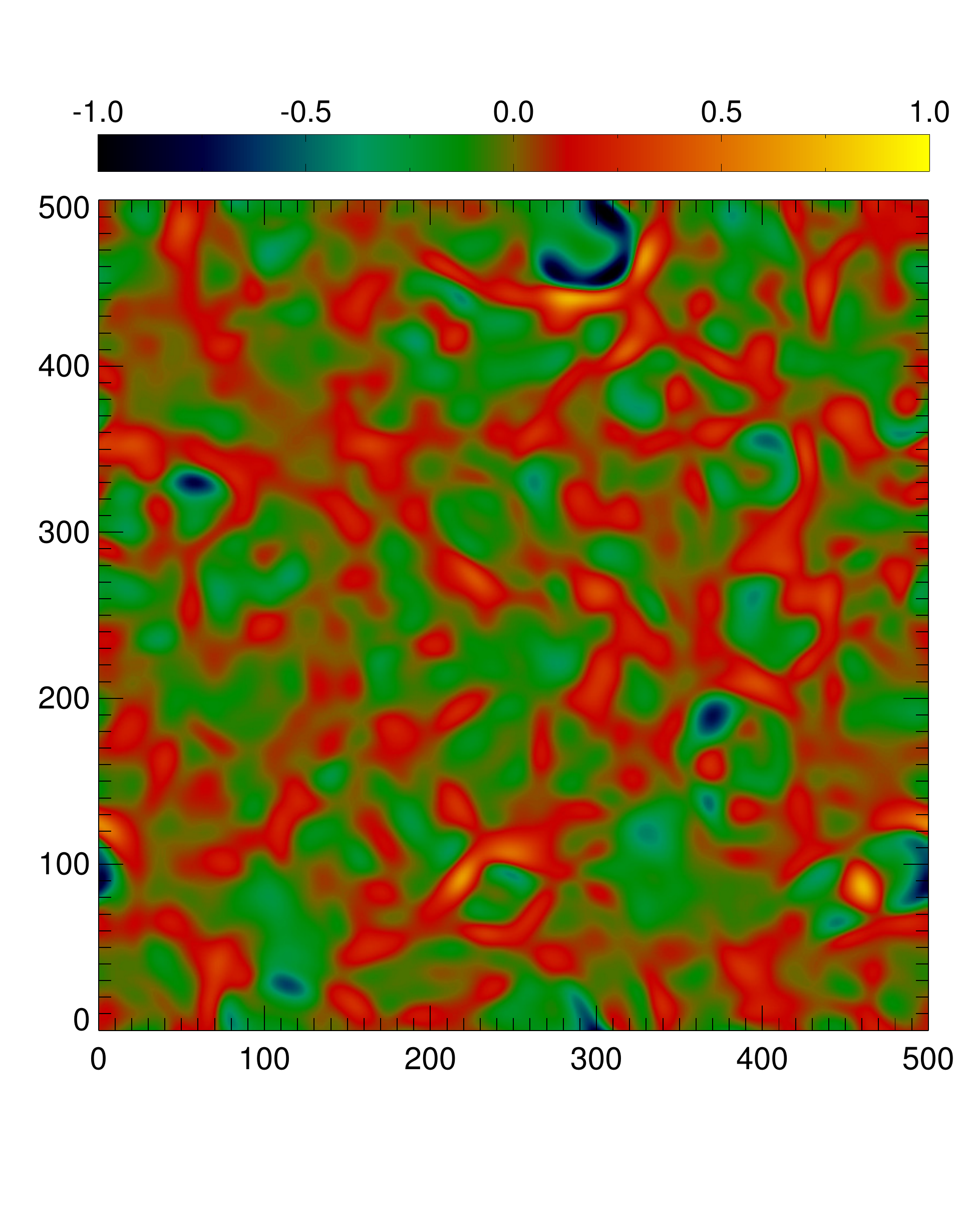}
\put(-175,117){\rotatebox[]{90}{{$Z$ [$h^{-1}\,$Mpc]}}}
\put(-105,15){{$X$ [$h^{-1}\,$Mpc]}}
\put(-140,220){{$\delta_D^{\rm Nbody}(\mbi x=\mbi q,z=127)$}}
\put(-110,210){{$-\delta^{\rm L}_{\rm ZELD}(\mbi q,z=0)$}}
\includegraphics[width=6.0cm]{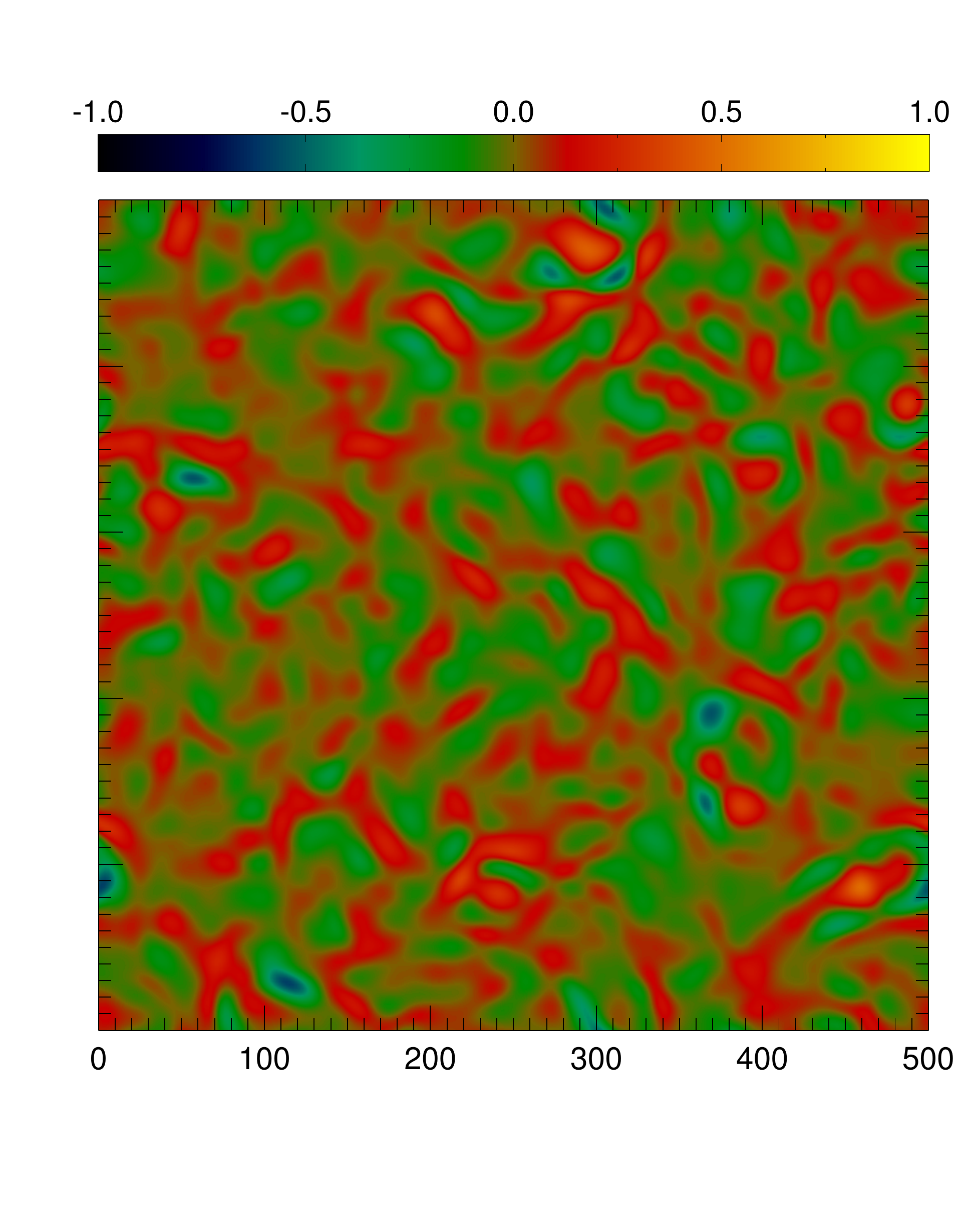}
\put(-105,15){{$X$ [$h^{-1}\,$Mpc]}}
\put(-140,220){{$\delta_D^{\rm Nbody}(\mbi x=\mbi q,z=127)$}}
\put(-110,210){{$-\delta^{\rm L}_{\rm GRAM}(\mbi q,z=0)$}}
\includegraphics[width=6.0cm]{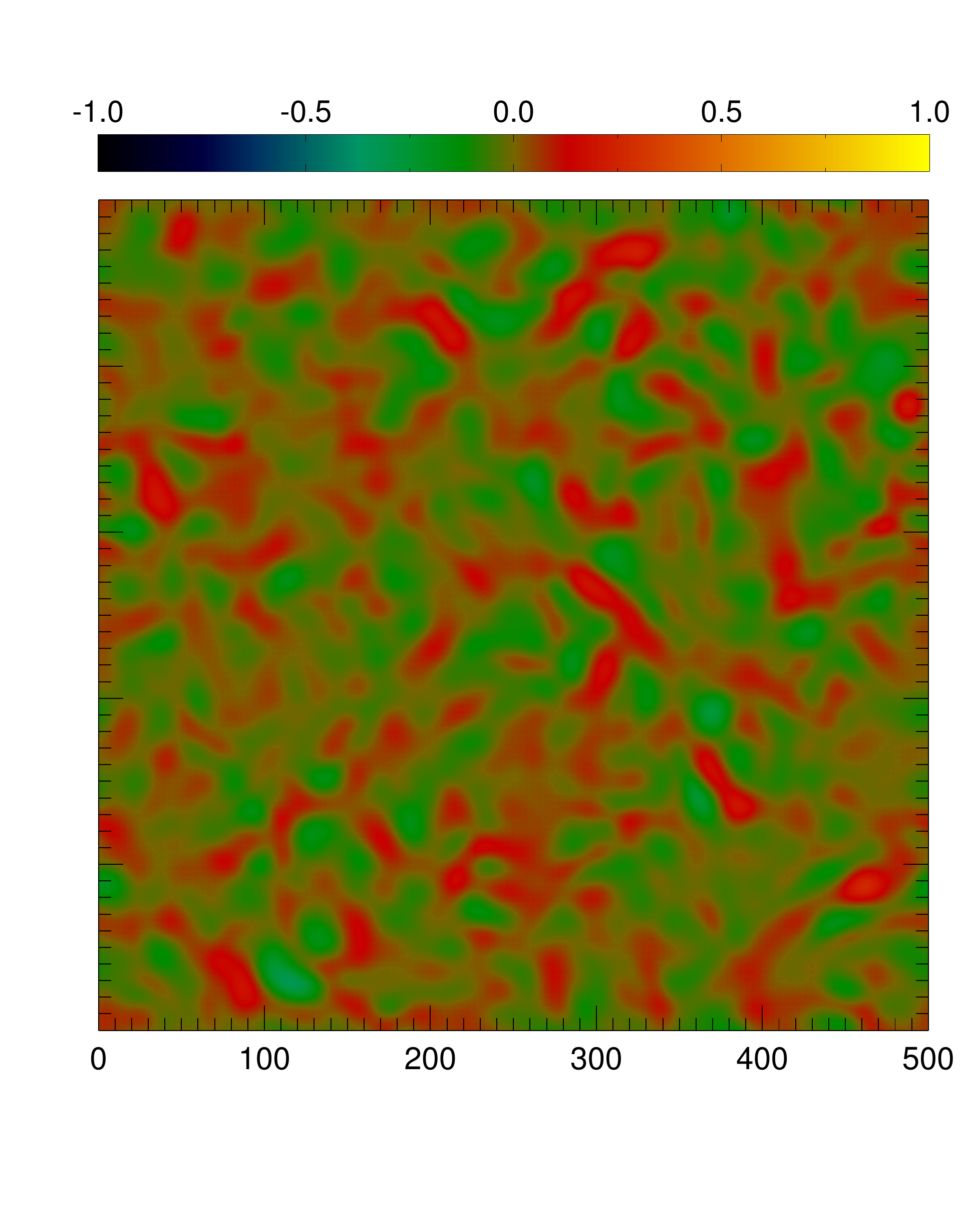}
\put(-105,15){{$X$ [$h^{-1}\,$Mpc]}}
\put(-140,220){{$\delta_D^{\rm Nbody}(\mbi x=\mbi q,z=127)$}}
\put(-110,210){{$-\delta^{\rm L}_{\rm ELPT}(\mbi q,z=0)$}}
\\
\includegraphics[width=5.5cm]{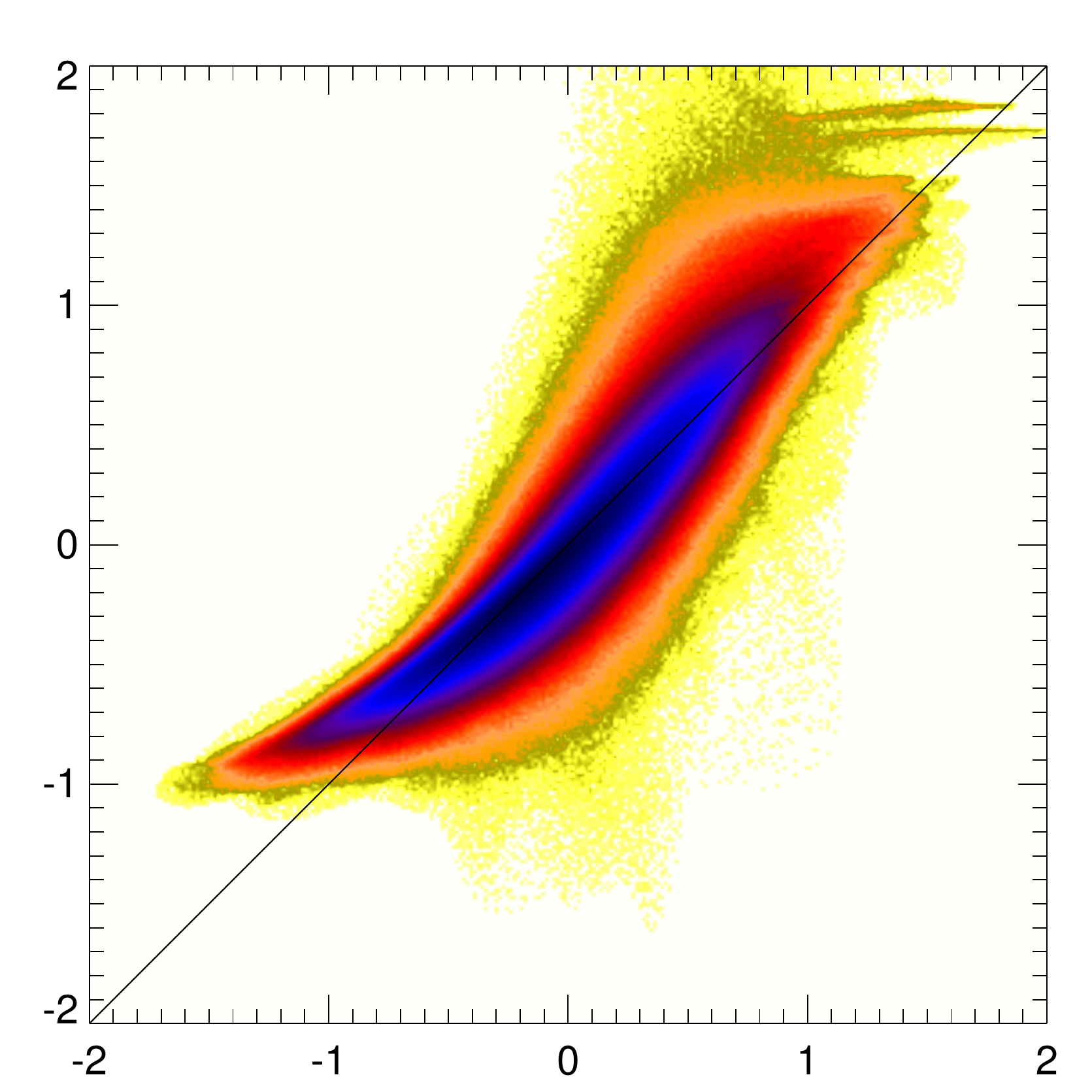}
\put(-160,80){\rotatebox[]{90}{$\delta^{\rm L}_{\rm rec}(\mbi q,z=0)$}}
\put(-100,20){\rotatebox[]{0}{$\delta^{\rm L}_{\rm rec}=\delta^{\rm L}_{\rm ZELD}(\mbi q,z=0)$}}
\put(-135,130){{\large $r_{\rm S}^0$=10 $h^{-1}\,$Mpc}}
\put(-135,-10){{$\delta^{\rm L}(\mbi q,z=0)=\delta_D^{\rm Nbody}(z=127)$}}
\hspace{0.2cm}
\includegraphics[width=5.5cm]{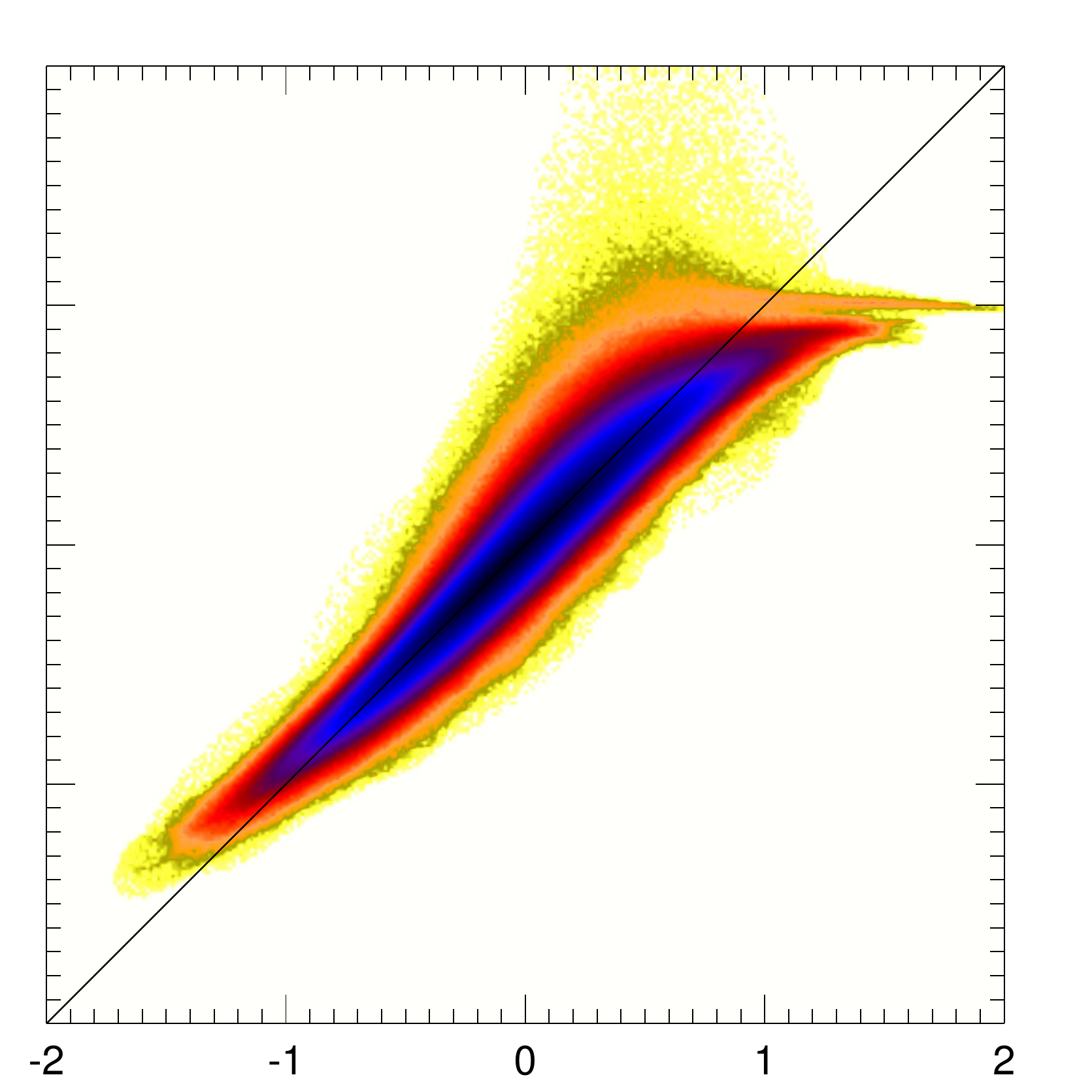}
\put(-110,20){\rotatebox[]{0}{$\delta^{\rm L}_{\rm rec}=\delta^{\rm L}_{\rm GRAM}(\mbi q,z=0)$}}
\put(-140,-10){{$\delta^{\rm L}(\mbi q,z=0)=\delta_D^{\rm Nbody}(z=127)$}}
\put(-137,130){{\large $r_{\rm S}^0$=10 $h^{-1}\,$Mpc}}
\includegraphics[width=5.5cm]{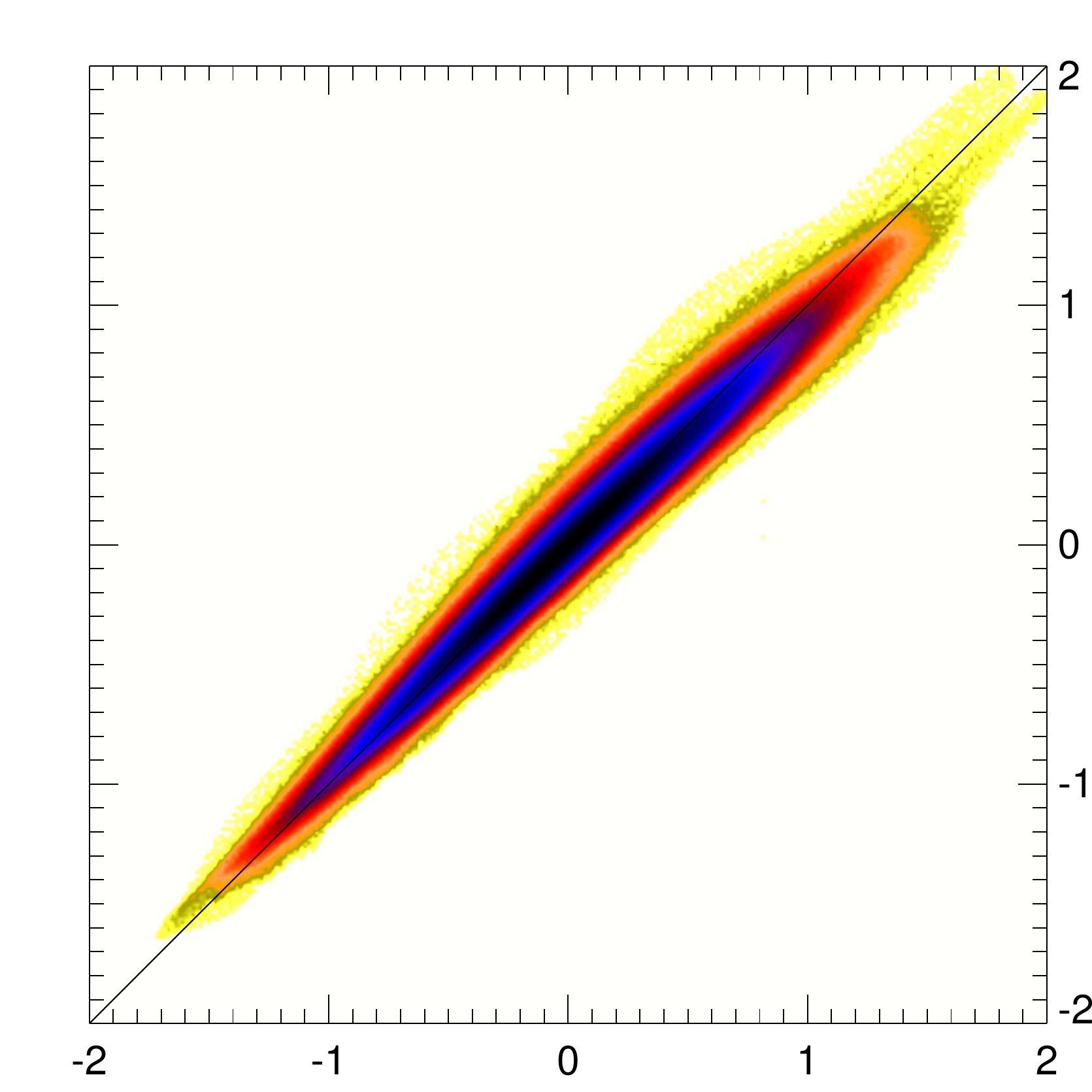}
\put(-100,20){\rotatebox[]{0}{$\delta^{\rm L}_{\rm rec}=\delta^{\rm L}_{\rm ELPT}(\mbi q,z=0)$}}
\put(-140,-10){{$\delta^{\rm L}(\mbi q,z=0)=\delta_D^{\rm Nbody}(z=127)$}}
\put(-135,130){{\large $r_{\rm S}^0$=10 $h^{-1}\,$Mpc}}
\vspace{0.5cm}
\end{tabular}
\caption{\label{fig:ic} Upper panels: slices through the reconstructed initial conditions  after Gaussian smoothing with $r^0_{\rm S}=10$ $h^{-1}\,$Mpc using Left: Zeldovich approximation, Middle: Gramann approximation, Right: this work. Middle panels: difference fields between the reconstruction and the actual initial field. Lower panels: cell-to-cell comparison between the reconstruction and the actual initial field. }
\end{figure*}

\begin{figure*}
\begin{tabular}{cc}
\includegraphics[width=8.cm]{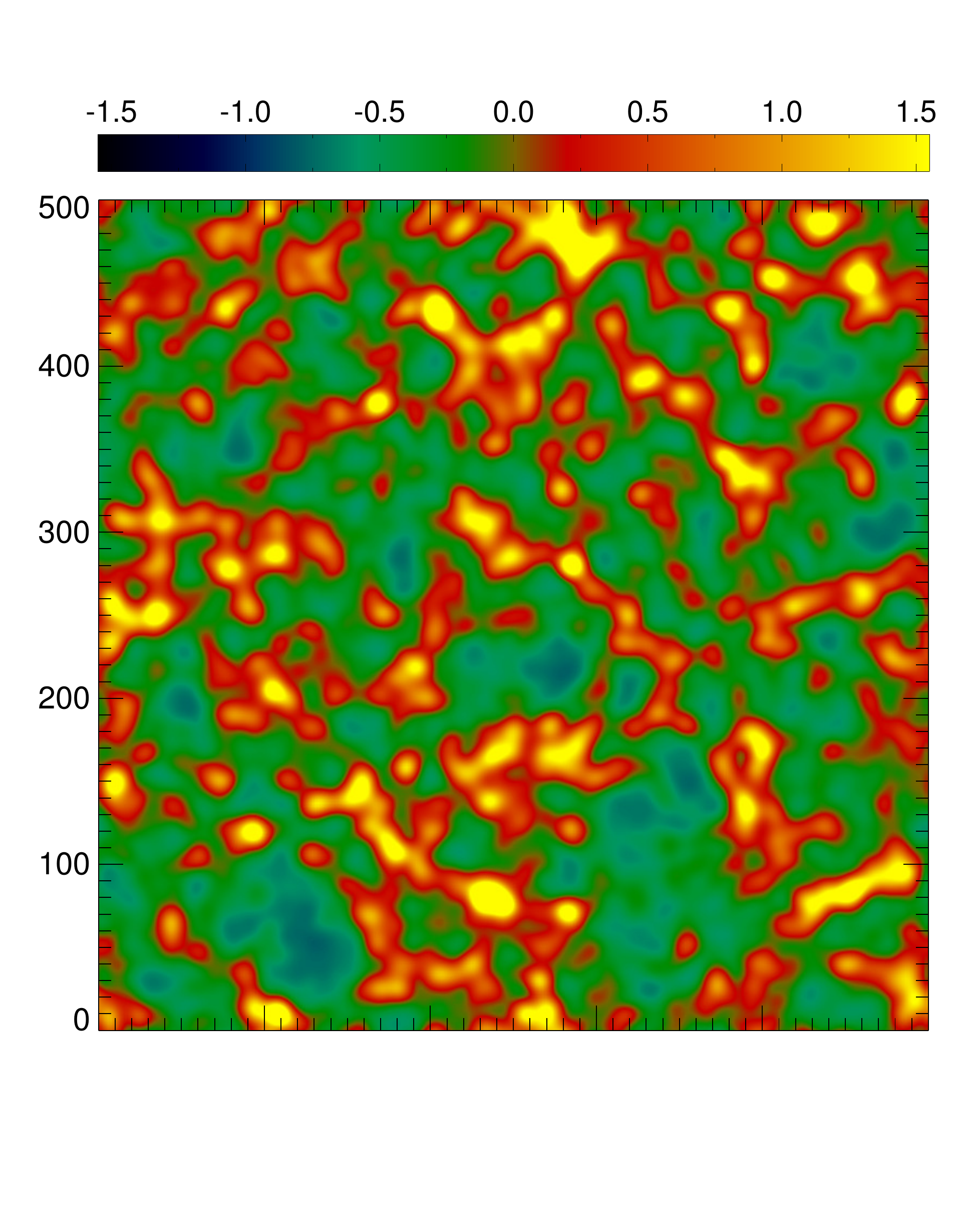}
\put(-230,140){\rotatebox[]{90}{{$Y$ [$h^{-1}$Mpc]}}}
\hspace{-.7cm}
\includegraphics[width=8.cm]{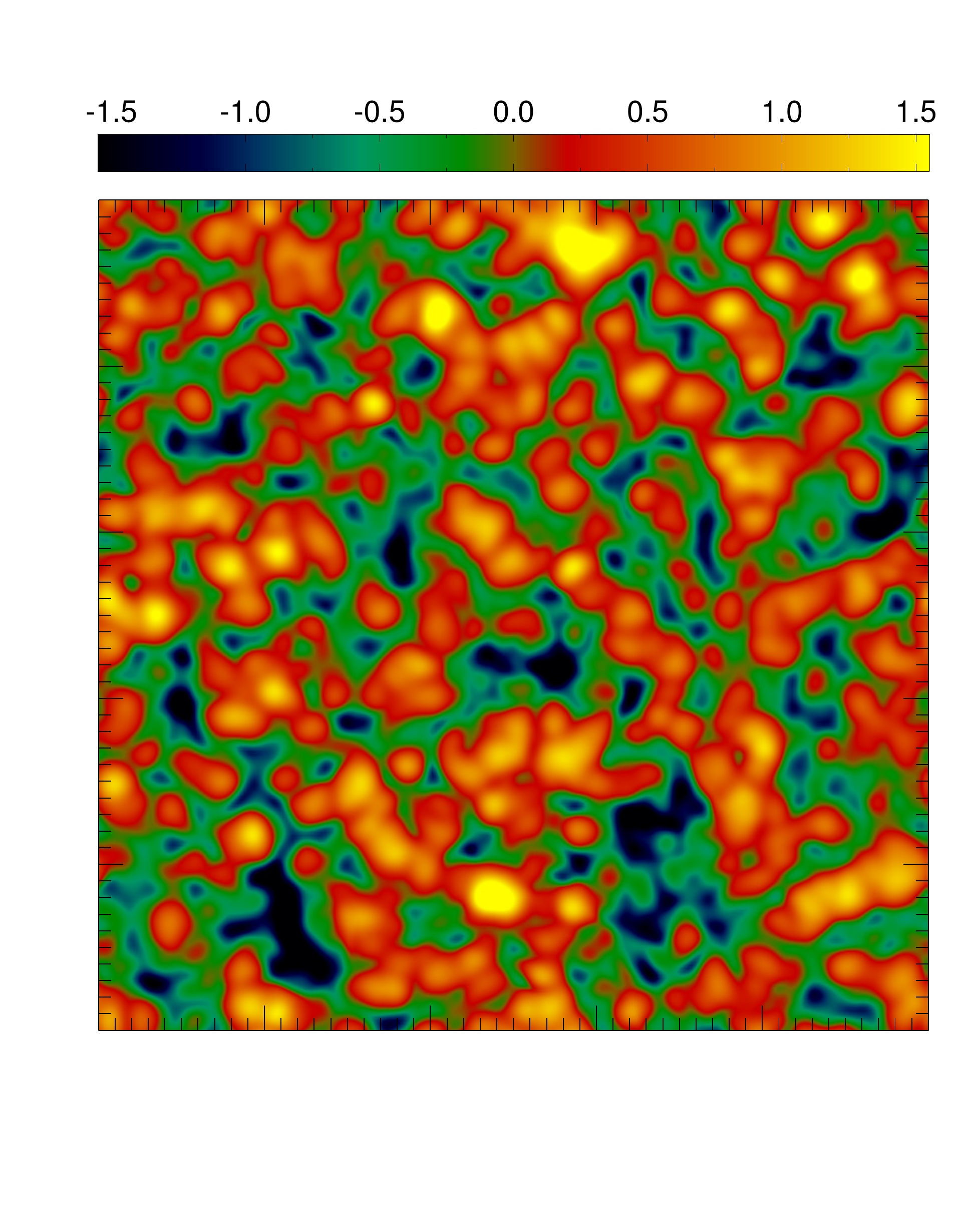}
\vspace{-2.9cm}
\\
\includegraphics[width=8.cm]{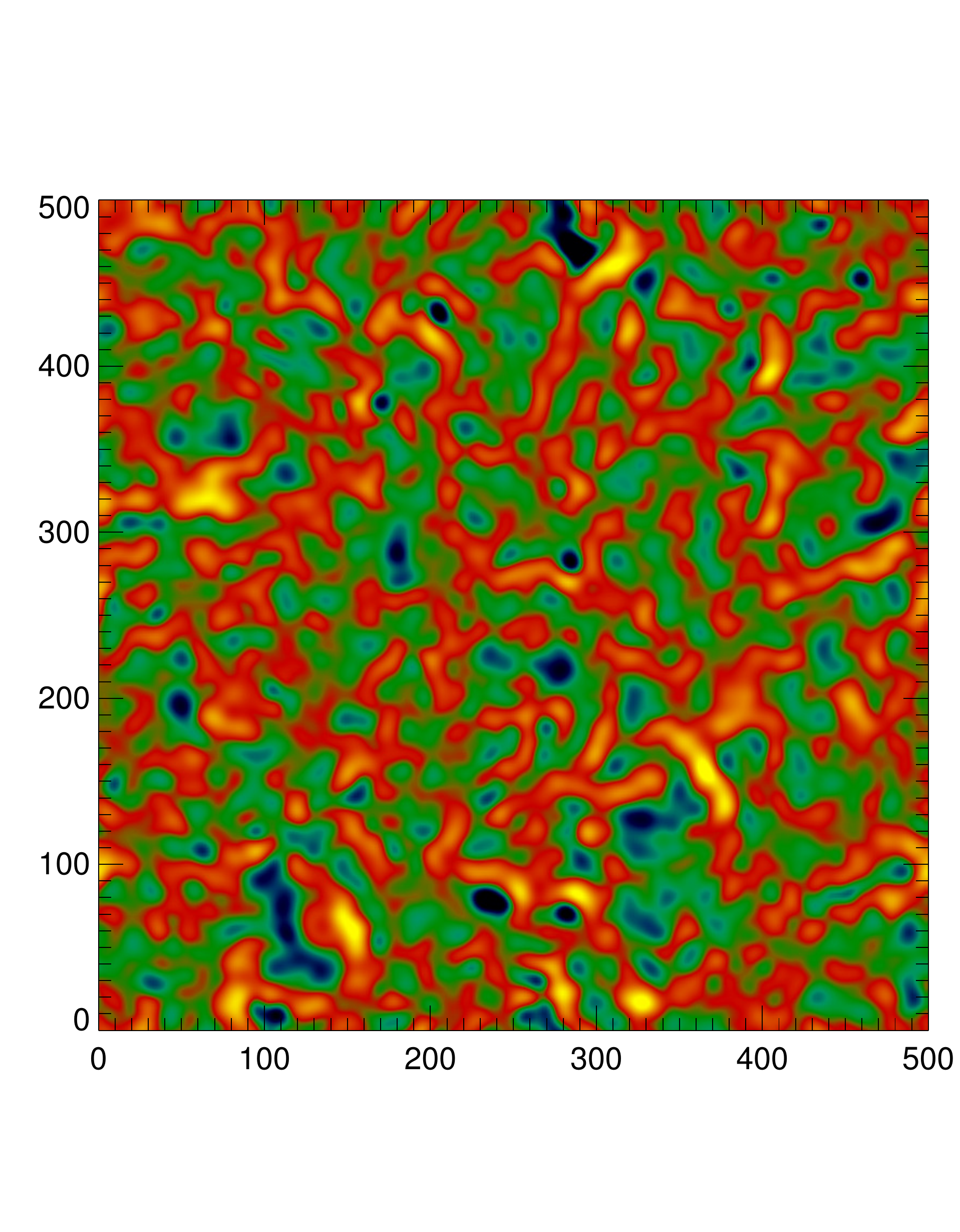}
\put(-230,140){\rotatebox[]{90}{{$Y$ [$h^{-1}$Mpc]}}}
\put(-130,25){{$X$ [$h^{-1}$Mpc]}}
\hspace{-.7cm}
\includegraphics[width=8.cm]{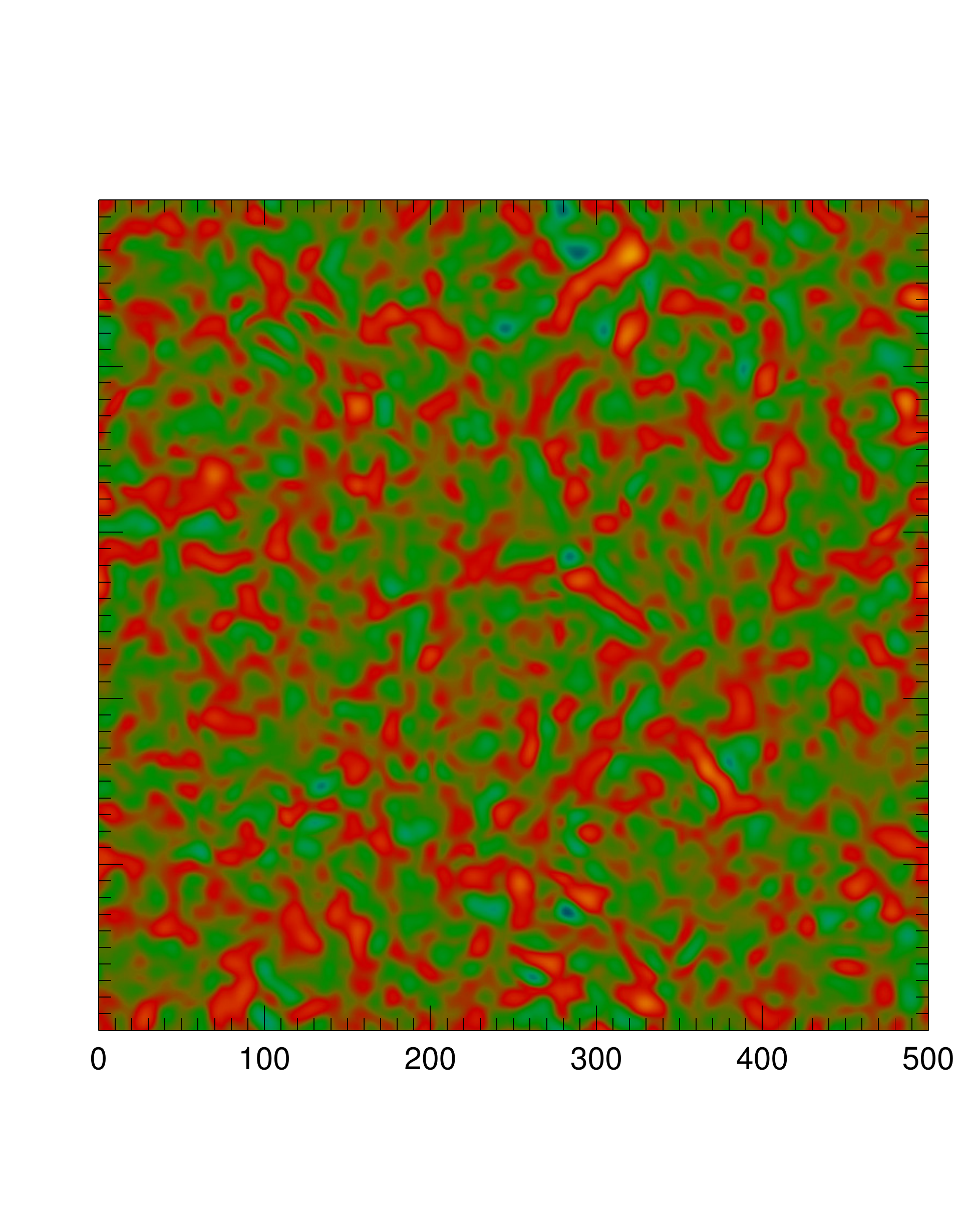}
\put(-130,25){{$X$ [$h^{-1}$Mpc]}}
\end{tabular}
\caption{\label{fig:icr5}Upper left panel: slice through the simulation at $z=0.5$ with Gaussian smoothing of $r_{\rm S}^0$=5 $h^{-1}\,$Mpc. Upper right panel: reconstructed initial condition. Lower panels: difference fields between the corresponding fields in the upper panels and the simulation at $z=127$.}
\end{figure*}

\begin{figure*}
\begin{tabular}{cc}
\includegraphics[width=7.cm]{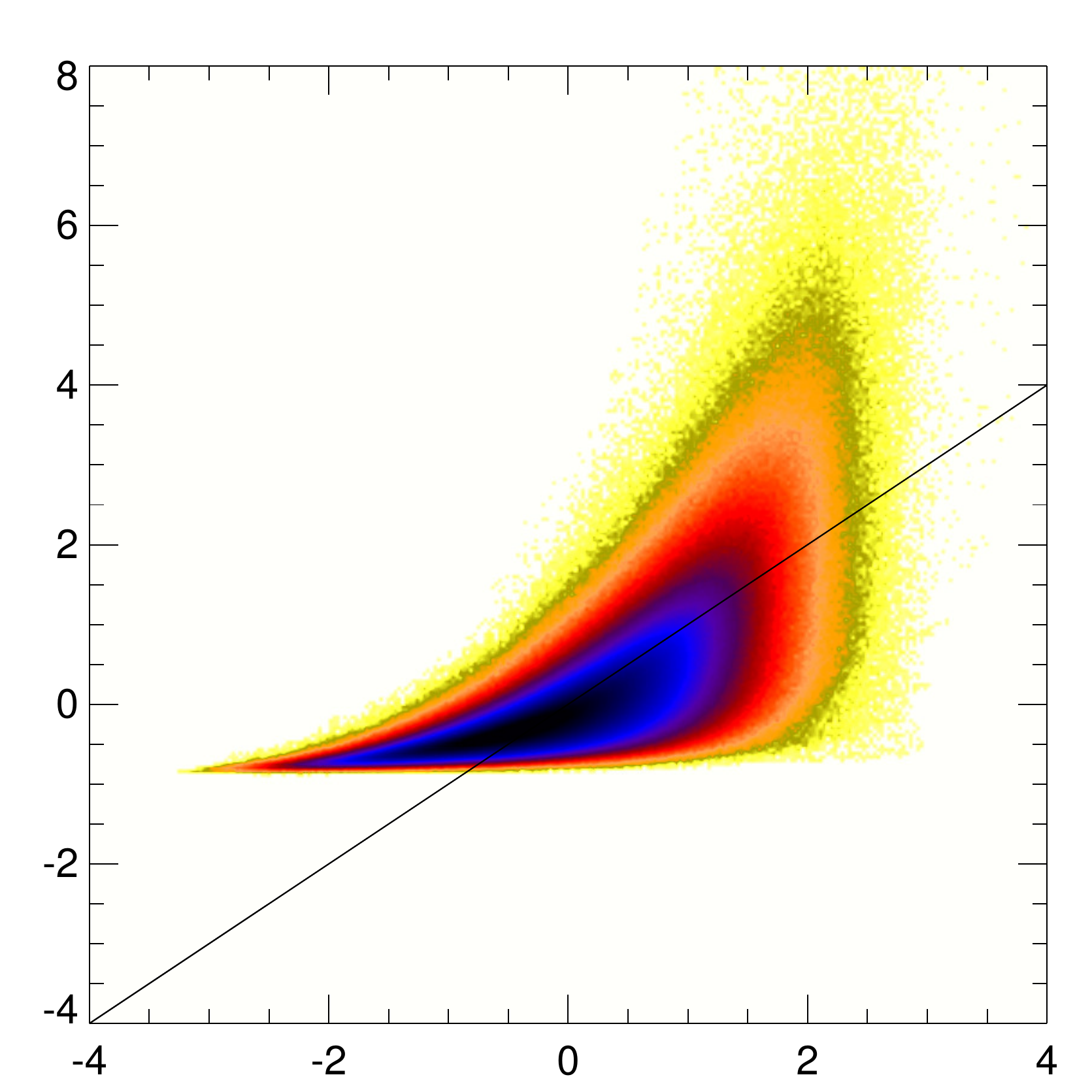}
\put(-210,95){\rotatebox[]{90}{$\delta_D^{\rm Nbody}(\mbi x,z=0.5)$}}
\put(-170,160){{\large $r_{\rm S}^0$=5 $h^{-1}\,$Mpc}}
\put(-150,-10){{$\delta^{\rm L}(\mbi q,z=0)=\delta_D^{\rm Nbody}(z=127)$}}
\hspace{0.cm}
\includegraphics[width=7.cm]{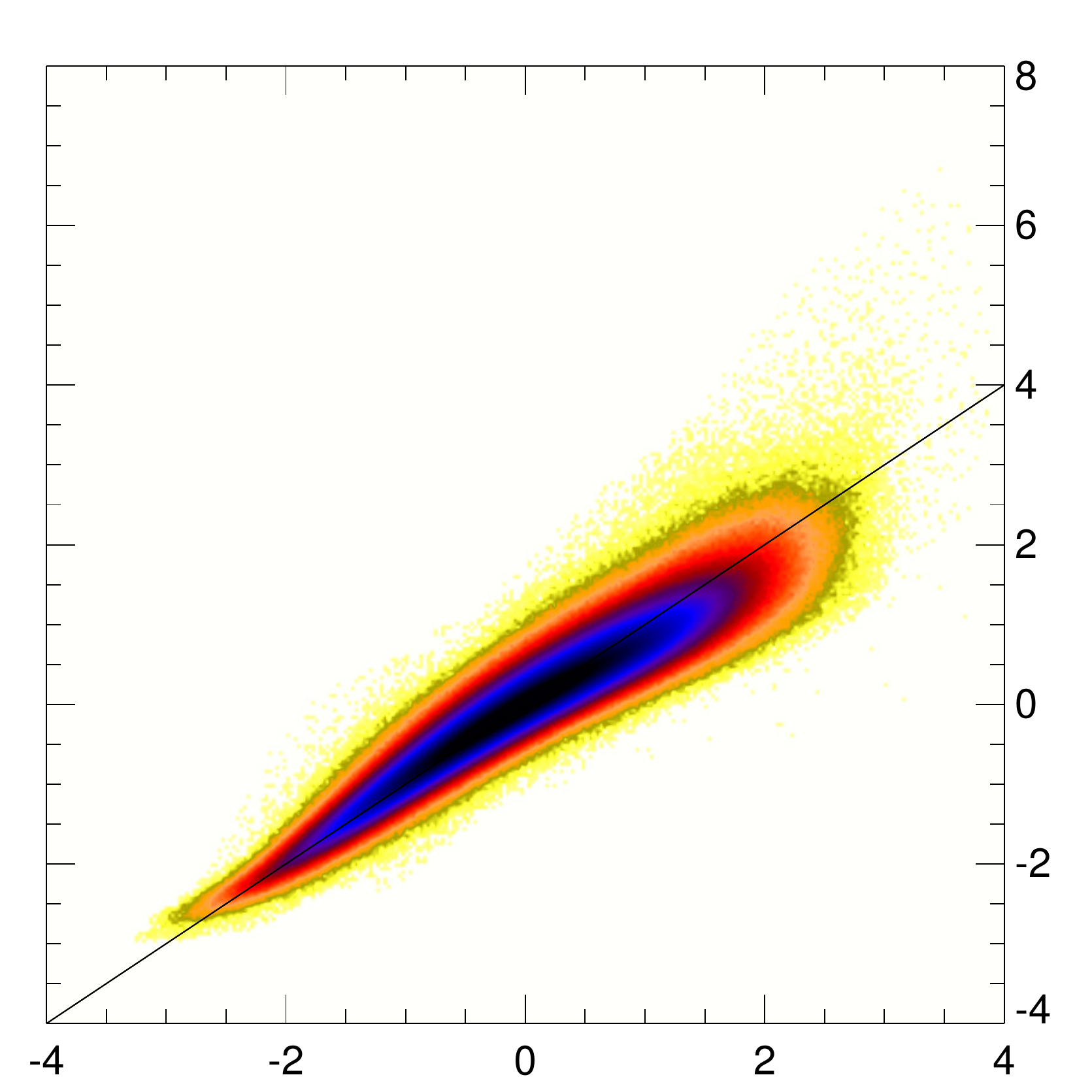}
\put(10,95){\rotatebox[]{-90}{$\delta^{\rm L}_{\rm rec}(\mbi q,z=0)$}}
\put(-150,-10){{$\delta^{\rm L}(\mbi q,z=0)=\delta_D^{\rm Nbody}(z=127)$}}
\put(-170,160){{\large $r_{\rm S}^0$=5 $h^{-1}\,$Mpc}}
\end{tabular}
\caption{\label{fig:c2cr5} Cell-to-cell comparison between the simulation  after Gaussian smoothing with $r^0_{\rm S}=5$ $h^{-1}\,$Mpc at $z=127$ and Left: the simulation  at $z=0.5$, Right:  the reconstruction of the initial field.}
\end{figure*}

\section{Numerical solution schemes}
\label{sec:solution}

Here we present our numerical approach to iteratively solve Eqs.~\ref{eq:expa2}
and \ref{eq:timer}.  Note that the first equation determines the linear
component of the field: $\phi_g(\mbi x,z)\rightarrow\phi^{(1)}(\mbi x,z)$ (the
arrow indicates that $\phi^{(1)}$ is calculated from $\phi_g$) and the second
equation traces that component back in time yielding an estimate of the full
component to an earlier cosmic time: $\phi^{(1)}(\mbi
x,z)\rightarrow\phi_g(\mbi x,z+\Delta z)$ (where $\Delta z<0$ in our case of
study and in this case the arrow indicates that $\phi_g$ at an earlier time is
computed from $\phi^{(1)}$).

\begin{enumerate}

\item $\phi_g(\mbi x,z)\rightarrow\phi^{(1)}(\mbi x,z)$

We propose to solve Eq.~(\ref{eq:main}) iteratively by updating the nonlinear
component which depends on the linear potential $\phi^{(1)}$:

\ba
\lefteqn{ \phi^{(1)}_{i+1}= \phi_g}\\
&&\hspace{-.5cm}+ \tau_{i}\left(\frac{D_2}{D}\phi^{(2)}\left[\phi^{(1)}_i, r_{\rm S}^i\right]-\frac{1}{D}\left(\phi^{(2)}+\phi^{(3)}\right)[\Theta(\phi^{(1)}_{i}), r_{\rm S}^i]\right) \nonumber\,,
\ea

\noindent with $r_{\rm S}^j$ being a scale at iteration $i$ which stabilises
the solution. Here we use a Gaussian filter with decreasing smoothing radii.

\item $\phi^{(1)}(\mbi x,z)\rightarrow\phi_g(\mbi x,z+\Delta z)$
To compute the time-reversal solution we follow \citet[][]{1992ApJ...391..443N,1993ApJ...405..449G} and integrate the equation with finite time differences
\ba
\label{eq:bw}
\lefteqn{ \phi^{j+1}_g(\phi^{(1)}_{j+1})= \phi^{(1)}_{j}}\\
&&\hspace{-0.5cm}-\frac{D_{2j}}{D_j}\phi^{(2)}[\phi^{(1)}_{j}]+\frac{1}{D_j}\left(\phi^{(2)}[\Theta(\phi^{(1)}_{j})]+\phi^{(3)}[\Theta(\phi^{(1)}_{j})]\right)\nonumber\\
&&\hspace{-0.5cm}+ \Delta D_{j}\left( \frac{1}{2}\left(v(\phi^{(1)}_{j})\right)^2+2\phi^{(2)}[\phi_{v}(\phi^{(1)}_{j})]-\frac{1}{D_{j}}\phi_{gv}(\phi^{(1)}_{j})\right.\nonumber\\
&&\hspace{-0.5cm}\left.-\nabla^{-2}\nabla\cdot \left(\delta_{gv}(\phi^{(1)}_{j})v(\phi^{(1)}_{j}) \right)\right)  \nonumber\,.
\ea
\end{enumerate}

We should mention here that more adequate integration solvers are possible
which are mass conserving (solvers for hyperbolic partial differential
equations). However, for the studies we are performing in this work the simple
scheme presented above is adequate.  One can notice that the form of the
continuity equation as given by Eq.~\ref{eq:bw} is time-reversal as it remains
invariant under the transformation: $\phi^{(1)}\rightarrow-\phi^{(1)}$ and
$D\rightarrow-D$.

\section{Results}
\label{sec:results}

We carry out our numerical experiments using the Millennium Run. This simulation
tracks the nonlinear evolution of more than 10 billion particles, in a box of
comoving side-length $500\,h^{-1}{\rm Mpc}$ \citep[][]{Springel-05}. In
particular, we consider the simulation at  different redshifts
($z=0,0.5,1,2,3,127$) gridded with nearest-grid-point (NGP) on a $256^3$ mesh.
To iteratively solve the combined Eulerian-Lagrangian set of equations
described above, we have developed a parallel code that uses Fast Fourier
Transforms to evaluate Laplacian operators and a finite differences method for
divergence operators. We have dubbed this code as \textsc{Kigen}\footnote{KInetic
GENeration of initial conditions (in Japanese: {\it origin}).}.

\subsection{Linearisation}

\label{sec:lin}

First, we show how Eq.~\ref{eq:expa2} can be used to  decompose the full
nonlinear density field into a linear component and a non-linear one.  Since
LPT breaks down when shell-crossing becomes dominant, we have to smooth the
density field to suppress the power on small scales. We apply here a
Gaussian-kernel with different smoothing radii 5 and 10 $h^{-1}$Mpc. We also
note that the operation of convolution does not commute with the linearisation.
Therefore, we need to ensure that this does not seriously affect our results by
comparing with the true linear field, i.~e.~with the initial conditions of the
simulation as we show below.  In the upper panels of Fig.~\ref{fig:dec} we
solve Eq.~\ref{eq:expa2}  forwards given a linear density field taking the
first snapshot of the simulation at $z=127$ which we define as the linear
component in Lagrangian coordinates: $\delta^{\rm L}(\mbi q,z=0)=\delta_D^{\rm
Nbody}(\mbi x=\mbi q,z=127)$ (middle panel) and computing the nonlinear
component shown in the right panel $\delta^{\rm NL}(\mbi q,z=0)$. Adding both
components we get an estimate of the full nonlinear density field at $z=0$:
$\delta(\mbi q,z=0)$. We can see that the nonlinear component is positive both
in the high and the low density regions in such a way that the peaks get more
clustered  as can be seen in the left panel. On the contrary, the voids become
less deep. This effect is only apparent since the linear component has been
multiplied by the relative growth factor as explained above. However, one
should note that all the quantities are in the same (Lagrangian) coordinates.
The panels in the second row of Fig.~\ref{fig:dec} show analogous plots but
starting from the full gravitationally evolved overdensity field at $z=0$ on
the left $\delta(\mbi x,z=0)$. Here the linear and nonlinear components are
computed by numerically solving Eq.~\ref{eq:expa2}, as described in \S3. Both upper and
middle sets of panels look very similar, however, a careful inspection shows
that the structures are shifted. This is more clearly shown in the lower panels
in which the differences between both corresponding panels are shown. The
reason for the shift is that while the upper panels show the different components in
Lagrangian coordinates the middle panels show them in Eulerian ones.

The upper panels in Fig.~\ref{fig:stats} show the decomposition of the fields
into a linear and a nonlinear component in a more quantitative way for two
smoothing scales; 5 and 10 $h^{-1}$Mpc. We show the PDF for the matter in the
simulation at $z=0$ is shown (black line), the corresponding linear (red line)
and nonlinear (blue line) components calculated with LPT (red line) and the
lognormal linearisation (green line). We can see that the linearised fields are
closely Gaussian distributed with low skewness ($S$) and kurtosis ($K$),
whereas the full field, and even more dramatically the nonlinear component,
have considerably large values for $S$ and $K$. A careful inspection of the
plots shows that the nonlinear component does not have a symmetric PDF. This is
better shown in Fig.~\ref{fig:c2c}.  The lower panels show the convergent
behaviour of our numerical scheme, demonstrating its stable approach to the a
solution with progresively smaller skewness and kurtosis.

To further see the effects of the LPT and lognormal linear mappings we compute
the cell-to-cell correlation between the simulation at $z=0$: $\delta_D^{\rm
Nbody}(\mbi x,z=0)$ and the linear component $\delta^{\rm L}(\mbi x,z=0)$. This
can be seen in the upper left panel of Fig.~\ref{fig:c2c}. We find that the
relation between both fields is highly nonlinear and that the lognormal mapping
is in good agreement with the LPT linearisation. However, in the LPT case we
see a scatter showing that the relation is nonlocal. We can see the
non-Gaussian nature of the full nonlinear field in the x-axis, starting with an
overdensity $\delta\approx-1$ and reaching moderately large overdensities
$\delta>6$. The linearised field, shown in the y-axis, presents overdensities
in the range $-2<\delta<2$. The comparison between the simulation at $z=0$:
$\delta_D^{\rm Nbody}(\mbi x,z=0)$ and the simulation at $z=127$:
$\delta_D^{\rm Nbody}(\mbi x=\mbi q,z=127)$ shows a similar relation with a
larger scatter. This is due to the fact that apart from the gravitational
effects described in the upper panels of Fig.~\ref{fig:dec} there is a
transformation from Lagrangian to Eulerian coordinates. It is remarkable how
well the lognormal transformation traces the mean mapping between both fields.
Additionally, the right panels in Fig.~\ref{fig:c2c} show the corresponding
nonlinear components. Here we can see how the nonlinear field gets positive
both in the underdense and in the overdense regions compensating for the
overestimation of the deepness of voids in linear theory and largely increasing
the power in the high density regions. 

\subsection{Evolving the linear component back in time}

The purpose of this section is to show that the linear component  can be
translated from Eulerian to Lagrangian coordinates. We will make such a
demonstration by solving Eq.~\ref{eq:timer} as presented in \S \ref{sec:solution}.
In the numerical experiments of this section we take a starting redshift of
$z=0.5$ which compensates for thte high value of  $\sigma_8$ employed in the
MS \citep{2010MNRAS.405..143A}. 

The results for different redshifts are shown in Fig.~\ref{fig:c2c2}.  The
left panels show the cell-to-cell comparison between the simulation at $z=0.5$:
$\delta_D^{\rm Nbody}(\mbi x,z=0.5)$ and the simulation at different redshifts
$\delta_D^{\rm Nbody}(\mbi x,zj)$ with $zj=1,2,3$. One can see in these plots
how the relation between the fields gets increasingly more biased as expected.
The central panels show the nearly unbiased cell-to-cell correlation between
the simulation at different redshifts  $\delta_D^{\rm Nbody}(\mbi x,zj)$ and
the time-reversal reconstruction of the full nonlinear density field at the
same redshift $\delta_D^{\rm ELPT}(\mbi x,z=z_j)$. This demonstrates the
success in recovering the full nonlinear field at scales of 10 $h^{-1}$Mpc. The
right panels show the linear component $\delta^{\rm L}_{\rm ELPT}(\mbi x,zj)$
and the tight correlation with the actual initial conditions from the
simulation. This correlation becomes larger with increasing smoothing scale. We have
denoted the time-reversal reconstructed fields with the superscript ELPT
standing for Eulerian-Lagrangian perturbation theory due to the combination of
both approaches.  We should note at this point that we have tried the less
time-consuming approach of estimating the linear field at each time-step with
the lognormal approximation. Sorrowfully, systematic errors propagate yielding
significantly poorer solutions. We analyze this issue in more detail in
\citet[][]{kitvel}.

In Fig.~\ref{fig:ic} we show the performance of various grid based methods to
recover the initial conditions including the Eulerian-Zeldovich approximation
on the left \citep[][]{1992ApJ...391..443N}, \citet[][]{1993ApJ...405..449G} in
the middle and the one presented in this work on the right (see \S
\ref{sec:evo} for a derivation of the schemes). Here we integrate the ELPT
equations back in time up to a redshift of $z=10$. For higher redshifts we do
not observe an appreciable shift in the structures, neither an improvement in
the correlation to the initial field. Moreover, getting stable solutions of the
linear component becomes more difficult, since the field we are trying to
linearise is already quite linear.  The bare eye inspection of the plots
already shows that the structures are less smooth, the voids deeper, and the
peaks better confined with increasing order of the continuity equation (see
upper panels).  As a consequence, the difference between the reconstructed
fields and the initial conditions becomes smaller (see middle panels).  From
the comparison between the upper and the middle panels  we can conclude that
the largest differences are in the high density regions. The accuracy of the
reconstruction is assessed in a more quantitative way in the cell-to-cell
correlations (see lower panels). Clearly, the biases present in the
Eulerian-Zeldovich approach are considerably reduced by taking higher order
terms in the continuity equation.

The power of the Eulerian-Lagrangian reconstruction of the initial conditions
is more clearly shown with smaller smoothing scales. Although our approach will
break down at scales in which shell-crossing becomes important (as it is the
case of LPT in general), we find that it is still extremely accurate even for
scales $\gsim$5 $h^{-1}$Mpc \citep[this is further demonstrated in a companion
paper][]{kitvel}. In Fig.~\ref{fig:icr5} we compare a slice through the
simulation at $z=0.5$ with Gaussian smoothing of $r_{\rm S}^0$=5 $h^{-1}\,$Mpc
(upper left panel) with the Eulerian-Lagrangian reconstructed initial field
(upper right panel). There, we can see how the clustered regions move away from
each other and the voids become as strong as the peaks when going back in time
-- a signature of Gaussian fields. The lower panels hints on the correctness of
our approach by showing the difference between the corresponding reconstructed
fields and the simulation at $z=127$. We can now appreciate how dipoles, caused
by the incorrect position of large overdensities, are dramatically reduced with
ELPT. 

In Fig.~\ref{fig:c2cr5} we show the cell-to-cell correlation for the simulation
(after Gaussian smoothing with $r^0_{\rm S}=5$ $h^{-1}\,$Mpc) at $z=127$ and
both the simulation at $z=0.5$ (left panel) and the reconstruction of the
initial field with ELPT (right panel). This quantifies the accuracy of our
time-reversal reconstruction. Here the improvement provided by the ELPT scheme in capturing
the highly nonlinear and nonlocal relation between the initial and final fields
is evident.  The correlation between the reconstructed and the
actual initial field gets significantly tighter and closely unbiased.

 We
define the true linear field $\delta^{\rm LIN}(\mbi k)$ as the one given by the
first snapshot in the $N$-body simulation $\delta^{\rm Nbody}_D(\mbi k,z_{\rm
init})$: $\delta^{\rm LIN}(\mbi k)\equiv \delta^{\rm Nbody}_D(\mbi k,z_{\rm
init})$ ($z_{\rm init}=127$ in the case of the Millennium Run). Accordingly,
the linear power-spectrum is given by:  $P^{\rm LIN}(k)\equiv P^{\rm
Nbody}_D(k,z_{\rm init})$.  We should note that the recovered initial fields
$\hat{\delta}^{\rm rec}_D(\mbi k,z)$ are smoothed. We need
thus to deconvolve the fields to compare them to the unsmoothed linear field
$\delta^{\rm LIN}(\mbi k)$. For this reason we define a spherically averaged
kernel $\mathcal K(k,z)=\sqrt{\frac{P^{\rm LIN}(k)}{P^{\rm rec}_D(k,z)}}$ which
permits us to deconvolve the fields: 

\be
\hat{\delta}^{\rm dec}_{{\rm rec}}(\mbi k,z)=\mathcal K(k,z)\hat{\delta}^{\rm rec}_D(\mbi k,z)\,.
\ee

\noindent We define a normalised cross-correlation between two fields by

\be
G(k,z)=\frac{\langle\hat{\delta}^{\rm dec}_{\rm rec}(\mbi k,z)\overline{\hat{\delta}^{\rm LIN}}(\mbi k)\rangle}{P^{\rm LIN}(k)}=\frac{\langle\hat{\delta}^{\rm rec}_D(\mbi k,z)\overline{\hat{\delta}^{\rm LIN}(\mbi k)}\rangle}{\sqrt{P^{\rm rec}_D(k,z)}\sqrt{P^{\rm LIN}(k)}}\,.
\ee

\begin{figure}
\begin{tabular}{cc}
\includegraphics[width=8.0cm]{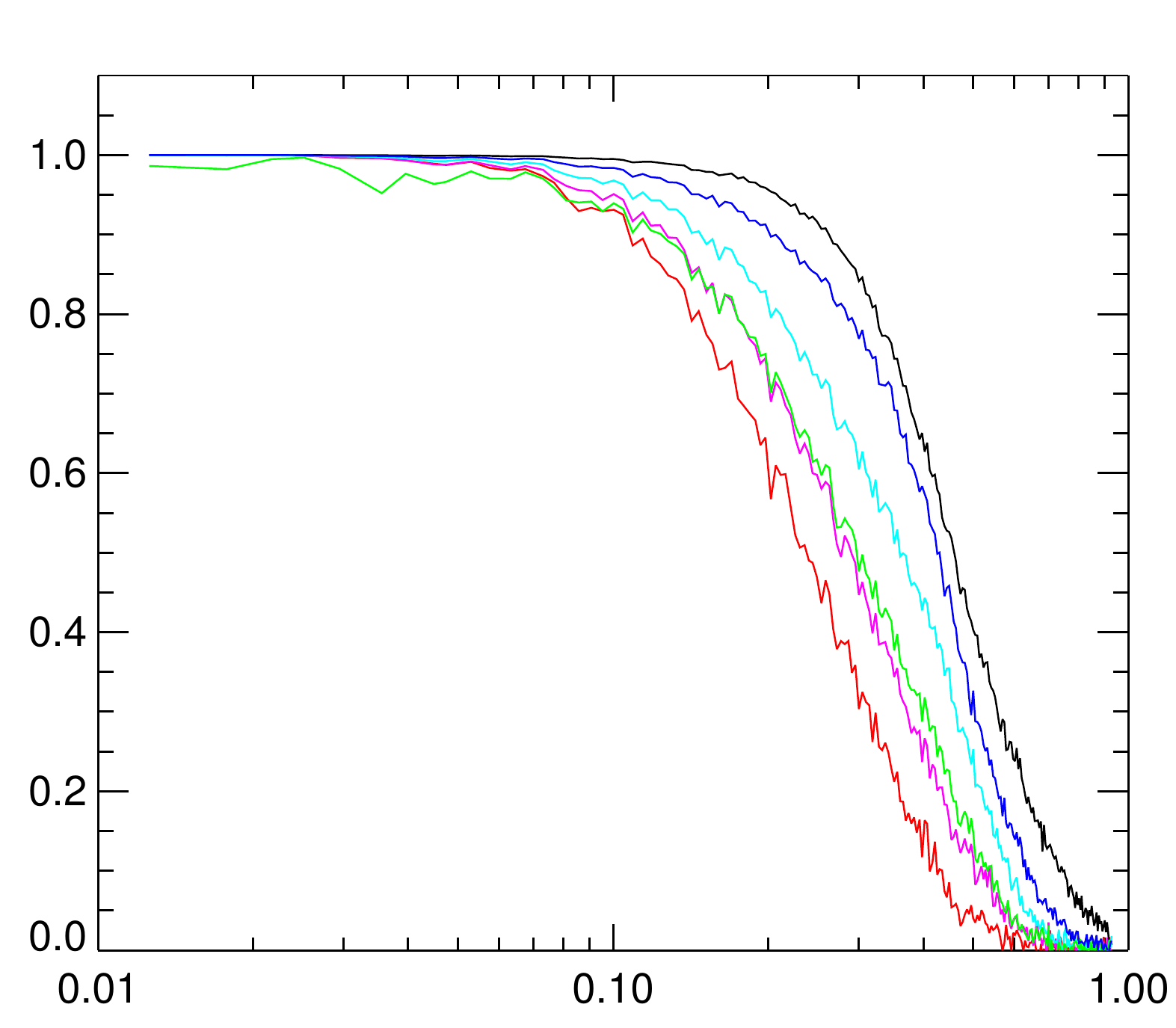}
\put(-130,-5){{$k$ [$h\,$Mpc$^{-1}$]}}
\put(-240,100){\rotatebox[]{90}{{$G(k,z)$}}}
\put(-190,140){\color{black}ELPT z=10}
\put(-190,130){\color{blue}ELPT z=2}
\put(-190,120){\color{cyan}ELPT z=1}
\put(-190,110){\color{magenta}ELPT z=0.5}
\put(-190,100){\color{green}$\ln(1+\delta)-\mu$ z=0.5}
\put(-190,90){\color{red}$N$-body z=0.5}
\end{tabular}
\caption{\label{fig:prop} Normalised cross-correlation between the simulation at redshift $127$ and: the simulation at $z=0.5$ (red curve), the lognormal transformation (green curve) and the ELPT reconstructions at $z=0.5,1,2,10$ (magenta, cyan, blue and black, respectively). }
\end{figure}
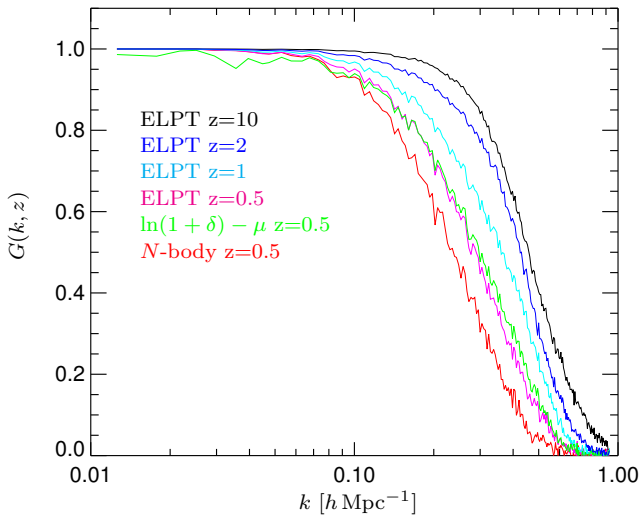

\noindent We compute $G(k,z)$ for the simulation at redshift $z=0.5$ represented by
the red curve in Fig.~\ref{fig:prop}. Any reconstruction should yield larger
values than this curve. We find that the lognormal reconstruction  (shown by the green line)
gains information, however, it introduces  small systematic deviations on large
scales (deviations from the unity at scales $k\lsim0.05$).  A similar result is
obtained by linearising the field at $z=0.5$ with LPT (magenta curve) as
described in \S \ref{sec:solution}. The advantage of this
linearisation is that it does not introduce systematic effects on large scales.
We then trace this field backward in time with ELPT to different redshifts:
$z=1$ (cyan), $z=2$ (blue) and $z=10$ (black).  We can see that there is an
important gain of information between $z=0.5$ and $z=2$. However, this gain
becomes rather moderate when going to even larger redshifts.  We have checked
that the Baryon Acoustic Oscillations are significantly recovered as expected
from the cross-correlation results. 
{\color{black} Here systematic effects due to nonlinear evolution \citep[see][]{Angulo2008}  are corrected by undoing gravitation within 2LPT \citep[the original idea was based on the Zel'dovich approximation, see][]{2007ApJ...664..675E}.}
   We will  present a more detailed study 
 based on a large volume $N$-body simulation in a forthcoming paper.

\section{Conclusions}

In this work we have investigated the linearisation of cosmic density fields
with nonlocal Lagrangian perturbation theory and local rank ordering mapping by
a lognormal transformation. 

Let us summarise the implications of our findings in a series of points:

\begin{enumerate}

\item Linearisation of cosmic density fields generate estimations of the initial
conditions of the Universe in Eulerian coordinates, i.~e.~in the coordinates in
which structures are located at present.

\item The relation between the density field and its linear component is
nonlinear and nonlocal. Local mappings  like the lognormal
transformation  can introduce fluctuations  on
large scales that not present in the original fields.

\item The linear component in Eulerian coordinates can be used to estimate the
peculiar velocity field or the displacement field using Lagrangian perturbation
theory. We further demonstrate this in a companion paper \citep[][]{kitvel}.
Note that the use of the lognormal approximation to obtain an estimate of the
linear displacement field has been investigated in an independent recent work
\citep[][]{2011arXiv1111.4466F}.

\item The linear component is more correlated with the initial conditions than
the full gravitationally evolved density field and has a potential use to
better constrain cosmological parameters. This has already been pointed out by
\citet[][]{2011ApJ...731..116N} for the lognormal case. The LPT linearisation
should be even more accurate as it takes the nonlocal tidal field component
into account. This point remains to be further studied and quantified. 

\item The linear component can be accurately traced back in time on large-scales $\gsim$ 5 $h^{-1}$Mpc  from Eulerian
to Lagrangian coordinates yielding fields that are more correlated with
the initial conditions than the Eulerian representation. This implies that
Eulerian grid-based methods \citep[as opposed to particle based methods,
see][]{Peebles89,NB00,BEN02,2007ApJ...664..675E,LMCTBS08} could be used to
recover Baryon Acoustic Oscillations or other physical signals. 

\item We have demonstrated that one can compute the nonlinear component from
the linear component with LPT. It was shown in \citet[][]{kitaura_lyman} (see
appendix A) how to do that in the lognormal approximation even for the case in
which the mean field is not known. This can be useful for various reasons. It
is easier to obtain estimates of the linear component than of the full
nonlinear density field from observational data. The reason being that modeling
the power-spectrum (or two-point correlation function) in the reconstruction
method is easier than including higher-order correlation functions
\citep[see][]{kitaura_skewlog}. It was demonstrated in
\citet[][]{kitaura_lyman} how to transform the density field into its linear
component to apply a Gaussian prior and determine the power-spectrum
iteratively. One could use the same concept with more complex relations between
the density field and its linear component like the one provided by LPT
discussed  in this work.

In summary, we have shown how to apply higher order Lagrangian perturbation
theory to gravitationally evolved fields and discussed the manifold of applications which can be further developed based on this approach.

\end{enumerate}

\section*{Acknowledgments}
We thank  Adi Nusser and S.~D.~M.~White for encouraging conversations.
Warm thanks to Volker M\"uller for his comments on the manuscript.
We are indebted to the  German Astrophysical Virtual Observatory (GAVO) and the MPA facilities  for providing us the Millennium Run simulation data. 
The work of REA was supported by
Advanced Grant 246797 "GALFORMOD" from the European Research Council.

{\small
\bibliographystyle{mn2e}
\bibliography{lit}
}

\end{document}